\documentclass[%
preprint,
nofootinbib,
 amsmath,amssymb,
 aps,
]{revtex4-1}

\pdfoutput=0

\usepackage{comment}
\usepackage{graphicx}% Include figure files
\usepackage{dcolumn}% Align table columns on decimal point
\usepackage{bm}% bold math
\usepackage{color}
\usepackage{appendix}
\begin{document}

\title{Quantum Monte Carlo Calculations for Carbon Nanotubes}% Force line breaks with \\

\author{Thomas Luu}
\email{t.luu@fz-juelich.de}

\author{Timo A. L\"ahde}
\email{t.laehde@fz-juelich.de}
\affiliation{
 Institute for Advanced Simulation,
 Institut f\"ur Kernphysik, and J\"ulich Center for Hadron Physics,\\
 Forschungszentrum J\"ulich, D--52425 J\"ulich, Germany}

\date{\today}% It is always \today, today,

\begin{abstract}
We show how lattice Quantum Monte Carlo can be applied to the electronic properties of carbon nanotubes in the presence of strong electron-electron correlations.
We employ the path-integral formalism and use methods developed within the lattice QCD community for our numerical work. Our lattice Hamiltonian is closely related to the
hexagonal Hubbard model augmented by a long-range electron-electron interaction. We apply our method to the single-quasiparticle spectrum of the (3,3) armchair nanotube configuration,
and consider the effects of strong electron-electron correlations. Our approach is equally applicable to other nanotubes, as well as to other carbon nanostructures. We benchmark our 
Monte Carlo calculations against the two- and four-site Hubbard models, where a direct numerical solution is feasible.
\end{abstract}

\maketitle

\tableofcontents
\newpage

\section{Introduction\label{sect:intro}}

Carbon nanotubes have proven to be a prime testing ground of our knowledge of quantum 
many-body physics~\cite{nature08918,Charlier:2007zz,PhysRevB.55.R11973,PhysRevLett.101.246802}.  Viewed as ``rolled-up" sheets of its ``parent material'' graphene~\cite{GrapheneFound1,GrapheneFound2}, their electronic properties  are closely related to those of 
graphene~\cite{GeimNovoselovReview,RevModPhys.81.109}, and
depend on how the graphene sheet has been compactified. The allowed momentum modes in a carbon nanotube, for example, are quantized within the two-dimensional Brillouin zone
of the graphene sheet (with appropriate use of zone folding). 
In the absence of electron-electron interactions, graphene exhibits a
linear dispersion in the vicinity of the ``Dirac points'' which are characterized by a Fermi velocity
of $v_F^{} \simeq c/300$, where $c$ is the speed of light in vacuum~\cite{GrapheneDirac1,GrapheneDirac2}. 
Depending on its geometry, a nanotube can also inherit these Dirac points within its dispersion. The remarkable electronic properties of nanotubes, coupled with the 
their excellent mechanical and thermal properties, has spurred interest in using them as a replacement for silicon in future electronic applications.

The low dimensionality of graphene (2D), and particularly nanotubes (quasi-1D), provides a good environment for investigating strong-interaction phenomena.  
For example, the enhanced electron correlation and interaction effects in 1D systems has motivated the Luttinger liquid description of the electronic
ground state of nanotubes, where
the low-energy excitations consist of bosonic waves of charge and spin~\cite{Kane:1997zz,bockrathCobdennature}. In contrast, the 
properties of 3D~metals can often be well described in terms of a Fermi liquid of weakly interacting 
quasiparticles similar to non-interacting electrons. The possibility of an interaction-induced Mott gap at the Dirac points~\cite{Khveshchenko2004323,PhysRevB.81.075429,PhysRevB.59.R10457}, particularly in the case of nanotubes, opens the possibility of using these systems as field-effect transistors.  Many other phenomena due to strong electron-electron correlations in graphene and nanotubes have been predicted~\cite{RevModPhys.84.1067,PhysRevLett.97.146401,PhysRevB.86.115447,PhysRevLett.90.016401,PhysRevLett.93.157402,PhysRevLett.96.196405}. 

Because of electron screening due to underlying substrates and/or surrounding gates, the empirical observation of interaction-driven phenomena in these systems has been surprisingly difficult and for the vast field of applications inspired by these systems, the non-interacting, or tight-binding, picture has proven sufficient.  However, experiments with ``cleaner'' environments 
(\emph{e.g.}\ ``suspended'' graphene) provide a growing body of empirical evidence for strong electron-electron correlations~\cite{BolotinFQHE,EliasVf1,PhysRevLett.110.146802,Yu26022013,BilayerFeldman,Weitz05112010,Mayorov12082011,PhysRevLett.108.076602,VelascoBilayer,PhysRevLett.101.086402} including, to our knowledge, the only tentative evidence for an interaction-induced gap in the absence of an external magnetic field~\cite{PhysRevLett.102.176804}.  In~\cite{Deshpande02012009}, for example, gaps were observed and measured by means of transport spectroscopy in ``ultra-clean'' samples of nanotubes.  Such gaps could not be attributed to curvature effects, and therefore the ground states of nominally metallic carbon nanotubes were identified to be Mott insulators with induced gaps of $10-100$~meV, with the largest diameter tubes exhibiting
the smallest energy gaps.  Just as interesting, bound ``trions'' were observed in doped nanotubes in~\cite{PhysRevLett.106.037404}.  In all these cases, the non-perturbative effects of electron-electron correlations cannot be ignored and, at the very least, must be placed on equal footing with other electronic couplings~\cite{nature2007}.

Monte Carlo methods are 
well suited for strongly interacting quantum mechanical many-body problems, as exemplified by lattice Quantum Chromodynamics (LQCD). The great
advantage offered by the Monte Carlo treatment of the path-integral formalism is that quantum mechanical and thermal fluctuations are fully accounted for, without the need for uncontrolled or 
{\it ad hoc} approximations. For a given Lagrangian or Hamiltonian theory, the Monte Carlo results are regarded as fully {\it ab initio}. The systematical errors in any 
such calculation are due to discretization (non-zero spatial or temporal lattice spacing) or finite volume effects (when studying an infinite system). These errors can be systematically reduced by
use of multiple lattice spacings and volumes, and by means of ``improved'' lattice operators.  Monte Carlo methods have been applied to graphene, using either 
 a ``quasi-relativistic'' low-energy theory of Dirac fermions  valid 
near the Dirac K~points for monolayer~\cite{Drut:2008rg,Drut:2009aj,Drut:2009zi,Hands:2008id,Armour:2009vj,Armour:2011hf} and bilayer~\cite{Armour:2013yk,Armour:2015mta} systems, 
or applied directly to ``tight-binding'' models formulated
on the physical, underlying honeycomb lattice of graphene, supplemented by a long-range Coulomb interaction which may or may not be screened at short 
distances~\cite{PhysRevLett.106.236805,Ulybyshev:2013swa,Smith:2014tha}.
The former approach is
attractive in the sense of being independent of the details of the tight-binding approximation, while the latter appears more amenable to connect with the framework of applied graphene research,
and is furthermore closely related to the hexagonal Hubbard model, of which many lattice Monte Carlo studies 
exist~\cite{PhysRevB.72.085123,MengGraphene,SorellaGraphene,PhysRevLett.109.126402}.
Notably, in the absence of short-range screening of the Coulomb interaction,
both methods predict the opening of a mass gap around a graphene fine-structure constant 
of $\alpha_g \simeq 300\alpha \simeq 1$, which may be attainable in 
suspended graphene, unaffected by a supporting dielectric substrate.
The AC and DC conductivities of graphene~\cite{Buividovich:2012uk,Buividovich:2012nx}, its dispersion relation~\cite{Buividovich:2013ak}, and the effects of an external 
magnetic field and strain~\cite{PhysRevB.89.245404,PhysRevLett.114.246801,PhysRevLett.115.186602} have also been studied using lattice Monte Carlo methods.

As opposed to graphene, where a gap opens if the coupling $\alpha_g$ is above some critical value, for 1D nanotubes it is expected that a gap is induced for any positive value of the coupling (at half-filling).  Therefore a non-perturbative Monte Carlo method for nanotubes is quite appropriate. 
This motivates our introduction of the Monte Carlo method for carbon nanotubes, where we consider (in this paper) the spectrum of a single quasiparticle.
While our method is completely applicable to any nanotube configuration (and in
principle to other carbon nanostructures as well), we benchmark it for the ``(3,3) armchair'' tube, which does not exhibit an energy gap in the non-interacting limit. We model our electron-electron interaction by the screened Coulomb interaction of 
Wehling {\it et al.}~\cite{PhysRevLett.106.236805}, although we emphasize that
a wide variety of other choices are feasible, including a pure contact interaction and an unscreened Coulomb interaction. 
In previous Monte Carlo calculations of graphene, the existence of an interaction-induced gap
was probed by means of a condensate $\langle \bar{\psi}\psi\rangle$ (see, e.g., Refs.~\cite{Ulybyshev:2013swa,Smith:2014tha}) or some equivalent order parameter. Here, we show how lattice QCD methods can be used to directly compute the dispersion
relation at the K (or Dirac) point. 
Furthermore, we compute the dispersion relation at \emph{all allowed} momenta points in
the first Brillouin zone which include, for instance, the high-symmetry $\Gamma$ and $M$ points.  To our knowledge this has not been attempted before using lattice Monte Carlo methods within both condensed matter physics and lattice QCD. 

Previous studies of carbon nanotubes using Density Functional Theory (DFT)~\cite{PhysRevLett.72.1878,PhysRevB.67.121406,Barone2004} have shown that curvature can significantly distort the band structure of small-radius carbon nanotubes, including changing the electronic properties from semiconducting to insulating and vice versa (for a recent review, see~\cite{Barone2012}). Such effects become significant for tube radii $<$ 10 Angstroms, which includes the (3,3) nanotube we consider here. However, for armchair nanotube configurations, their symmetry protects them from developing a bandgap due to curvature effects. Moreover, our ultimate objective is to describe large-radius nanotubes, where a Mott insulating state has been experimentally identified~\cite{Deshpande02012009}. Nevertheless, we discuss how curvature effects can be included into the tight-binding Hamiltonian.

%As is well documented, the effects of curvature can also induce band gaps~\cite{Charlier:2007zz}, particularly for small diameter tubes like the one we investigate.   As we do not include the effects of curvature within our calculations, this may call into question the validity of our gap calculations.  However, for armchair configurations, regardless of their tube diameter, their symmetry protects them from inducing a band gap due to curvature alone~\cite{Charlier:2007zz}.  Despite not including curvature effects, we comment in Section~\ref{sect:PI} on how one can generalize our method to include curvature effects. }

The rest of our paper is structured as follows: In Section~\ref{sect:tube description}, we summarize the mathematical description of a nanotube, with emphasis on the (3,3) armchair configuration.
The path-integral formalism for nanotubes is given in Section~\ref{sect:PI}, along with the lattice formulation that we use for our Monte Carlo calculations.  In Section~\ref{sect:NI} we discuss the non-interacting (tight-binding) solution in the context of our path-integral formalism, and its various approximations in discretized form.  Section~\ref{sect:interactions} provides details of our implementation of the long-ranged Coulomb interactions, and the consequences of using this interaction within small dimensions.  It also describes the momentum projection method that we use to extract the dispersion energies. As this paper serves as an initial description of the Monte Carlo method applied to nanotubes, we invest significant time in its description in Sections~\ref{sect:PI},~\ref{sect:NI}, and~\ref{sect:interactions}.  The reader interested instead in the results could skip to Section~\ref{sect:results}, where we present our results for the dispersion relation of the (3,3) armchair nanotube.  Here we also discuss our analysis techniques and demonstrate in detail our continuum-limit and infinite-volume extrapolations for the Dirac point energy. We conclude with a recapitulation of our methods and results, and comment on possible future applications.  We provide benchmark results of our code in Appendix~\ref{sect:benchmark}.

\section{Nanotube Geometry \label{sect:tube description}}

\begin{figure}
\includegraphics[width=.45\columnwidth]{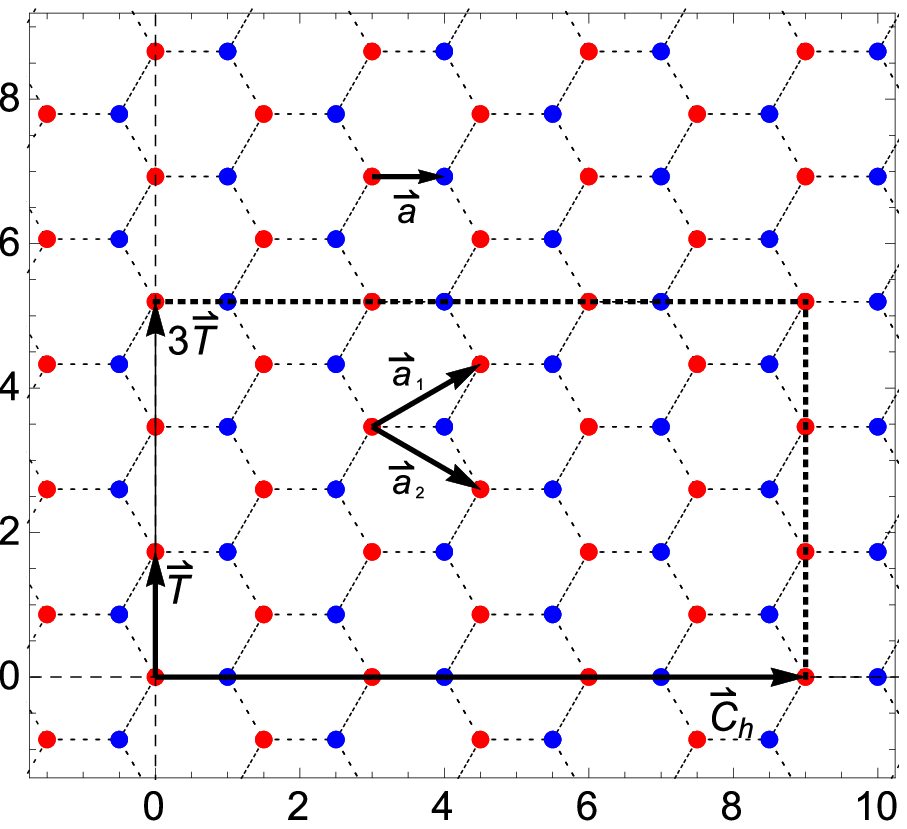}\includegraphics[width=.35\columnwidth]{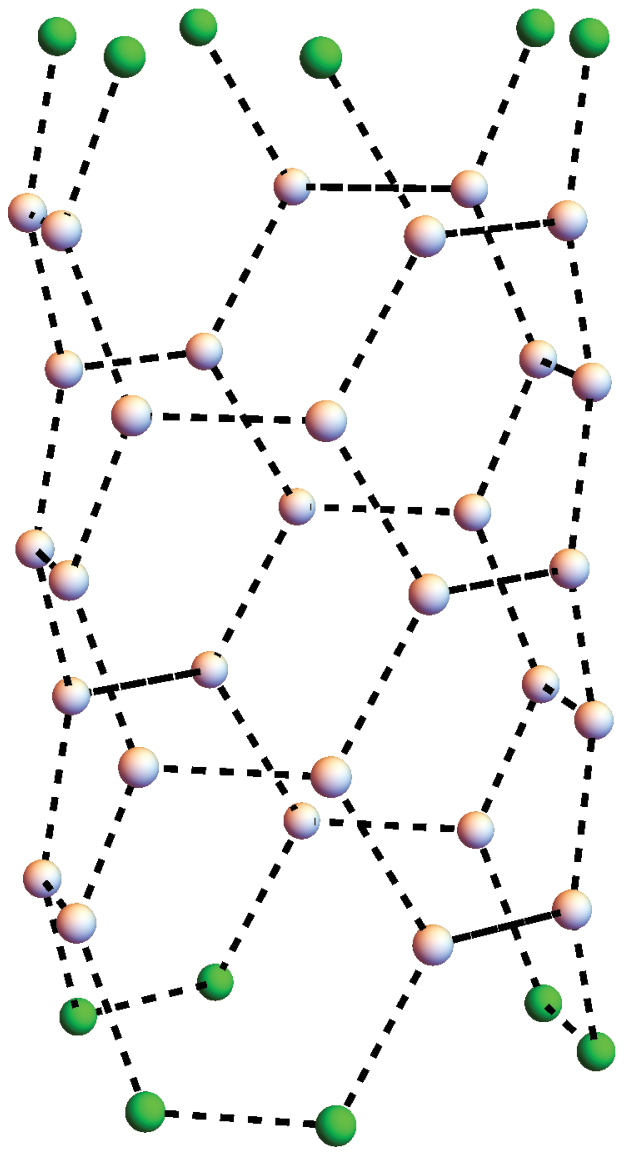}
\caption{Construction of nanotubes from a planar hexagonal lattice. The left panel shows the hexagonal lattice from which the tubes are formed.
The vectors $\vec{T}$ and $\vec{C}_h$ are shown for the $(3,3)$ chirality. Also shown are the hexagonal unit vectors $\vec{a}_1$ and $\vec{a}_2$. The hexagonal lattice can be described in 
terms of two triangular lattices~A~and~B (colored red and blue, respectively) shifted by the vector $\vec{a}$. The rectangle defined by the vectors 
$\vec{C}_h$ and $3\vec{T}$ can be cut and rolled along the longitudinal direction to form a 
nanotube with $N_L=3$, shown in the right panel. The ends of the tube are identified, due to the 
periodic boundary conditions applied in the longitudinal direction.
\label{fig:33 tube}}
\end{figure}

We shall first review the construction of nanotubes from a planar hexagonal lattice, with emphasis on the ``(3,3) nanotube'' which we shall later use in our lattice Monte Carlo calculations.
The geometry of the (3,3) nanotube can be obtained by first considering a planar graphene (honeycomb) lattice, shown in the left panel of Fig.~\ref{fig:33 tube}.  
Each point on the graphene lattice can be obtained by integer combinations of the unit vectors
\begin{eqnarray}
\vec{a}_1&\equiv&\left(\frac{3}{2},\frac{\sqrt{3}}{2}\right)a\ ,\\
\vec{a}_2&\equiv&\left(\frac{3}{2},-\frac{\sqrt{3}}{2}\right)a\ ,
\end{eqnarray}
where $a=1.42$~\AA \ is the physical lattice spacing (lattice constant) of graphene. We find
\begin{eqnarray}
\vec{b}_1&\equiv&\left(\frac{1}{3},\frac{1}{\sqrt{3}}\right)\frac{2\pi}{a}\ ,\\
\vec{b}_2&\equiv&\left(\frac{1}{3},-\frac{1}{\sqrt{3}}\right)\frac{2\pi}{a}\ .
\end{eqnarray}
for the reciprocal lattice vectors.
The hexagonal lattice can also be described in terms of two triangular lattices (labeled A and B), separated by the vector $\vec{a}\equiv(\vec{a}_1+\vec{a}_2)/3$ as shown in 
Fig.~\ref{fig:33 tube}. Such a description of the graphene lattice will be useful for our path integral formulation in Section~\ref{sect:PI}.  

A general nanotube of ``chirality'' $(n,m)$ is given in terms of the ``chiral vector'' $\vec{C}_h$, 
\begin{equation}
\vec{C}_h \equiv n\vec{a}_1+m\vec{a}_2,
\end{equation}
where $n$, $m$ are integers with $0\le|m|\le n$. The ``translation vector'' $\vec{T}$ perpendicular to the chiral vector $\vec{C}_h$ is defined as
\begin{equation}\label{eqn:T vector}
\vec{T} \equiv t_1\vec{a}_1+t_2\vec{a}_2,
\end{equation} 
with 
\begin{eqnarray}
t_1& \equiv &\frac{2m+n}{d_R} \\
t_2& \equiv &-\frac{2n+m}{d_R},
\end{eqnarray}
where $d_R \equiv \mathrm{gcd}(2m+n,2n+m)$ (greatest common divisor).  
These vectors are shown in Fig.~\ref{fig:33 tube} for the case of $(n,m)=(3,3)$.  

In order to construct a $(3,3)$ nanotube, we cut from the graphene lattice the rectangle formed by the chiral and translation vectors. Next, we roll the rectangle along the $\vec{C}_h$ 
vector, in order to form a nanotube. Thus, we identify $\vec{C}_h$ as the vector that points along the circumferential direction of the tube, 
while the vector $\vec{T}$ points along the longitudinal direction of the tube. This construction represents one ``unit cell'' 
of a nanotube of length $|\vec{T}|$. The number of hexagons $N$ within this nanotube unit is 
\begin{equation}
\label{eqn:num N}
N_U=\frac{|\vec{C}_h\times\vec{T}|}{|\vec{a}_1\times\vec{a}_2|},
\end{equation}  
and for the $(3,3)$ tube, this gives $N_U=6$ and $|\vec{T}|=\sqrt{3}a$.  

The length of the tube can be increased by adding additional unit cells to its ends. We denote by $N_L$ the number of unit cells along the longitudinal direction, giving an overall tube 
length of $N_L |\vec{T}|$ and a total number of hexagons $N_L N_U$. In our lattice Monte Carlo studies of the 
$(3,3)$ nanotube, we use \mbox{$N_L$ = 3, 6, and 9.} In the right panel of Fig.~\ref{fig:33 tube}, we show a $(3,3)$ tube with $N_L = 3$ unit cells. 
In Table~\ref{tab:tube properties} we summarize the other properties of the nanotubes under consideration.

\begin{table}
\center
\caption{Overview of the properties of the (3,3) nanotubes used in our lattice Monte Carlo calculations. 
All lengths are given in units of the graphene lattice constant $a=1.42$ \AA .\label{tab:tube properties}}
\vspace{.3cm}
\begin{tabular}{c|c|c|c|c}
%\hline
$N_L$ & diameter $|\vec{C}_h|/\pi$ & length $N_L|\vec{T}|$ & \# of hexagons $N=N_L N_U$ & \# of ions \\
\hline\hline
1 & 9/$\pi$ & $\sqrt{3}\ (=|\vec{T}|)$ & $6\ (=N_U)$ & 12\\
3 & 9/$\pi$ & $3\sqrt{3}$ & 18 & 36\\
6 & 9/$\pi$ & $6\sqrt{3}$ & 36 & 72\\
9 & 9/$\pi$ & $9\sqrt{3}$ & 54 &108\\
\hline\hline
\end{tabular}
\end{table}

\section{Path Integral Formalism \label{sect:PI}}

We note that detailed treatments of the
path integral formalism for a graphene monolayer in the tight-binding description have already been given in Refs.~\cite{Brower:2012zd,Smith:2014tha}. Hence, our main objectives
are to give a cursory overview intended to introduce notation, and to highlight the differences encountered in the application to carbon nanotubes.
The Hamiltonian $H$ of the carbon nanotube system consists of the tight binding Hamiltonian $H_{tb}$ that describes the interaction of the electrons with the carbon ions, and
of the interaction Hamiltonian $H_I$, responsible for electron-electron correlations.  We write this in the form
\begin{eqnarray}
\label{eqn:H1}
H &\equiv& H_{tb}+H_I \\ \nonumber
&\equiv& -\kappa\sum_{\langle x,y\rangle,s} a^\dag_{x,s}a_{y,s}^{}+\frac{1}{2}\sum_{x,y} V_{x,y} \, q_x q_y,
\end{eqnarray}
where $x$ and $y$ denote sites on the honeycomb lattice, $\kappa \simeq 2.7$~eV is the nearest-neighbor hopping 
amplitude for electrons in graphene, and $V_{x,y}$ is the electron-electron potential matrix (see Section~\ref{sect:screened coulomb}). 
Further, $\langle x,y \rangle$ denotes summation over nearest neighbors, and $s$ assumes the values ($\uparrow$, ``spin up'') or ($\downarrow$, ``spin down'').
Also, $q_i \equiv a^\dag_{i,\uparrow}a_{i,\uparrow}^{}+a^\dag_{i,\downarrow}a_{i,\downarrow}^{}-1$ is the charge operator at position~$i$, shifted by ($-1$) to ensure 
overall neutrality (``half-filling''). In contrast to Ref.~\cite{Smith:2014tha}, we do not introduce a ``staggered mass'' term to our Hamiltonian (see Eqn.~10 of Ref.~\cite{Smith:2014tha}).  As mentioned in the introduction, we do not include the effects of curvature in our Hamiltonian, which induces geometrical tilting of $\pi$ orbitals and hybridization of $\sigma$ bonds~\cite{PhysRevLett.72.1878}.   We note that the effects of curvature can be incorporated into our calculations by using hopping parameters that are dependent on the direction of the three nearest neighbor bonds relative to the tube and azimuthal directions, i.e., $\kappa_i$ for $i=1,2,3$, as described in~\cite{PhysRevB.64.113402}.

In order to recast the Hamiltonian in a form more amenable to Quantum Monte Carlo calculations, we define the ``hole'' operators
for spin $\downarrow$ electrons,
\begin{equation}
b^\dag_{x,\downarrow}\equiv a_{x,\downarrow},\qquad b_{x,\downarrow}\equiv a^\dag_{x,\downarrow},
\end{equation}
and similarly for spin $\uparrow$ electrons. In terms of these new operators, Eqn.~\eqref{eqn:H1} becomes
\begin{equation}
\label{eqn:H2}
H=-\kappa\sum_{\langle x,y \rangle}\left(a^\dag_{x,\uparrow}a_{y,\uparrow}^{} - b^\dag_{x,\downarrow}b_{y,\downarrow}^{}\right)
+\frac{1}{2}\sum_{x,y}V_{x,y} \, q_x q_y,
\end{equation}
with the charge operator $q_i=a^\dag_{i,\uparrow}a_{i,\uparrow}^{}-b^\dag_{i,\downarrow}b_{i,\downarrow}^{}$.  
Finally, we flip the sign of the operators $b$ and $b^\dag$ on one of the sublattices. This impacts only the nearest-neighbor hopping term of the Hamiltonian and leaves the dynamics of the
system invariant, as the anticommutation relations of the hole operators remain unchanged. However, this last step is essential in ensuring a positive definite probability measure for
our Monte Carlo calculations, as we shall discuss below. The Hamiltonian now becomes
\begin{equation}
\label{eqn:H3}
H=-\kappa\sum_{\langle x,y \rangle}\left(a^\dag_{x}a_{y}^{}+b^\dag_{x}b_{y}^{}\right)+\frac{1}{2}\sum_{x,y} V_{x,y} \, q_x q_y,
\end{equation}
where the superfluous spin indices have been dropped. 

The basis of our Monte Carlo calculations is Eqn.~\eqref{eqn:H3}, and
we are interested in calculating expectation values of operators $O$ (or time-ordered products of operators),
\begin{multline}
\label{eqn:O}
\langle O(t)\rangle \equiv \frac{1}{Z}\text{Tr}\left[O(t)e^{-\beta H}\right] \\
= \frac{1}{Z}\int\left[\prod_\alpha d\psi^*_\alpha d\psi_\alpha d\eta^*_\alpha d\eta_\alpha\right]
e^{-\sum_\alpha(\psi^*_\alpha\psi_\alpha+\eta^*_\alpha\eta_\alpha)}\langle -\psi,-\eta|O(t)e^{-\beta H}|\psi,\eta\rangle,
\end{multline}
where $Z \equiv \mathrm{Tr}\left[e^{-\beta H}\right]$ 
is the partition function. The Grassmann-valued fields $\psi$ and $\eta$ represent electrons and holes, respectively. 
Their products and sums (denoted by $\alpha$) are over all fermionic degrees of freedom. Here, $\beta$ is an inverse temperature and is identified with the
temporal extent of our system.  

If we now divide $e^{-\beta H}$ into $N_t$ ``time slices'' according to
\begin{equation}
\label{eqn:factor}
e^{-\beta H} \equiv e^{-\delta H}e^{-\delta H}\cdot\cdot\cdot e^{-\delta H},
\end{equation}
where $\delta \equiv \beta/N_t$, we may insert a complete set of fermionic coherent states,
\begin{displaymath}
\openone = \int\left[\prod_\alpha d\psi^*_\alpha d\psi_\alpha d\eta^*_\alpha d\eta_\alpha\right]
e^{-\sum_\alpha(\psi^*_\alpha\psi_\alpha^{}+\eta^*_\alpha\eta_\alpha^{})}|\psi,\eta\rangle\langle \psi,\eta|,
\end{displaymath}
between each of the factors on the RHS of Eqn.~\eqref{eqn:factor}. One then arrives at the 
following expression for the partition function,
\begin{multline}
\label{eqn:Z function}
Z=\text{Tr}\left[e^{-\beta H}\right] = \\
\int \prod_{t=0}^{N_t-1}\left\{\left[\prod_\alpha d\psi^*_{\alpha,t} d\psi_{\alpha,t} d\eta^*_{\alpha,t} d\eta_{\alpha,t}\right]
e^{-\sum_\alpha(\psi^*_{\alpha,t+1}\psi_{\alpha,t+1}^{}+\eta^*_{\alpha,t+1}\eta_{\alpha,t+1}^{})}\langle \psi_{t+1},\eta_{t+1}|e^{-\delta H}|\psi_t,\eta_t\rangle\right\},
\end{multline}
which depends on the Grassmann fields only. In order to account for the minus sign in the Grassmann fields generated by the trace in 
Eqn.~\eqref{eqn:O}, we identify $\psi_{N_t}=-\psi_0$ and $\eta_{N_t}=-\eta_0$, which corresponds to 
anti-periodic boundary conditions in the temporal dimension.  

We now introduce an ``auxiliary field'' $\phi$ by means of a Hubbard-Stratonovich~(HS) transformation in the matrix element on the RHS of Eqn.~\eqref{eqn:Z function},
\begin{multline}
\label{eqn:HS}
\langle \psi_{t+1},\eta_{t+1}|e^{-\delta H}|\psi_t,\eta_t\rangle=
\langle \psi_{t+1},\eta_{t+1}|e^{\delta \kappa\sum_{\langle x,y \rangle}\left(a^\dag_{x}a_{y}+b^\dag_{x}b_{y}\right)-\frac{1}{2}\sum_{x,y}\delta V_{x,y}q_xq_y}|\psi_t,\eta_t\rangle\\
\propto\int \prod_x d\tilde{\phi}_x \langle \psi_{t+1},\eta_{t+1}|e^{\tilde{\kappa}\sum_{\langle x,y \rangle}\left(a^\dag_{x}a_{y}+b^\dag_{x}b_{y}\right)-\frac{1}{2}
\sum_{x,y}[\tilde{V}]^{-1}_{x,y}\tilde{\phi}_x \tilde{\phi}_y+\sum_x i\tilde{\phi}_x q_x}|\psi_t,\eta_t\rangle,
\end{multline}
where we have introduced the dimensionless variables
\begin{displaymath}
\tilde{\kappa} \equiv \delta \kappa,\quad
\tilde{V} \equiv \delta V,\quad
\tilde{\phi} \equiv \delta \phi,
\end{displaymath}
and we note that Eqn.~\eqref{eqn:HS}
is valid up to an irrelevant overall constant and rescaling. We note that the stability of this transformation 
in a Monte Carlo calculation relies on $V_{x,y}^{-1}$ being positive definite.     

\begin{comment}
Using the fact that
\begin{equation}
\langle \psi,\eta| F\left(a^\dag_\alpha,a_\alpha,b^\dag_\alpha,b_\alpha\right)|\psi',\eta'\rangle = F\left(\psi^*_\alpha,\eta^*_\alpha,\psi'_\alpha,\eta'_\alpha\right)
e^{\sum_\gamma\left(\psi^*_\gamma\psi'_\gamma+\eta^*_\gamma\eta'_\gamma\right)}\ ,
\end{equation}
where $F$ is some function of normal-ordered creation and annihilation operators,
\end{comment}

We now apply the identity~\cite{Luscher:1976ms}
\begin{equation}
\langle \psi |\exp\left\{\sum_{x,y} a^\dag_x A_{x,y} a_y\right\}|\psi'\rangle \equiv \exp\left\{\sum_{x,y} \psi^*_x[e^{A}]_{x,y}\psi'_y\right\},
\end{equation}
where $A_{x,y}$ is a matrix of c-numbers, to the interaction term. We then obtain~\cite{Buividovich:2012nx,Ulybyshev:2013swa}
\begin{multline}
\langle \psi_{t+1},\eta_{t+1}|e^{-\delta H}|\psi_t,\eta_t\rangle=\int \prod_x d\tilde{\phi}_{x,t} \, e^{-\frac{1}{2}\sum_{x,y}[\tilde{V}]^{-1}_{x,y}\tilde{\phi}_{x,t} \tilde{\phi}_{y,t}}\\
\times \exp\left\{\tilde{\kappa}\sum_{\langle x,y\rangle}\left(\psi^*_{x,t+1}\psi_{y,t}+\eta^*_{x,t+1}\eta_{y,t}\right)+\sum_{x}\left(e^{i\tilde{\phi}_{x,t}}\psi^*_{x,t+1}\psi_{x,t}
+e^{-i\tilde{\phi}_{x,t}}\eta^*_{x,t+1}\eta_{x,t}\right)\right\}+\mathcal{O}(\delta^2),
\end{multline}
where we have introduced a ``time index'' $t$ for the auxiliary field $\phi_{x,t}$. If we insert this
expression into Eqn.~\eqref{eqn:Z function}, we find
\begin{multline}
\label{eqn:Z function 2}
Z=\int \mathcal{D}\tilde{\phi}\mathcal{D}\psi^*\mathcal{D}\psi\mathcal{D}\eta^*\mathcal{D}\eta 
\, e^{-\frac{1}{2}\sum_{x,y,t}[\tilde{V}]^{-1}_{x,y}\tilde{\phi}_{x,t} \tilde{\phi}_{y,t}}
\exp\bigg\{\tilde{\kappa}\sum_{\langle x,y\rangle,t}\left(\psi^*_{x,t+1}\psi_{y,t}+\eta^*_{x,t+1}\eta_{y,t}\right)\\
-\sum_{x,t}\left(\psi^*_{x,t+1}(\psi_{x,t+1}-e^{i\tilde{\phi}_{x,t}}\psi_{x,t})+\eta^*_{x,t+1}(\eta_{x,t+1}-e^{-i\tilde{\phi}_{x,t}}\eta_{x,t})\right)\bigg\},
\end{multline}
where $\mathcal{D}\tilde{\phi}$ is a shorthand notation for $\prod_{x,t=0}^{N_t-1} d\tilde{\phi}_{x,t}$ (and similarly for the other fields). 
The motivation for the HS~transformation is now clear:  Only quadratic powers of the fermion fields appear in the argument of the exponent (without the HS transformation, quartic 
powers would also appear). We are now in a position to perform the Gaussian-type integrals over the fermion fields. Up to irrelevant overall factors, the partition function becomes
\begin{equation}
\label{eqn:Z function 3}
Z=\int \mathcal{D}\tilde{\phi}\det[M(\tilde{\phi})]\det[M^*(\tilde{\phi})]\exp\left\{-\frac{1}{2}\sum_{x,y,t=0}^{N_t-1}[\tilde{V}]^{-1}_{x,y}\tilde{\phi}_{x,t} \tilde{\phi}_{y,t}\right\},
\end{equation}
where the fermion matrix $M$ is a functional of $\tilde{\phi}$,
\begin{equation}
\label{eqn:M}
M(x,t;y,t';\tilde{\phi}) \equiv \delta_{x,y}\left(\delta_{t,t'}-e^{i\tilde{\phi}_{x,t'}}\delta_{t-1,t'}\right)-\tilde{\kappa}\,\delta_{\langle x,y\rangle}\delta_{t-1,t'},
\end{equation}
where $\delta_{\langle x,y\rangle}$ equals unity if $x$ and $y$ are nearest-neighbor sites, and zero otherwise. This is referred to as the ``compact formulation'' of the 
path integral for the interacting, hexagonal tight-binding system.

The feasibility of a Monte Carlo evaluation of the 
path integral relies on the generation of configurations of $\tilde{\phi}$ that follow the probability distribution
\begin{eqnarray}
\label{eqn:probability}
P(\tilde{\phi})&\equiv&\frac{1}{Z}\det[M(\tilde{\phi})]\det[M^*(\tilde{\phi})]\exp\left\{-\frac{1}{2}\sum_{x,y,t=0}^{N_t-1}[\tilde{V}]^{-1}_{x,y}\tilde{\phi}_{x,t} \tilde{\phi}_{y,t}\right\}\\
&=&\frac{1}{Z}\det[M(\tilde{\phi})M^\dag(\tilde{\phi})]\exp\left\{-\frac{1}{2}\sum_{x,y,t=0}^{N_t-1}[\tilde{V}]^{-1}_{x,y}\tilde{\phi}_{x,t} \tilde{\phi}_{y,t}\right\},
\end{eqnarray}
which is positive definite as long as $V^{-1}_{x,y}$ is positive definite. Also, $\det[M(\tilde{\phi})M^\dag(\tilde{\phi})] \ge 0$ for any $\phi$. 
We use global Hybrid Monte Carlo~(HMC) lattice updates in order to generate the necessary ensembles of configurations, which we denote by $\{\tilde{\phi}\}$. 
For a thorough discussion of the HMC algorithm and related issues, see for example Refs.~\cite{Smith:2014tha,gattringer2009quantum}. Given an ensemble $\{\tilde{\phi}\}$, the Monte Carlo estimate of the expectation value of 
any operator $O$ is given by
\begin{equation}
\langle O\rangle \approx \frac{1}{N_\mathrm{cf}}\sum_{i=1}^{N_\mathrm{cf}} O[\tilde{\phi_i}],
\end{equation}
where $\tilde{\phi}_i\in\{\tilde{\phi}\}$ and $N_\mathrm{cf}$ is the number of configurations within the ensemble. Each such estimate carries with it an
associated uncertainty which (in principle) can be arbitrarily reduced with increased statistics (\emph{i.e.}\ by taking $N_\mathrm{cf}\rightarrow\infty$).
In this first study, we are interested in computing the single quasi-particle spectrum, 
which can be accessed by taking $O=a_x(\tau)a^\dag_y(0)$,
\begin{equation}
\langle a_x(\tau)a^\dag_y(0) \rangle = \langle M^{-1}(x,\tau;y,0)\rangle \approx\frac{1}{N_\mathrm{cf}}\sum_{i=1}^{N_\mathrm{cf}} M^{-1}(x,\tau;y,0;\tilde{\phi}_i),
\end{equation}
and by analyzing the temporal behavior of the resulting correlator.

We finally note that the fermion fields can be recast in terms of two-component fields, 
with one component for the underlying $A$~sublattice and the other one for the $B$~sublattice.  For instance, the electron fields can be written as
\begin{equation}
\Psi(x,t) =
\begin{pmatrix}
\Psi_A(x,t) \\
\Psi_B(x,t)
\end{pmatrix}
=
\begin{pmatrix}
\psi_{x,t} \\
\psi_{x+\vec{a},t}
\end{pmatrix},
\end{equation}
where $x$ in this case represents the location of a given hexagonal unit cell. In this manner, the ion on site~$A$ associated with this particular hexagonal unit cell is 
located at position $x$, while the ion on site $B$ is located at $x+\vec{a}$. An analogous definition can be made for the auxiliary HS~field,
\begin{equation}
\Phi(x,t) =
\begin{pmatrix}
\Phi_A(x,t) \\
\Phi_B(x,t)
\end{pmatrix}
=
\begin{pmatrix}
\tilde{\phi}_{x,t} \\
\tilde{\phi}_{x+\vec{a},t}
\end{pmatrix},
\end{equation}
and the matrix $M(x,t';y,t)$ acting on the two-component fermion field is now given by
\begin{multline}
\label{eqn:M matrix original}
M(x,t';y,t)\Psi(y,t)=\\
\begin{pmatrix}
\delta_{x,y}\left(\delta_{t',t} -e^{i\Phi_A(x,t')}\delta_{t-1,t'}\right) & -\tilde{\kappa}\ \delta_{\langle x, y\rangle}\delta_{t-1,t'}\\
-\tilde{\kappa}\ \delta_{\langle x, y\rangle}\delta_{t-1,t'} & \delta_{x,y}\left(\delta_{t',t} -e^{i\Phi_B(x,t')}\delta_{t-1,t'}\right)
\end{pmatrix}
\begin{pmatrix}
\Psi_A(y,t) \\
\Psi_B(y,t)
\end{pmatrix},
\end{multline}
where the 
coordinates $x$ and $y$ now represent locations of hexagonal unit cells (and not of the ions themselves), so that the definition of $\delta_{\langle x, y\rangle}$ must be slightly modified to account 
for all pairs of unit cell locations $x$ and $y$ that share nearest neighbor ions. While we stress that the matrix notation for $M(x,y;t)$ in Eqn.~\eqref{eqn:M matrix original} is equivalent to Eqn.~\eqref{eqn:M}, 
the underlying $A/B$ sublattice structure has now been made explicit. We find this representation convenient in analyzing the non-interacting limit of our theory, as discussed 
Section~\ref{sect:NI}, and also in our zero-mode analysis in Section~\ref{sect:zero modes} and Appendix~\ref{app:zero modes}.

%%%%%%%%%%%%%%%%%%%%%%%%%

\section{Non-interacting system\label{sect:NI}}

Before we present results of calculations that include electron-electron correlations, 
it is highly instructive to recall the non-interacting (tight-binding) theory and to compare with the results of our path-integral calculations in this regime. 
Not only does this exercise allow us to emphasize some salient features of our formalism, but it also allows us to find an accurate way of representing temporal finite differences
on the lattice in a way which avoids the infamous ``doubling problem'', where spurious high-momentum modes contribute in the continuum limit.

\subsection{Zero-temperature continuum limit}

The non-interacting case is obtained by setting $\tilde{\phi}=0$ in our expressions for the path integral.
In the $\delta\rightarrow 0$ (continuous time) limit, it is straightforward to show that Eqn.~\eqref{eqn:M matrix original} becomes
\begin{equation}\label{eqn:M matrix}
M(x,y;t)\Psi(y;t)=
\begin{pmatrix}
\delta_{x,y}\ \partial_t  & -\kappa\ \delta_{\langle x, y\rangle}\\
-\kappa\ \delta_{\langle x, y\rangle} & \delta_{x,y}\ \partial_t 
\end{pmatrix}
\begin{pmatrix}
\Psi_A(y,t) \\
\Psi_B(y,t)
\end{pmatrix},
\end{equation}
when expressed in terms of dimensionful quantities. We shall first consider the zero-temperature limit, followed by the case of finite temperature.
We now move to Fourier space in the $\beta\rightarrow\infty$ (zero temperature) limit, expressing Eqn.~\eqref{eqn:M matrix} as
\begin{equation}\label{eqn:M kw}
M(x,y;t)=
\frac{1}{2\pi}\int_{-\infty}^{\infty}d\omega \, e^{i \omega t}
\frac{1}{N_U}\sum_{i=0}^{N_U-1}\frac{|\vec{T}|}{2\pi}\int d\vec{k}_{||} e^{i(\vec{k}_{||}+\vec{k}_{\bot,i})\cdot(\vec{x}-\vec{y})}
\,\widetilde{M}(\vec{k}_{\bot}+\vec{k}_{\bot,i};\omega),
\end{equation}
where 
\begin{equation}\label{eqn:M tilde}
\widetilde{M}(\vec{k};\omega)=
\begin{pmatrix}
i\omega & -\kappa f(\vec{k}) \\
-\kappa f^*(\vec{k}) & i \omega 
\end{pmatrix},
\end{equation}
and
\begin{equation}
f(\vec{k}) = e^{iak_x /\sqrt{3}}+2e^{-iak_x /(2\sqrt{3})}\cos(ak_y/2),
\end{equation}
following Ref.~\cite{Saito1998}. In Eqn.~\eqref{eqn:M kw} we have introduced the momentum variables $\vec{k}_{||}$ and $\vec{k}_{\bot,i}$ which satisfy
\begin{displaymath}
\vec{k}_{||}\cdot\vec{k}_{\bot,i}=0\quad,\quad\vec{T}\cdot\vec{k}_{\bot,i}=0\ ,
\end{displaymath}
where $\vec{T}$ (Eqn.~\eqref{eqn:T vector}) is parallel to the tube axis.  
Since we assume that the tube is infinitely long, $\vec{k}_{||}$ is continuous within an interval of length $2\pi/|\vec{T}|$\footnote{We note that the 
interval of integration over $\vec k_{||}$ depends, in general, on the choice of $\vec k_{\bot,i}$.}. However, the momentum $\vec{k}_{\bot}$ is discrete due to the finite circumference of the 
tube. These discrete momenta $\vec{k}_{\bot,j}$ are given by~\cite{Saito1998}
\begin{equation}
\vec{k}_{\bot,j} \equiv \frac{j}{N_U}(t_1\vec{b}_2-t_2\vec{b}_1),
\end{equation}
where the $t_i$ are translation vector components and $\vec{b}_i$ the reciprocal lattice vectors, as discussed in Section~\ref{sect:tube description}. Also, $N_U$ is given by 
Eqn.~\eqref{eqn:num N} and $j\in[0,N_U-1]$.

\begin{figure}
\center
\includegraphics[width=0.8\columnwidth]{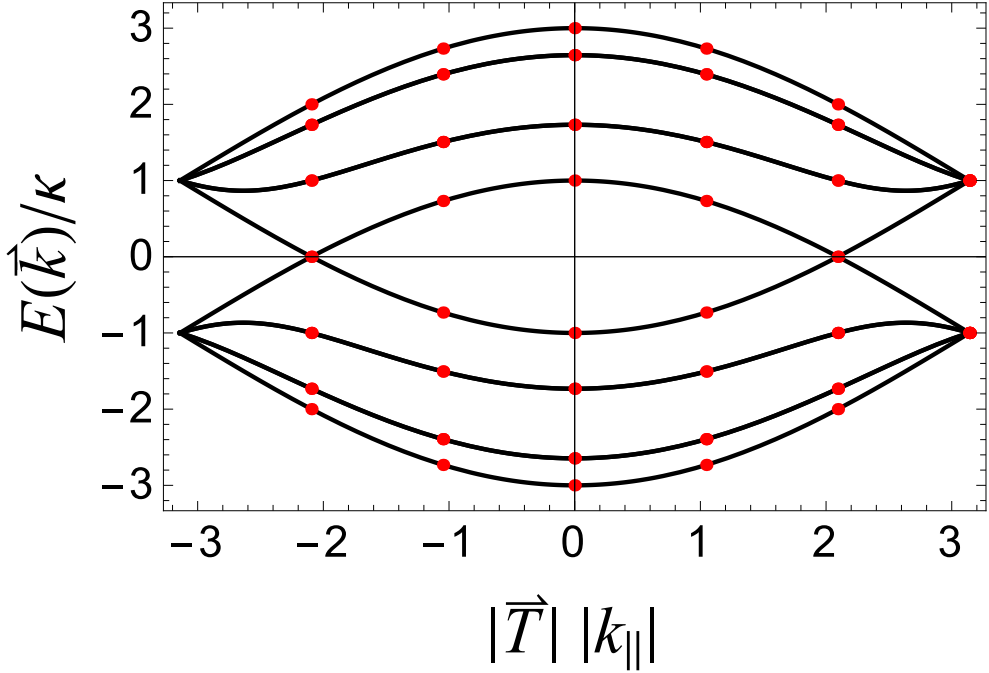}
\caption{Non-interacting (tight-binding) dispersion relation for a $(3,3)$ nanotube of infinite length (solid black lines) and one with 
$N_L=6$ unit cells (red points). The abscissa shows the momentum
$|\vec{T}| |k_{||}|$ parallel to the tube axis, while the ordinate shows the energy (in units of $\kappa$) for a single quasiparticle. 
Positive energies denote particles, and negative energies denote holes.
\label{fig:dispersion33}}
\end{figure}

To determine 
the zero-temperature dispersion relation for a single quasiparticle in the non-interacting limit, 
it suffices to study the pole structure of $\widetilde{M}^{-1}$. 
This is equivalent to finding simultaneous values of $\omega$ and $\vec{k}$ that satisfy the quantization condition
\begin{equation}
\label{eqn:QC}
\det[\widetilde{M}(\vec{k};\omega)]=0,
\end{equation}
which admits the solution 
\begin{equation}\label{eqn:wk}
E(\vec{k})=i\omega(\vec{k}) = \pm \kappa | f(\vec{k}) |,
\end{equation}
for the energy $E(\vec{k})$ of the quasiparticle.
In Fig.~\ref{fig:dispersion33}, we show the dispersion relation as function of $\vec{k}_{||}$ for the $(3,3)$ tube. 
Because of the discrete momenta perpendicular to the tube direction, the dispersion relation consists of bands of energy curves. 
Note that the point with the largest magnitude of the energy occurs at the $\Gamma$~point $(|k_{||}|,|k_{\bot,i}|)=(0,0)$, 
while the zero-energy Dirac point~$K$ occurs at non-zero momentum $|\vec{T}|(|k_{||}|,|k_{\bot,i}|)=(\frac{2\pi}{3},\frac{2\pi}{\sqrt{3}})$. 

\subsection{Dispersion for a tube of finite length\label{sect:finite length}}

For reasons of computational practicality, our Monte Carlo calculations are performed with tubes of finite length, with periodic boundary conditions at the ends of the tube.
As shown in the right panel of Fig.~\ref{fig:33 tube}, the top (green) lattice points are (from the point of view of the Monte Carlo calculation) identical to the bottom (green) lattice points, 
by virtue of the periodic boundary conditions. This implies that the momenta $\vec{k}_{||}$ in the direction parallel to the tube axis will also be discrete, with wave vectors 
separated by $2\pi/(N_L |\vec T|)$, where $N_L |\vec T|$ is the overall tube length. For the non-interacting case, the dispersion relation becomes a series of points that 
coincide with the continuous lines shown in Fig.~\ref{fig:dispersion33}. The density of points and the exact functional form of the lines depends on the length and chirality of the tube.  

In Fig.~\ref{fig:dispersion33}, the discrete dispersion points are shown for the specific case of the $(3,3)$ tube with $N_L=6$ unit cells. 
It should be noted that some of these coincide with the Dirac \emph{K} points. A shift of the energy away from this point (for instance due to interactions) would indicate the existence of an energy gap
at the Dirac point. In general, given an $(n,m)$ tube that exhibits a Dirac point, the number of unit cells should be a multiple of three in order for the discrete dispersion to access 
the Dirac point~\cite{Saito1998}. In other words, the discrete momentum modes should include a subset of $|\vec{T}|(|k_{||}|,|k_{\bot,i}|)=(\frac{2\pi}{3},\frac{2\pi}{\sqrt{3}})$ 
and/or $|\vec{T}|(|k_{||}|,|k_{\bot,i}|)=(0,\frac{4\pi}{3})$. This condition is the reason why we focus on tubes with $N_L = 3$, 6, and 9 unit cells.

\subsection{Finite temperature}

\begin{comment}
\begin{figure}
\includegraphics[width=0.8\linewidth]{figures/BZ40.pdf}\nobreak
\caption{\label{fig:BZ40} Points within shaded region indicate the allowed momenta within the first Brillouin zone for the $(4,0)$ tube with five unit cells.  Also shown is the hexagonal Brillouin zone for graphene for comparison.} 
\end{figure}
\end{comment}

In addition to calculations with a finite tube length, the path integral formalism requires the introduction of a finite temporal extent $\beta$ (as discussed in Section~\ref{sect:PI}), 
which in turn can be viewed as an inverse temperature. This implies that the frequency integral should be replaced by the summation
\begin{equation}
\frac{1}{2\pi}\int_{-\infty}^{\infty}d\omega\ e^{i\omega\tau}\rightarrow\frac{1}{\beta}\sum_{n=-\infty}^{\infty}e^{i\omega_n\tau}\ ,
\end{equation}
where
\begin{equation}
\omega_n \equiv \frac{2\pi}{\beta}\left(n+\frac{1}{2}\right),
\end{equation}
are the Matsubara frequencies.
We note that the expression for the correlator
\begin{equation}
\label{eqn:correlatorT}
G(\vec{k}_i,\tau)\equiv\frac{1}{\beta}\sum_{n=-\infty}^{\infty}\  e^{i\omega_n t}\ \widetilde{M}^{-1}(\vec{k}_i;\omega_n),
\end{equation}
can be evaluated analytically using straightforward (though tedious) algebra. In the range \mbox{$0<\tau<\beta$}, we find
\begin{eqnarray}
\label{eqn:xxx}
G(\vec{k}_i,\tau)&=&
\frac{1}{2\cosh(\omega(\vec{k}_i) \beta/2)}
\begin{pmatrix}
\cosh(\omega(\vec{k}_i)(\tau-\beta/2)) & e^{i\theta_{k_i}}\sinh(\omega(\vec{k}_i)(\tau-\beta/2))\\
e^{-i\theta_{k_i}}\sinh(\omega(\vec{k}_i)(\tau-\beta/2)) & \cosh(\omega(\vec{k}_i)(\tau-\beta/2))
\end{pmatrix}
\quad \\ \label{eqn:C matrix}
&\equiv&
\begin{pmatrix}
G_{AA}(\vec{k}_i,\tau) & G_{AB}(\vec{k}_i,\tau)\\
G_{BA}(\vec{k}_i,\tau) & G_{BB}(\vec{k}_i,\tau)   
\end{pmatrix},
\end{eqnarray}
where $G_{BA}(\vec{k}_i,\tau)=G^*_{AB}(\vec{k}_i,\tau)$,
\begin{equation}
\theta_{k_i}\equiv\tan^{-1}(\text{Im} f(\vec{k}_i)/ \text{Re} f(\vec{k}_i)),
\end{equation}
and $\omega(\vec{k}_i)$ is given by the positive solution in Eqn.~\eqref{eqn:wk}. The form of Eqn.~\eqref{eqn:C matrix} is due to the underlying $A/B$ sublattice 
structure\footnote{This is equivalent to a system that consists of a unit cell plus one basis function.}, and 
admits two linearly independent correlator solutions (see for instance Ref.~\cite{terakura2013interatomic}),
\begin{eqnarray}\label{eqn:analytic correlator} 
G_\pm(\vec{k}_i,\tau)&\equiv&\frac{1}{2}\left[G_{AA}(\vec{k}_i,\tau) + G_{BB}(\vec{k}_i,\tau) \pm (G_{AB}(\vec{k}_i,\tau) + G_{BA}(\vec{k}_i,\tau) )\right] \\
&=& \frac{1}{2\cosh(\omega(\vec{k}_i)\beta/2)}\left[\cosh(\omega(\vec{k}_i)(t-\beta/2))\pm\cos(\theta_{k_i})\sinh(\omega(\vec{k}_i)(t-\beta/2))\right], \quad
\end{eqnarray}
which for $t\ll\beta$ behave as
\begin{equation}
G_{\pm}(\vec{k}_i,\tau)\propto e^{\pm\omega(\vec{k}_i)\tau},
\end{equation}
which shows that the ``leading'' exponential behavior of these correlators provides access to the (non-interacting) spectrum of the theory. As we show in 
Section~\ref{sect:results}, we use this aspect of the correlators when we compute the spectrum in the presence of electron-electron correlations.  

\subsection{Discretization of time}

\begin{comment}
\begin{figure}
\includegraphics[width=0.5\linewidth]{figures/44_6.pdf}\nobreak
\caption{\label{fig:44_6} Dispersion relation calculated with a discretised non-interacting path integral formalism, using the backward-difference (blue points), forward-difference (black points), and mixed back/forward-difference (red points) equations, compared to the tight-binding result (solid line). Calculations were done on a (4,4) armchair tube with 6 unit cells, a Dirac mass of .25 eV, $\beta=2$ eV$^{-1}$, and 24 time slices. The breaking of the degeneracy at the Dirac point is due to the Dirac mass and not due to numerical discretization errors, and occurs for all difference formulations.} 
\end{figure}
\end{comment}

We now consider the case where the temporal dimension is also discretized.
Given a temporal extent $\beta$ divided into $N_t$ time steps of equal width $\delta=\beta/N_t$, the allowed Matsubara frequencies $\omega_n=\frac{2\pi}{T}(n+1/2)$ are those that fall 
within the first Brillouin zone $[-\pi/\delta,\pi/\delta)$, which corresponds to $-N_t/2\le n< N_t/2$\footnote{We assume here that $N_t$ is even.}.  
The time derivative in Eqn.~\eqref{eqn:M matrix} should now be approximated using these discrete steps. 
As we show below, analytic expressions are still obtainable for the non-interacting case.
In what follows, we make use of the representation,
\begin{equation}
\delta_{t_i,t_j}=\frac{1}{N_t}\sum_{n=-N_t/2}^{N_t/2-1}e^{i\omega_n(t_i-t_j)},
\end{equation}
where $t_i=i\delta$ and $t_j=j\delta$ are lattice time sites ($i$ and $j$ are integers with $0\le i,j < N_t$).

\subsubsection{Forward difference}
We first consider the case of forward discretization to approximate the derivative
\begin{eqnarray}
\partial_\tau f(t)&\rightarrow&\frac{1}{\delta}\left(\delta_{\tau+\delta,t}-\delta_{\tau,t}\right)f(t)\\
&=&\frac{1}{N_t\delta}\left(\sum_{n}e^{i\omega_n(\tau+\delta-t)}-\sum_{n}e^{i\omega_n(\tau-t)}\right)f(t)\\
&=&\frac{1}{N_t}\sum_n e^{i\omega_n(\tau-t)}\frac{1}{\delta}e^{i\omega_n\delta/2}\left(e^{i\omega_n\delta/2}-e^{-i\omega_n \delta/2}\right)f(t)\\
&=&\frac{1}{N_t}\sum_n e^{i\omega_n(\tau-t)}\frac{2i}{\delta}e^{i\omega_n\delta/2}\sin(\omega_n\delta/2)f(t),
\end{eqnarray}
where $f(t)$ is an arbitrary function on the lattice. Under this differencing scheme, the
matrix $\widetilde{M}$ in Eqn.~\eqref{eqn:M matrix} in the momentum-frequency domain becomes 
\begin{equation}
\label{eqn:forward diff matrix}
\widetilde{M}(\vec{k};\omega_n)=
\begin{pmatrix}
\frac{2i}{\delta}e^{i\omega_n\delta/2}\sin(\omega_n\delta/2) & -\kappa f(\vec{k}) \\
-\kappa f^*(\vec{k}) & \frac{2i}{\delta}e^{i\omega_n\delta/2}\sin(\omega_n\delta/2)
\end{pmatrix},
\end{equation}
for which the quantization condition
\begin{equation}
\det(\widetilde{M}(\vec{k};\omega_n))=0\ ,
\end{equation}
gives the solution 
\begin{equation}
\label{eqn:qc forward}
\omega_n^2(1+i\omega_n\delta)+\mathcal{O}(\delta^2)=-\kappa^2|f(\vec{k})|^2,
\end{equation}
for small $\delta$. Hence, we expect our energies computed in this discretized scheme 
to be shifted by $\mathcal{O}(\delta)$ from the result in the (temporal) continuum limit. 
We note that for a backward time difference
\begin{equation}
\partial_\tau f(t)\rightarrow\frac{1}{\delta}\left(-\delta_{\tau-\delta,t}+\delta_{\tau,t}\right)f(t),
\end{equation}
an analogous derivation gives similar results, provided that the replacement
$i\omega_n\rightarrow-i\omega_n$ is made in Eqs.~\eqref{eqn:forward diff matrix} and~\eqref{eqn:qc forward}.
 
\subsubsection{Mixed difference}
Given our results for the forward and backward differences, a natural choice would be to consider a 
symmetric differencing scheme to approximate the time derivative according to
\begin{equation}
\partial_\tau f(t)\rightarrow\frac{1}{2\delta}\left(\delta_{\tau+\delta,t}-\delta_{\tau-\delta,t}\right)f(t),
\end{equation}
although it is well know that this 
admits spurious high-energy solutions that have no analog in the continuum limit 
(see for instance the discussion on the ``doubling problem'' in Ref.~\cite{gattringer2009quantum})\footnote{We have numerically confirmed the existence of such spurious states.}. 
Instead, we employ a ``mixed'' differencing scheme where we use a forward difference on $A$ sites and a backward difference on $B$ sites. We are free to do this, since the mixed scheme 
has the correct continuum limit. This idea was first pointed out in Ref.~\cite{Brower:2012zd}, and we shall use it here for our non-interacting system. With mixed differencing, our 
fermion matrix becomes
\begin{equation}
\label{eqn:mix diff matrix}
\widetilde{M}(\vec{k};\omega_n)=
\begin{pmatrix}
\frac{2i}{\delta}e^{i\omega_n\delta/2}\sin(\omega_n\delta/2) & -\kappa f(\vec{k}) \\
-\kappa f^*(\vec{k}) & \frac{2i}{\delta}e^{-i\omega_n\delta/2}\sin(\omega_n\delta/2)
\end{pmatrix},
\end{equation}
and the quantization condition gives
\begin{equation}\label{eqn:mixed forward}
\omega_n^2+\mathcal{O}(\delta^2)=-\kappa^2|f(\vec{k})|^2,
\end{equation}
which is ``$\mathcal{O}(\delta)$ improved'', in comparison with Eqn.~\eqref{eqn:qc forward}.  

\begin{figure}
\includegraphics[width=.5\columnwidth]{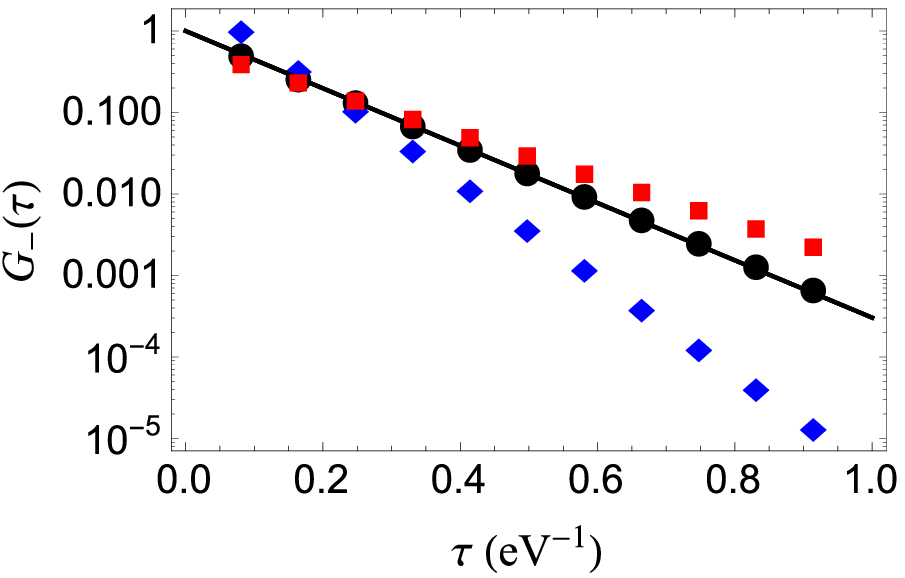}\includegraphics[width=.5\columnwidth]{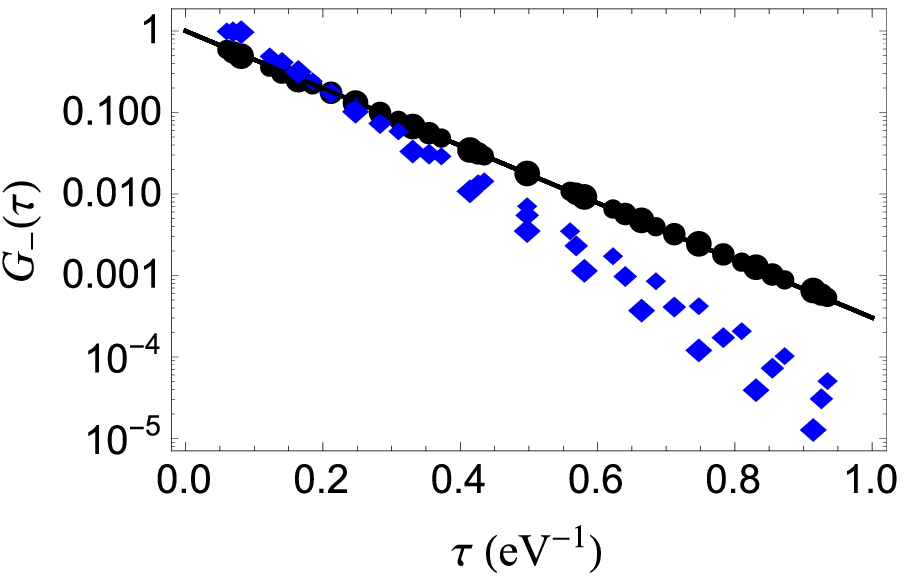}
\caption{Comparison of analytic $G_-(\tau)$ correlator at the $\Gamma$ point (black line) to its discretized form.  The left panel shows a calculation with $N_t=24$ discretized steps, where the (red) squares use the backward differencing scheme, (blue) diamonds use forward differencing scheme, and (black) circles use the mixed differencing scheme as described in text.   The right panel shows the convergent behavior of the mixed and forward differencing schemes, with $N_t=24$, 28, and 32 timesteps.  The decreasing pointsizes correspond to increasing $N_t$.  Similar behavior is observed for the backward differencing scheme, but is not shown to reduce clutter.  All calculations were performed with $\beta=2$ eV$^{-1}$. \label{fig:NI correlator}}
\end{figure}

We can visualize this improvement by direct inspection of the correlators. In Fig.~\ref{fig:NI correlator}, we compare the exact analytic correlator at the $\Gamma$ point to its discretized form, 
using the forward, backward, and mixed differencing schemes, noting that the time dependence of $\widetilde{M}$ is given by Eqn.~\eqref{eqn:correlatorT}. As can be seen from the 
left panel of Fig.~\ref{fig:NI correlator}, the mixed differencing scheme (black points) compares very well with the analytic result (black line) given by Eqn.~\eqref{eqn:analytic correlator}, 
whereas the forward (blue diamonds) and backward (red squares) differencing schemes have clear systematic errors. These calculations were performed with $\beta=2$ eV$^{-1}$ 
and $N_t=24$ time steps. The right panel of Fig.~\ref{fig:NI correlator} shows the convergence of the mixed and forward differencing schemes with increasing number of time steps: 
$N_t=24$, 28, and 32 (corresponding to decreasing symbol size). In this case, the improved convergence of the mixed differencing scheme is obvious, and indicates that extraction 
of spectra from the leading exponential behavior of the correlator is best done with the mixed differencing scheme. For the forward differencing scheme, we have confirmed that it 
does indeed converge to the analytical line as $N_t$ is increased.  However, in order to get comparable results to the $N_t=24$ mixed-differencing scheme, the forward differencing 
scheme requires $N_t=256$ or larger. In the presence of interactions, the fermion matrix in the mixed-differencing scheme becomes
\begin{equation}
\label{eqn:M3}
M(x,t';y,t;\Phi)=
\begin{pmatrix}
\delta_{x,y}\left(e^{-i\Phi_A(x,t')}\delta_{t+1,t'} -\delta_{t,t'}\right) & -\tilde{\kappa}\ \delta_{\langle x, y\rangle}\delta_{t,t'}\\
-\tilde{\kappa}\ \delta_{\langle x, y\rangle}\delta_{t,t'} & \delta_{x,y}\left(\delta_{t',t}-e^{i\Phi_B(x,t')}\delta_{t-1,t'}\right)
\end{pmatrix},
\end{equation}
and we note that our use of this expression is motivated by the improved performance of the mixed-differencing scheme in the non-interacting case.

Since the conclusions of this section were obtained for the non-interacting system, it is not guaranteed that this $\mathcal{O}(\delta)$ improvement (or equivalently $\mathcal{O}(\delta^2)$ scaling of results) persists in the presence of interactions. 
Recent studies related to explicit $\mathcal{O}(\delta^2)$ differencing schemes in Ref.~\cite{Buividovich:2012uk,Smith:2014tha} suggest that the $\mathcal{O}(\delta)$ improvement is 
maintained in the presence of interactions, at least in the vicinity of the Dirac \emph{K} point. As we show in Section~\ref{sect:dirac point}, our results for the Dirac point support this finding as well. 
For dispersion points away from the Dirac point, our studies cannot definitively differentiate between $\mathcal{O}(\delta)$ or $\mathcal{O}(\delta^2)$ scaling. However, for this initial study, 
we assume $\mathcal{O}(\delta^2)$ scaling to perform our continuum limit extrapolations.  Future calculations with additional values of $\delta$ should be able to clarify this scaling with certainty.

\section{Interacting system\label{sect:interactions}}

Having considered the non-interacting system in some detail, we now turn to the case with electron-electron interactions. In Monte Carlo calculations of graphene, the electrons and holes
propagate on the plane defined by the hexagonal graphene sheet, and thus the interaction between the particles is constructed to reflect this geometry. Furthermore, the spatial extent of the
system in graphene calculations is typically much larger. In the case of a nanotube, 
interactions between particles can occur when they are, for example, on opposite sides of the tube wall. Thus, the interaction is not confined to a plane, and the construction of 
the potential matrix $V_{xy}$ depends on the chirality $(n,m)$ and length of the tube. We now turn to the construction of the potential.

\subsection{Screened Coulomb potential}\label{sect:screened coulomb}
\begin{figure}
\includegraphics[angle=-90,width=.65\columnwidth]{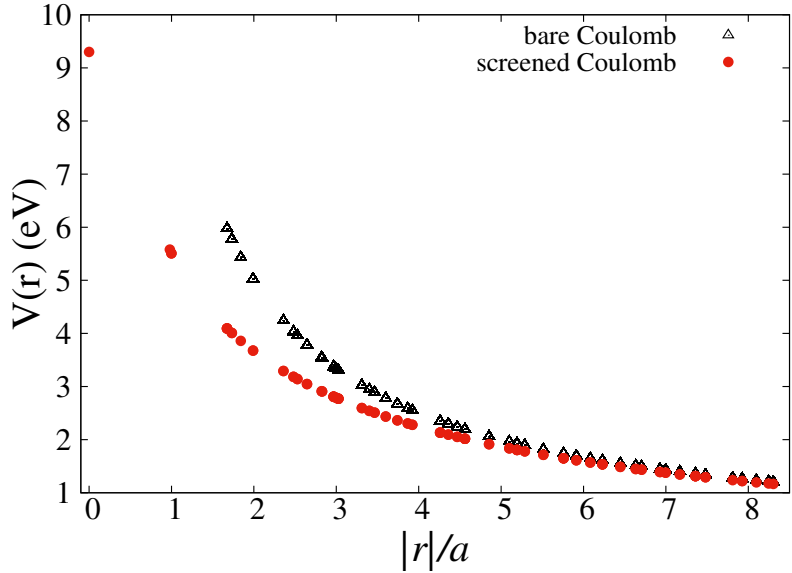}
\caption{Screened Coulomb potential matrix elements (dots) used in our $(3,3)$ tube simulations with nine unit cells. For comparison, the triangles show the bare Coulomb potential evaluated at the same distances.  The abscissa is plotted in units of the honeycomb lattice spacing.\label{fig:Vxy}}
\end{figure}

Our screened Coulomb interaction uses the results of RPA calculations performed by Wehling \emph{et al.} \cite{PhysRevLett.106.236805} for the onsite interaction $U_{00}$, nearest neighbor $U_{01}$, next-to-nearest neighbor $U_{02}$, and next-to-next-to-nearest neightbor $U_{03}$ interaction.  This interaction takes into account the short-distance screening due to the $\sigma$-band electrons (which are not dynamic in our calculations \emph{i.e.} do not hop).  We couple this interaction with a potential parameterized as in \cite{Smith:2014tha} that ensures the potential approaches the bare Coulomb potential at asymptotic distances.  Translational invariance of the potential is maintained by employing a procedure similar to one described in \cite{Smith:2014tha}:  For any two points $\vec{x}$ and $\vec{y}$ on the nanotube, we determine within the tube the shortest distance $r$ between these two points with the ends of the tube identified by periodic boundary conditions.  We then assign $V(r)$ as the potential matrix element $V_{xy}$ between these two points.    Note that because of periodic boundary conditions at the ends of the tube, there will be cases when $r^2<(\vec{x}-\vec{y})^2$ since the largest value or $r_{||}$, the component of $r$ parallel to the tube direction, is $a|\vec{T}|N_L/2$.  Due to the finite length of our tube calculations, the infrared divergence of the Coulomb potential is avoided.  In Fig.~\ref{fig:Vxy} we show the matrix elements of the screened Coulomb potential used in our calculation of the $(3,3)$ tube with nine unit cells (points).  For comparison, we also show the bare Coulomb potential evaluated at the same distances (triangles).  

\subsection{Momentum projection}

Unlike the non-interacting case, where the quasiparticle spectrum can be directly determined by analyzing the quantization conditions given by the determinant in Eqn.~\eqref{eqn:QC}, 
the spectrum of the interacting system must be determined by analyzing the temporal behavior of the appropriate correlator. To access the spectrum at a particular momentum, we must first 
project our correlator to the corresponding momentum. Such a procedure is routinely performed in lattice QCD calculations.  However, we discuss the formalism as it is applied to our system, 
in order to point out the specific differences to other lattice methods.

We denote the positions $\vec{x}_i$ of the unit cells of the tube collectively by $\{\vec{X}\}$. The momenta $\vec{k}_i$ conjugate to the unit cell cites are determined by the 
allowed reciprocal lattice vectors within the first Brillouin zone, which we denote collectively by $\{\vec{K}\}$. As our calculations use a finite number of unit cells, the 
allowed momenta in the $\vec{k}_{||}$ direction are also discrete, as discussed in Section~\ref{sect:finite length}. 
The unit cell positions and their conjugate momenta satisfy the orthogonality relations
\begin{eqnarray}
\delta_{\vec{k}_j,\vec{k}_l} &=& \frac{1}{N}\sum_{\vec{x}_i\in\{\vec{X}\}}e^{i \vec{x}_i \cdot(\vec{k}_j-\vec{k}_l)}, \\
\delta_{\vec{x}_j,\vec{x}_l} &=& \frac{1}{N}\sum_{\vec{k}_i\in\{\vec{K}\}}e^{-i \vec{k}_i \cdot(\vec{x}_j-\vec{x}_l)},
\end{eqnarray}
where $N$ is the number of unit cells (and not the number of ions).  
Given a function $f(\vec{x}_i)$ of the unit cell coordinates, the above relations can be used to define its Fourier and inverse Fourier transforms
\begin{eqnarray}
\label{eqn:inv fourier transform}
f(\vec{k}_i)&\equiv&\frac{1}{N}\sum_{\vec{x}_j\in\{\vec{X}\}}f(\vec{x}_j)\,e^{i\vec{x}_j\cdot\vec{k}_i}, \\
f(\vec{x}_i)&\equiv&\sum_{\vec{k}_j\in\{\vec{K}\}}\tilde{f}(\vec{k}_i)\,e^{-i\vec{x}_i\cdot\vec{k}_j}.
\end{eqnarray}

In addition to the unit cell locations $\vec{x}_i$, each unit cell also includes a basis vector $\vec{a}$ due to the 
two underlying sublattices $A$ and $B$. This basis vector connects the $A$ site to the $B$ site within each unit cell. Given a unit cell position $\vec{x}$ and its 
basis vector $\vec{a}$, it is convenient to express creation operators in two-component form and with the following linear combinations\footnote{These linear combinations correspond 
to ``bonding'' ($+$) and ``anti-bonding'' ($-$) orbitals.},
\begin{equation}\label{eqn:apm}
a_{\pm}^\dag(\vec{x}) \equiv
\frac{1}{\sqrt{2}}
\begin{pmatrix}
a^\dag_A(\vec{x})\\
\pm a^\dag_B(\vec{x})
\end{pmatrix}
=
\frac{1}{\sqrt{2}}
\begin{pmatrix}
a^\dag_{\vec{x}}\\
\pm a^\dag_{\vec{x}+\vec{a}}
\end{pmatrix}. 
\end{equation}
One can make an analogous definition for the hole operator $b^\dag_\pm(\vec{x})$. In momentum space, the electron correlators are given by
\begin{equation}\label{eqn:aa}
G_{\pm}(\vec{k}_i,\tau)\equiv\langle a_{\pm}(\vec{k}_i,\tau)a_{\pm}^\dag(\vec{k}_i,0)\rangle 
=\frac{1}{N^2}\sum_{\vec{x}_j,\vec{x}_k\in\{\vec{X}\}}e^{i \vec{k}_i\cdot(\vec{x}_j-\vec{x}_k)}\langle a_{\pm}(\vec{x}_j,\tau)a_{\pm}^\dag(\vec{x}_k,0)\rangle,
\end{equation}
and by inserting Eqn.~\eqref{eqn:apm} into Eqn.~\eqref{eqn:aa}, we find
\begin{eqnarray}
G_{\pm}(\vec{k}_i,\tau) &=&
\frac{1}{2N^2}\sum_{\vec{x}_j,\vec{x}_k\in\{\vec{X}\}}e^{i \vec{k}_i\cdot(\vec{x}_j-\vec{x}_k)}
\bigg\{\langle M^{-1}_{AA}(\vec{x}_j,\vec{x}_k;\tau)\rangle+\langle M^{-1}_{BB}(\vec{x}_j,\vec{x}_k;\tau)\rangle \nonumber \\
&& \pm\left(\langle M^{-1}_{AB}(\vec{x}_j,\vec{x}_k;\tau)\rangle+\langle M^{-1}_{BA}(\vec{x}_j,\vec{x}_k;\tau)\rangle\right)\bigg\} \\
&=& \frac{1}{2}\left[G_{AA}(\vec{k}_i,\tau) +G_{BB}(\vec{k}_i,\tau) \pm (G_{AB}(\vec{k}_i,\tau) + G_{BA}(\vec{k}_i,\tau))\right],
\end{eqnarray}
where
\begin{eqnarray}
G_{AB}(\vec{k}_i,\tau) &\equiv&
\frac{1}{N^2}\sum_{\vec{x}_j,\vec{x}_k\in\{\vec{X}\}}e^{i \vec{k}_i\cdot(\vec{x}_j-\vec{x}_k)}\langle a_A(x,\tau) a^\dag_B (y,0) \rangle \nonumber \\
&=&\frac{1}{N^2}\sum_{\vec{x}_j,\vec{x}_k\in\{\vec{X}\}}e^{i \vec{k}_i\cdot(\vec{x}_j-\vec{x}_k)}\langle M^{-1}_{AB}(\vec{x}_j,\vec{x}_k;\tau)\rangle,
\end{eqnarray}
with similar expressions for the other components of the correlator. Note the similarity of this expression to 
Eqn.~\eqref{eqn:analytic correlator}.

Finally, we emphasize a significant difference with respect to lattice QCD calculations.
In a discretized cubic box of length $L$ with $N^3$ lattice points $\vec{x}_i\equiv a(n_x,n_y,n_z)$, where $a$ is the lattice spacing and $n_i\in [0,N)$ integer, the conjugate momenta 
are $\vec{k}_i = \frac{2\pi}{a}(l_x,l_y,l_z)$ with $l_i\in [-\frac{N}{2},\frac{N}{2})$. The triplet of numbers $(l_x,l_y,l_z)$ are independent of each other, and thus the momenta in different spatial 
dimensions can be treated independently. This is in stark contrast to the nanotube case, where for a general tube chirality, the conjugate momenta in the different tube and 
azimuthal directions cannot be treated independently.  

\subsection{Zero-mode analysis\label{sect:zero modes}}

Even though the infrared behavior of the Coulomb interaction is regulated by the finite length of the nanotube, the long-distance nature of the 
interaction coupled with the small physical dimensions of our calculations provide a setting in which the zero momentum modes of the auxiliary field $\Phi$ can introduce 
non-perturbative contributions. The effects of such ``zero-modes'' have been investigated in the context of lattice QCD calculations with long-range 
electromagnetic interactions~\cite{Endres:2015gda}). 

In Appendix~\ref{app:zero modes}, we show for the ``quenched'' approximation (where $\det(M[\Phi]M^\dag[\Phi])=1$ in Eqn.~\eqref{eqn:Z function 3}) in the 
continuum ($\delta\to 0$) and low temperature ($\beta\to \infty$) limits, the zero-modes non-perturbatively induce a Gaussian time dependence in our correlators,
\begin{equation}
\label{eqn:gaussian correlator}
C(\vec{k}_i,\tau)\propto e^{-\alpha \tau^2}e^{-E(\vec{k}_i)\tau},
\end{equation}
where
\begin{equation}
\alpha\equiv\frac{(\hat{v}^{AA}_0+\hat{v}^{AB}_0)}{4\beta N}\ .
\end{equation}
Here $\hat{v}^{AA}_0$ and $\hat{v}^{AB}_0$ are particular matrix elements of the fourier-transformed potential evaluated at zero momentum, and are given by Eqn.~\eqref{eqn:ft V}, and $N$ is the number of hexagonal unit cells in our system.  When extracting the spectrum of our system from the time dependence of our correlators we must take into account the contribution due to the zero modes.  For example, the effective mass obtained by taking the logarithic derivative of the correlator,
\begin{eqnarray*}
-\frac{1}{\kappa}\frac{\partial}{\partial \tau} \log(C(\vec{k}_i,\tau)) &=& \frac{E(\vec{k}_i)}{\kappa}+\frac{(\hat{v}^{AA}_0+\hat{v}^{AB}_0)}{2\kappa\beta N}\tau\\
&=&\frac{E(\vec{k}_i)}{\kappa}+\frac{(\hat{v}^{AA}_0+\hat{v}^{AB}_0)}{2\kappa NN_T}\frac{\tau}{\delta}\ ,
\end{eqnarray*}
will have a linear dependence in $\tau$. 
\begin{table}
\caption{Values of the coefficient $(\hat{v}^{AA}_0+\hat{v}^{AB}_0)/(2\kappa N)$, which is proportional to the zero-mode induced Gaussian correlator term in the quenched approximation shown in Eqn.~\eqref{eqn:gaussian correlator}, as a function of the number of tube unit cells $N_L$.  \label{tab:zero mode slopes}}
\begin{tabular}{c|c|c|c}
\hline\hline
$N_L$ & 3 & 6 & 9\\
\hline
$\frac{\hat{v}^{AA}_0+\hat{v}^{AB}_0}{2\kappa N}$ & 1.30865 & 1.04809 & 0.875358\\
\hline\hline
\end{tabular}
\end{table}

In Table~\ref{tab:zero mode slopes} we give the values of $(\hat{v}^{AA}_0+\hat{v}^{AB}_0)/(2\kappa N)$ for the different systems we consider in this paper.  Though these values were obtained assuming a low-temperature quenched approximation, they nonetheless provide a scale of the expected size of the Gaussian term in our correlators. As we show in the next section, we indeed observe linear behavior in our calculated effective masses which we attribute to the zero-modes of our theory.  However, the slopes of the linear terms do not agree with those shown in Table~\ref{tab:zero mode slopes}, and in principle depend on the momentum state of the electron that we are considering.  This is to be expected as our numerical simulations are fully dynamical (\emph{i.e.} they include the determinant terms in Eqn.~\eqref{eqn:Z function 3}).    

\section{Results\label{sect:results}}

For our Monte Carlo calculations of the $(3,3)$ nanotube, we 
consider three different tube lengths, $N_L= 3, 6$, and $9$~units, which (in principle) allows us to perform an ``infinite volume'' (infinite tube length) extrapolation. For each tube length, 
we generated configurations with $N_t = 64, 80$, and $96$, which allowed us to perform a (temporal) continuum limit extrapolation. Every ensemble of configurations consists of 
30,000~HMC trajectories with 20~decorrelation steps between successive samples.
All calculations were performed with $\beta = 4$ (eV)$^{-1}$, which corresponds to an electron temperature of $0.25$~eV. We use the \texttt{PARDISO} package~\cite{pardiso1,pardiso2,pardiso3} 
to perform inversions of sparse matrices within our HMC algorithm.

\begin{figure}
\includegraphics[angle=-90, width=.48\columnwidth]{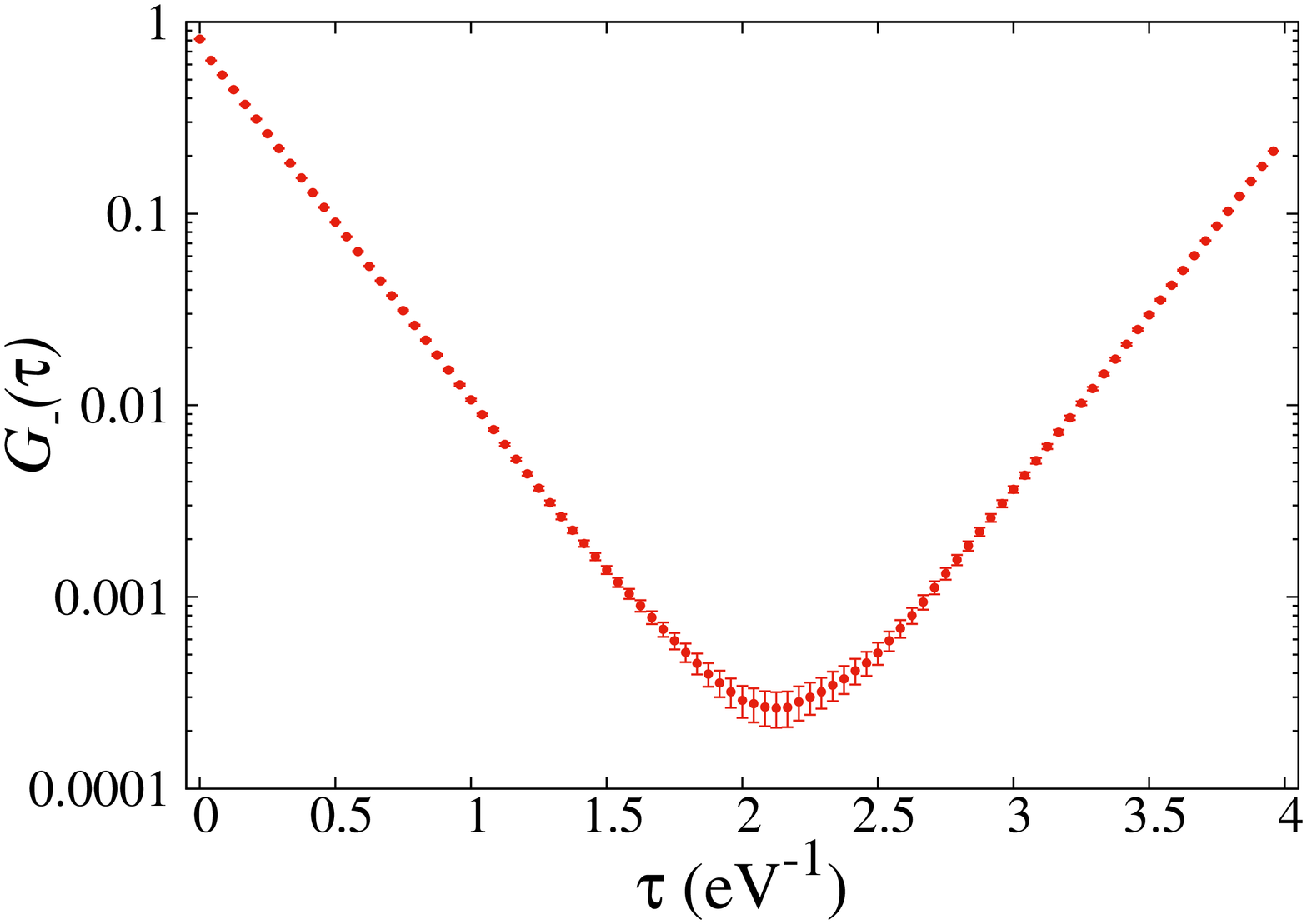} \includegraphics[angle=-90,width=.48\columnwidth]{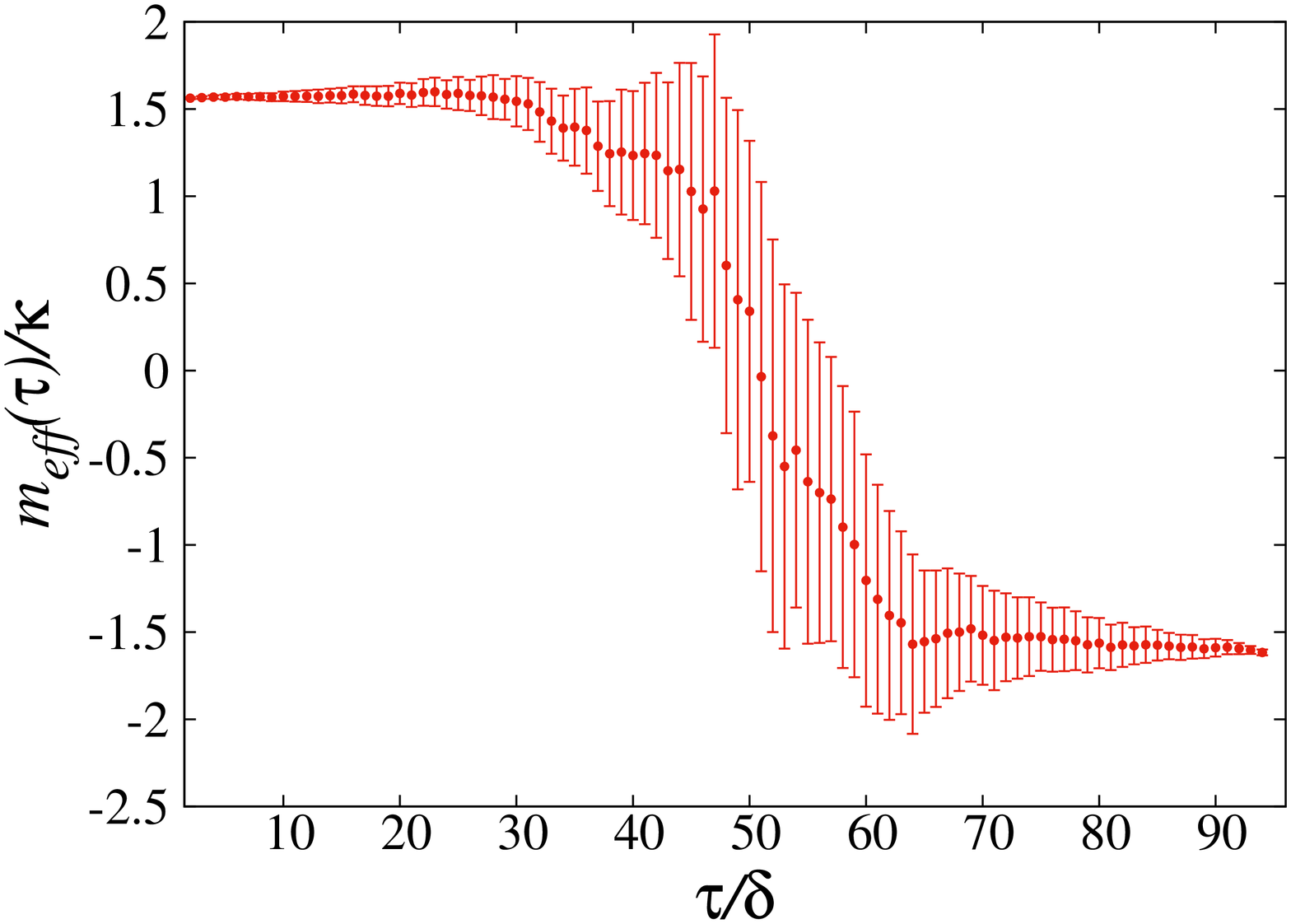}
\caption{Example of $G_-(\tau)$ correlator (left panel) with momentum $|\vec{T}|(|k_{||}|,|k_{\bot,i}|)=\left(\frac{2 \pi }{3},\frac{4 \pi }{3 \sqrt{3}}\right)$.  The corresponding effective mass plot (in units of hopping parameter $\kappa$) is given underneath with $\Delta=2$ (right panel).  Calculations were performed with $N_L=9$ and $N_t=96$. \label{fig:genCorr}}
\end{figure}

For the purpose of presentation \emph{only}, we make use of ``effective mass plots", defined by
\begin{equation}
\label{eqn:eff mass equation}
m_{\text{eff}}((\tau/\delta+\Delta)/2)=-\frac{1}{\Delta}\frac{\ln(G_-(\tau/\delta+\Delta))}{\ln(G_-(\tau/\delta))},
\end{equation}
where $\Delta$ is an integer parameter used for statistical analysis. Such a plot provides visual information on the argument of the exponential of the correlator $G_-(\tau)$. For example, 
in Fig.~\ref{fig:genCorr} we show the correlator $G_-(\tau)$ projected to momentum $|\vec{T}|(|k_{||}|,|k_{\bot,i}|)=\left(\frac{2 \pi }{3},\frac{4 \pi }{3 \sqrt{3}}\right)$ and the corresponding 
effective mass plot in units of $\kappa$. Unless otherwise noted, the uncertainties for all results and figures are obtained via the 
bootstrap procedure~\cite{Press:2007:NRE:1403886}. We also bin our data in order to reduce systematic errors due to autocorrelations. For the results presented below, we bin our data 
every~100 HMC trajectories.

\begin{figure}
\includegraphics[angle=-90, width=.5\columnwidth]{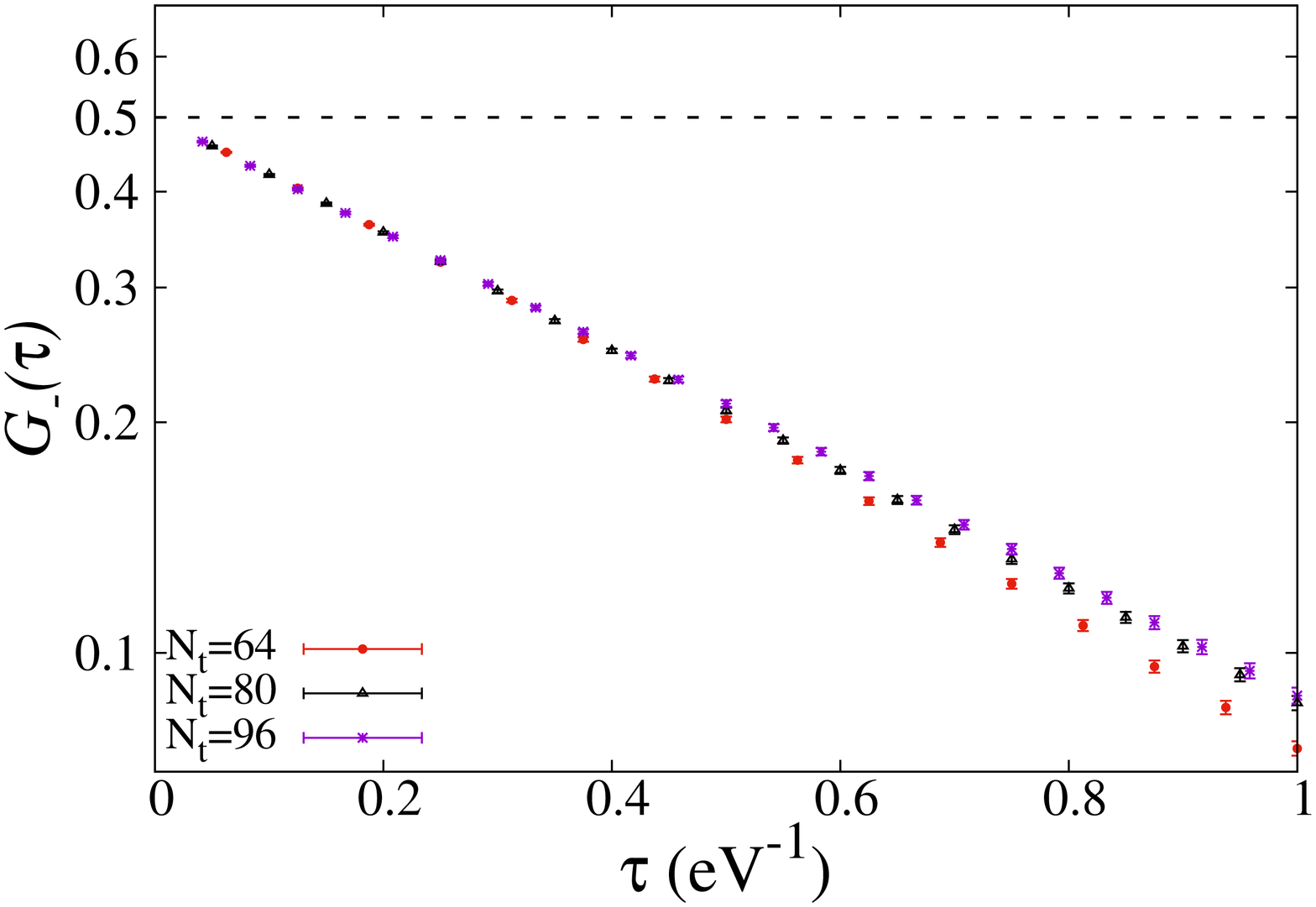}\includegraphics[angle=-90, width=.5\columnwidth]{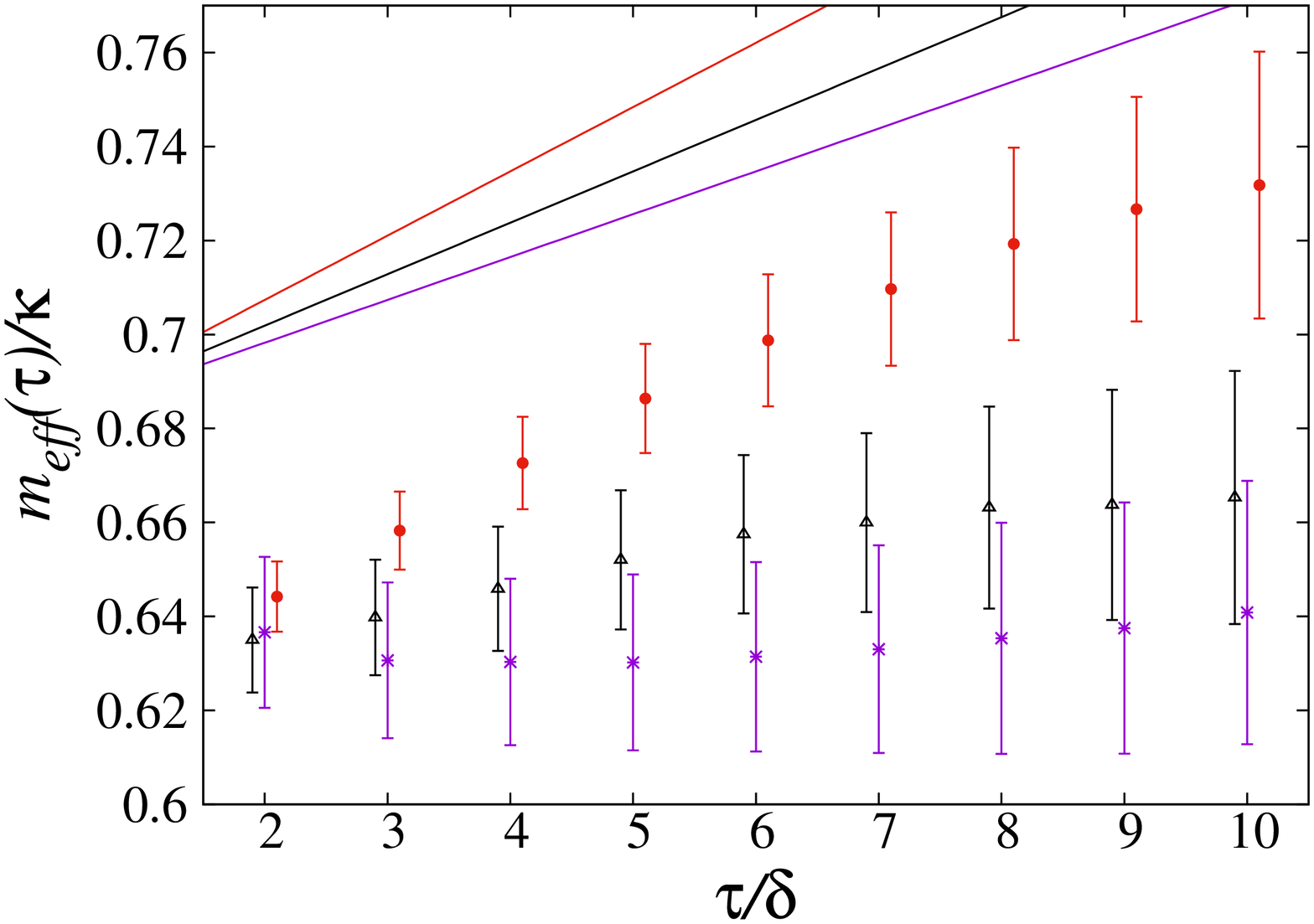}
\caption{The $G_-(\tau)$ correlator (left panel) at the Dirac point for a (3,3) armchair with $N_L=$9 unit cells, using different numbers of timesteps as shown in the figure.  Note that in this case $G_+(\tau)=G_-(\tau)$.  The dashed line is the non-interacting result.  The right panel shows the corresponding effective masses (points).  Also shown in the right panel are the calculated correlators (lines) in the quenched approximation.  To facilitate presentation, the quenched results have been shifted above the effective mass points such that their y-intercepts (at $\tau/\delta=0$) are .68.\label{fig:dirac}}
\end{figure}

We have also benchmarked our code to cases where analytic solutions are known, or where solutions can be obtained via direct numerical diagonalization, specifically the two- and four-site
Hubbard models. We discuss these benchmark calculations in Appendix~\ref{sect:benchmark}. 

\subsection{The Dirac K point\label{sect:dirac point}}

We now describe in detail our analysis for the Dirac K point. In the left panel of Fig.~\ref{fig:dirac}, we show the $G_{-}(\tau)$ correlator on a logarithmic scale at the Dirac point for different 
values of $N_t$. In the right panel of Fig.~\ref{fig:dirac}, we show the corresponding effective mass using $\Delta=2$. Also shown is the expected linear behavior of the effective mass 
for these calculations in the quenched approximation (lines), as discussed in Section~\ref{sect:zero modes}. 
Our dynamical calculations exhibit the same qualitative features as the quenched approximation. In particular, there is a clear linear behavior for the effective mass, particularly for smaller $N_t$. 
However, the slopes are not as steep as in the quenched approximation. Indeed, for $N_t = 96$, a linear contribution is hardly discernible in the dynamical case for the shown range of time steps, 
but is nevertheless statistically significant.

\begin{figure}
\includegraphics[width=.5\columnwidth]{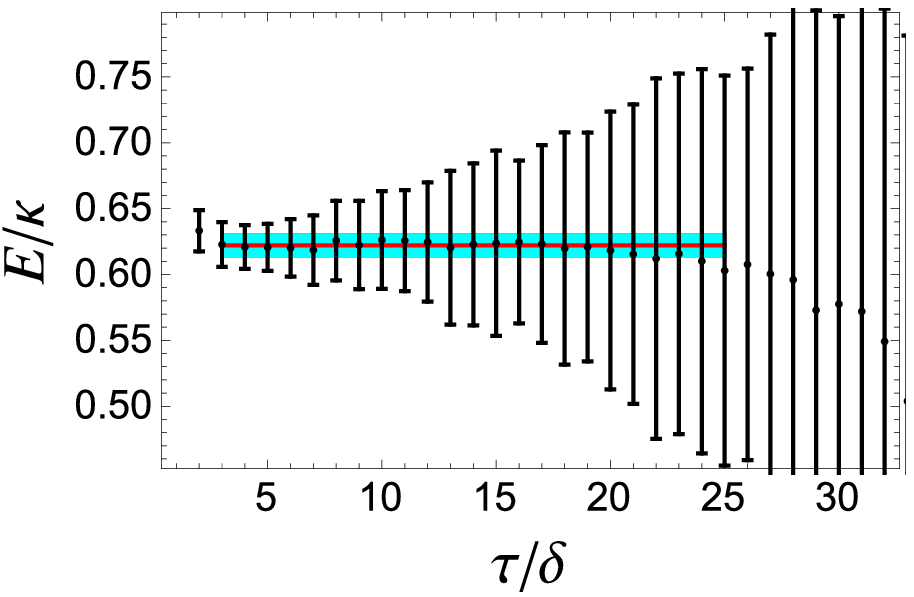}\includegraphics[width=.5\columnwidth]{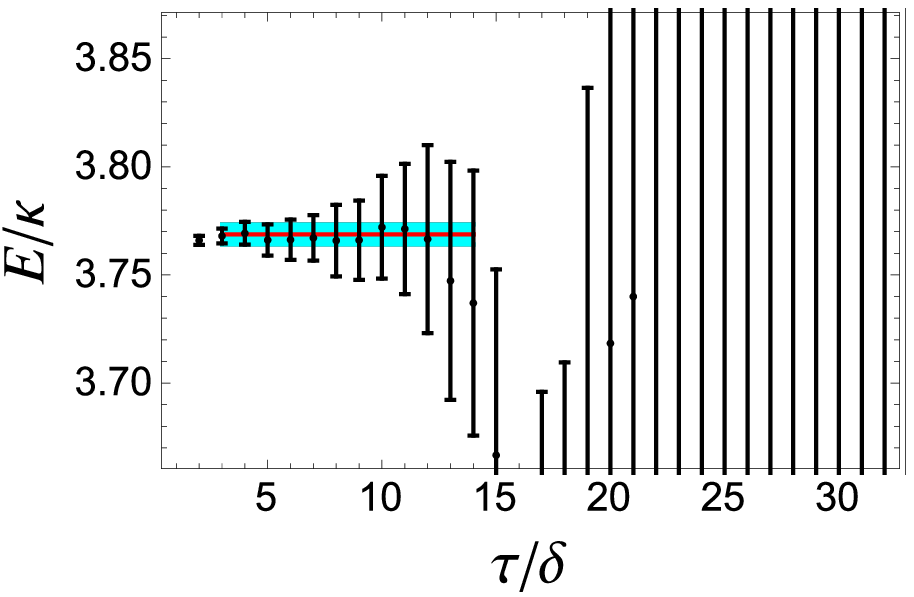}
\caption{The extracted energy at the Dirac point (left panel, cyan band) for the $N_L=9$, $N_t=96$ calculation using the fitting procedure described in the text.  The right panel shows extracted energy at the $\Gamma$ point.  The effective mass points are given by the black data points and are shown for comparison.  They were \emph{not} used in the fitting procedure (see text).  In both plots, the fitted Gaussian term has been subtracted from the effective mass points.   \label{fig:effmass_u9_constantFit}}
\end{figure}

As we do not know the analytic form of the slope in the dynamical case, we perform simultaneous fits of both the leading exponential term (which provides the energy) and Gaussian term (which is responsible for the slope) directly to our correlators within a specific time window to extract our spectrum.  We stress that we do not perform fits to the effective mass points (Eqn.~\eqref{eqn:eff mass equation}) themselves, but only to the correlator.
In the left panel of Fig.~\ref{fig:effmass_u9_constantFit} we show the extracted energy for the Dirac point determined by this fitting procedure.  Note that the Gaussian contribution has been subtracted from the effective mass points in this figure. The agreement between the effective mass points and our fitted energy (cyan band) provides a consistency check on our fitting routines.  The time window giving the optimal fit is given by the horizontal width of the band in the figure, whereas the the height provides the 1-$\sigma$ uncertainty, which in this case is the combination (in quadrature) of statistical and systematic errors.  We estimate systematic errors by analyzing the distribution of fit results performed with varying time-window widths.  

\begin{figure}
\includegraphics[angle=0, width=.5\columnwidth]{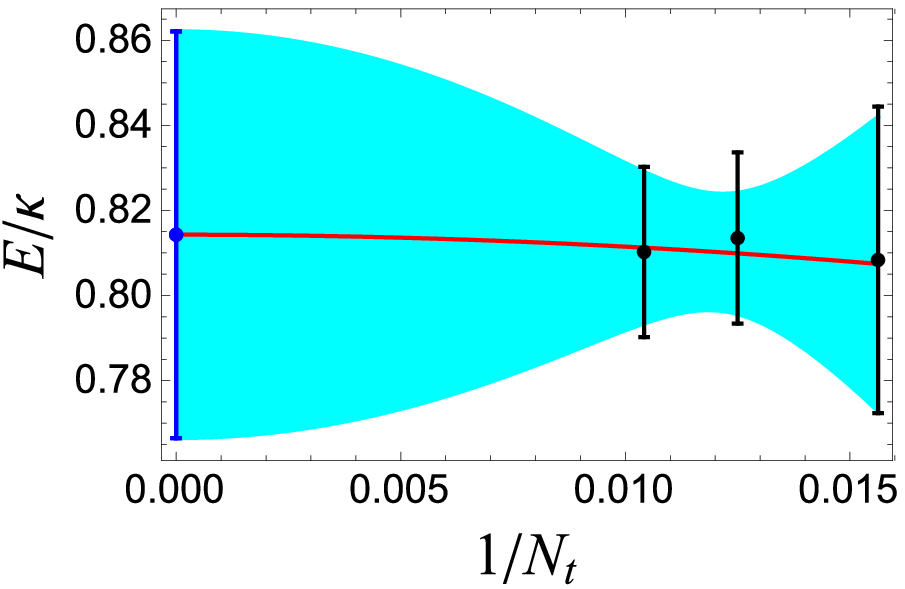}\includegraphics[angle=0, width=.5\columnwidth]{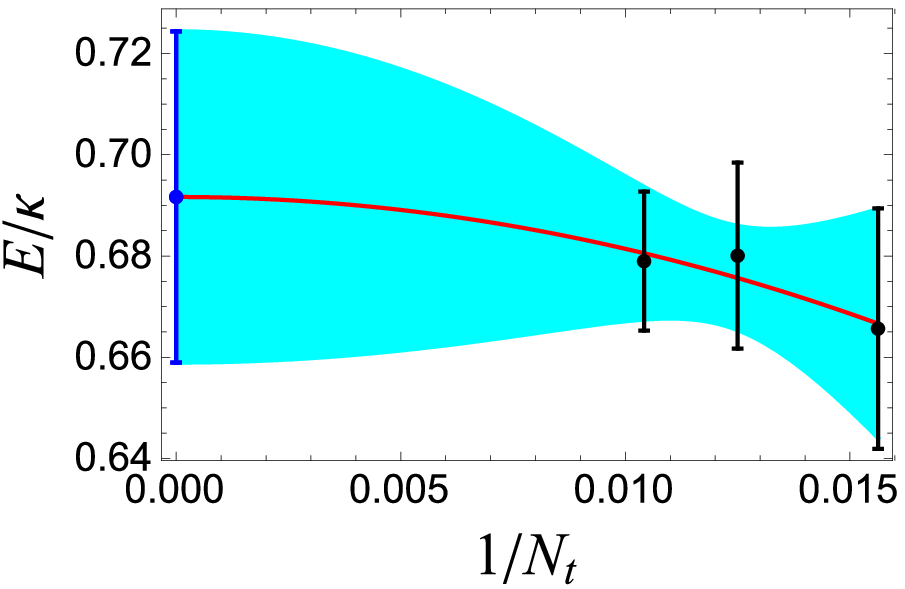}
\includegraphics[angle=0, width=.5\columnwidth]{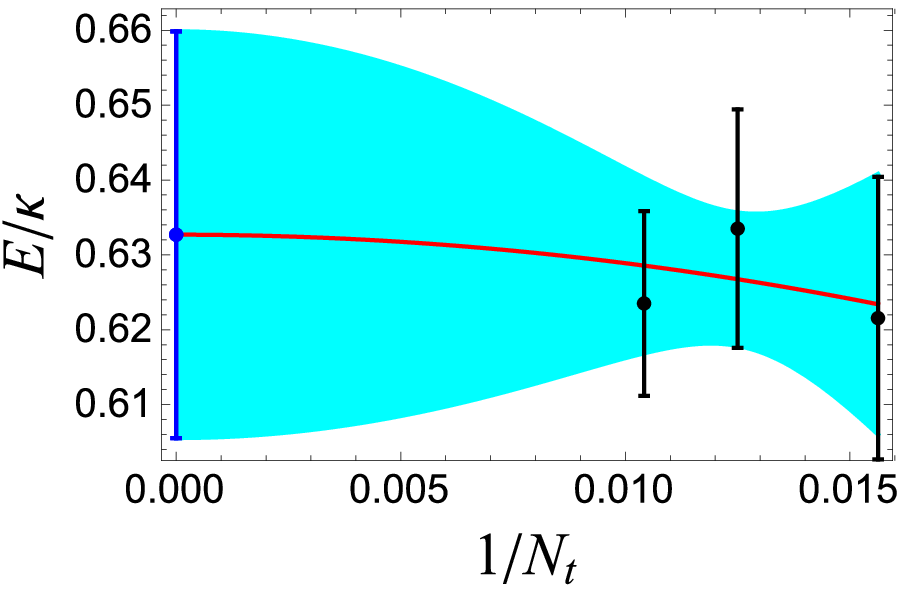}\includegraphics[angle=0, width=.5\columnwidth]{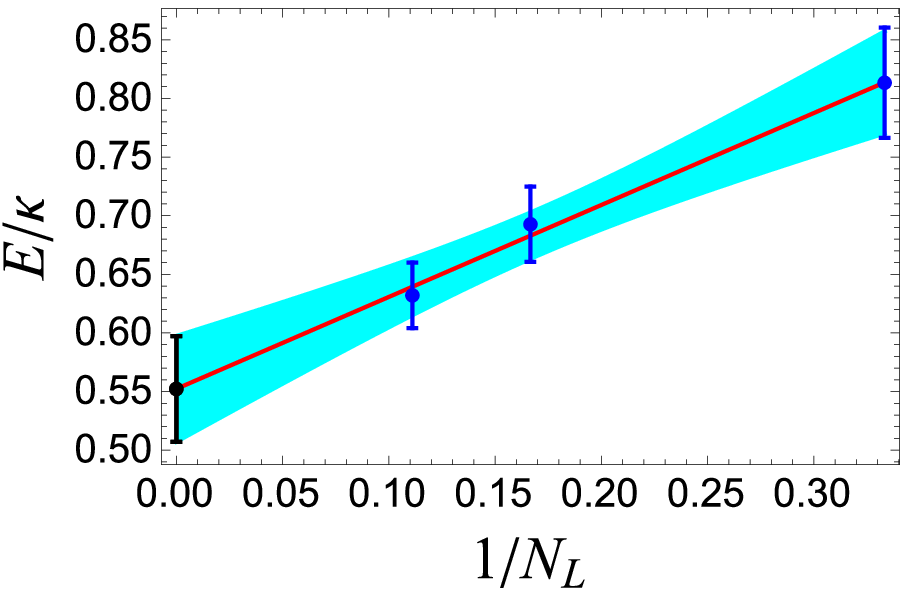}
\caption{Continuum and inifinite-volume extrapolations of the Dirac point.  The continuum-limit extrapolations for the $N_L=3$ (upper left), $N_L=6$ (upper right), and $N_L=9$ (bottom left) systems, using a scaling function quadratic in $\delta$.   The continuum-limit results are used in an infinite-volume extrapolation (bottom right), assuming a linear dependence in $1/N_L$. \label{fig:infinite volume extrapolation}}
\end{figure}

In Tabs.~\ref{tab:nl3}-\ref{tab:nl9}, we give the extracted energy at the Dirac point for each of our Monte Carlo calculations, as obtained from data with 
momentum $|\vec{T}|(|k_{||}|,|k_{\bot,i}|)=\left(\frac{2 \pi }{3},\frac{2 \pi }{\sqrt{3}}\right)$.  For each value of $N_L$, we perform a continuum ($\delta\to 0$) extrapolation using the functional 
form $E(\delta)=E_0+C_t \delta^2$. The extrapolation is determined by multiple fits of the data, where each data point is sampled according to a normal distribution given by the 
combined statistical and systematic errors reported in Tabs.~\ref{tab:nl3}-\ref{tab:nl9}. This distribution of fits is then used to estimate the uncertainty of the extrapolation. 
The results in the continuum limit, and their associated uncertainties, are given in the last columns of Tabs.~\ref{tab:nl3}-\ref{tab:nl9}. 
In Fig.~\ref{fig:infinite volume extrapolation}, we show this extrapolation for the Dirac point.  As can be seen from Fig.~\ref{fig:infinite volume extrapolation},
the data points at each value of $\delta$ are mutually consistent within uncertainties. This prevents us from determining with certainty that the discrete-time scaling is quadratic in $\delta$. 
Increased statistics, in addition to calculations at smaller values of $\delta$, would be needed.

Using the continuum-limit results at each $N_L$, we finally perform an infinite-volume, $N_L\rightarrow\infty$ (infinitely long tube) extrapolation.  We find that our data extrapolates well with a simple linear dependence on $1/N_L$, and therefore we use the following functional form to perform our extrapolation: $E(N_L)=E_\infty + C_L /N_L$.   Quoted uncertainties of our infinite-volume extrapolation are determined in a similar fashion as our continuum-limit extrapolations.  The extrapolation is shown in the bottom right panel of Fig.~\ref{fig:infinite volume extrapolation}.  We find that the energy at the Dirac point is $E_K/\kappa=.551(46)$.
We note that the true volume dependence of our calculations may be something other than linear (see \cite{Assaad:2013xua}, for example, for a discussion of finite-volume scaling within low-dimension systems).  However, our three points, and their associated uncertainties, are not sufficient to discern anything that deviates from linear dependence.  Future studies, with larger values or $N_L$, should provide answers to this question. 

\subsection{Spectrum of the (3,3) carbon nanotube}

\begin{figure}[t]
%\includegraphics[angle=0, width=.33\columnwidth]{u3nt64effmass.eps}\includegraphics[angle=0, width=.33\columnwidth]{u3nt80effmass.eps}\includegraphics[angle=0, width=.33\columnwidth]{u3effmass.eps}
%\\
%\includegraphics[angle=0, width=.33\columnwidth]{u6nt64effmass.eps}\includegraphics[angle=0, width=.33\columnwidth]{u6nt80effmass.eps}\includegraphics[angle=0, width=.33\columnwidth]{u6effmass.eps}
%\\
%\includegraphics[angle=0, width=.33\columnwidth]{u9nt64effmass.eps}\includegraphics[angle=0, width=.33\columnwidth]{u9nt80effmass.eps}\includegraphics[angle=0, width=.33\columnwidth]{u9effmass.eps}
\includegraphics[width=\columnwidth]{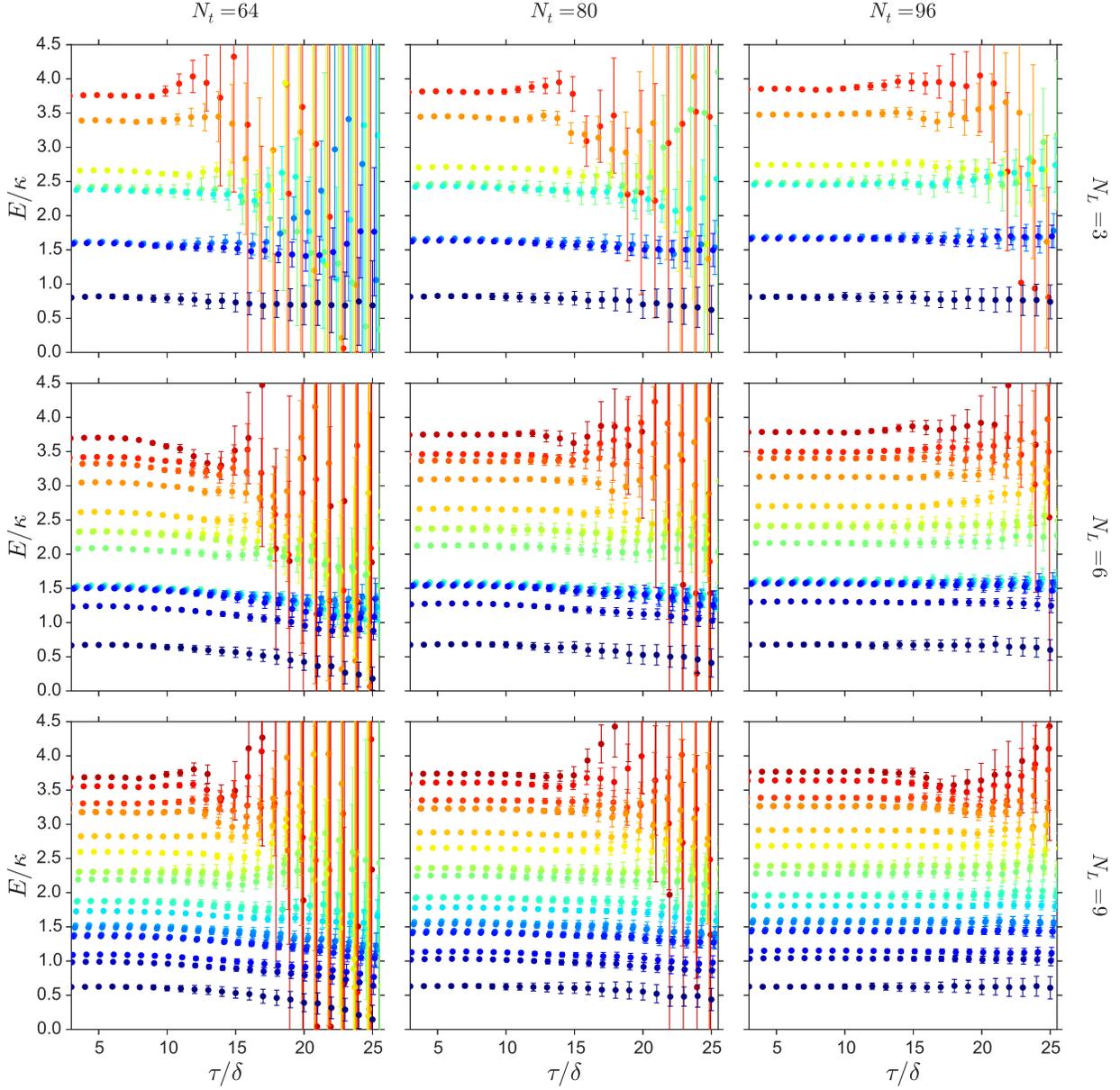}
\caption{The effective masses from all accessible momenta, with Gaussian term subtracted.  From top to bottom, the rows label $N_L=3$, 6, and 9 calculations, respectively.  From left to right, the columns represent $N_t=64$, 80, and 96 calculations, respectively. \label{fig:all effmasses}}
\end{figure}

With our analysis formalism described in the preceding section, we now present the results of the remaining spectrum points for the (3,3) tube.  In Fig.~\ref{fig:all effmasses} we show the effective masses (with Gaussian term subtracted) for all accessible momenta for each of our calculations.  Note that the $N_L=3$ case has less effective mass points than $N_L=6$, which in turn has less effective mass points than $N_L=9$.  This is due to the fact that the number of accessible momenta increases as $N_L$ increases.  In generating these figures, only the $G_-(\vec{k},\tau)$ electron correlators were analyzed since statistics for the $G_+(\vec{k},\tau)$ correlators were too poor for analysis.  Also, correlators with degenerate energies were combined to increase statistics.  A close-up of the effective mass plot for the $\Gamma$ point, as well as the fitted energy, is shown in the right panel of Fig.~\ref{fig:effmass_u9_constantFit} for the $N_L=9$, $N_t=96$ case.

\begin{figure}
\includegraphics[angle=0, width=.5\columnwidth]{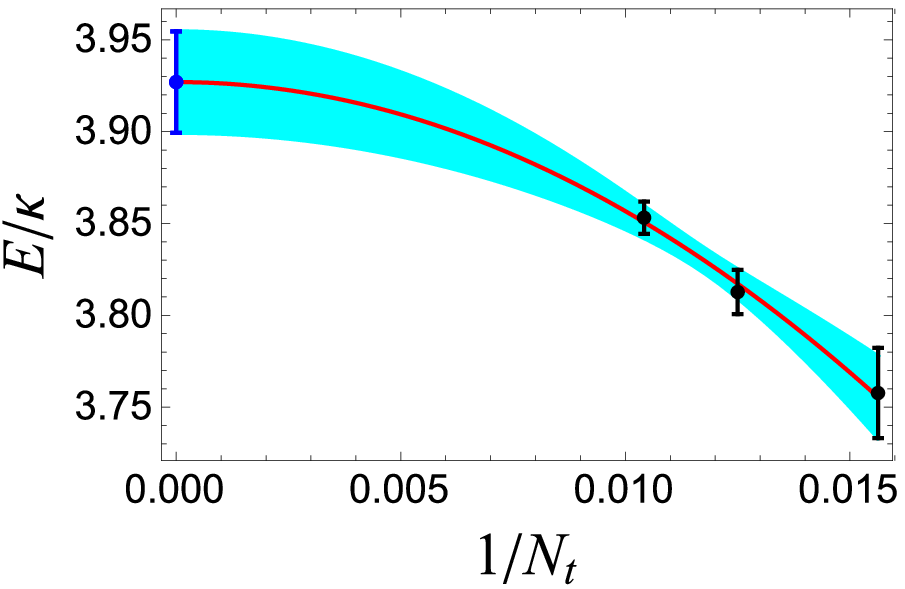}\includegraphics[angle=0, width=.5\columnwidth]{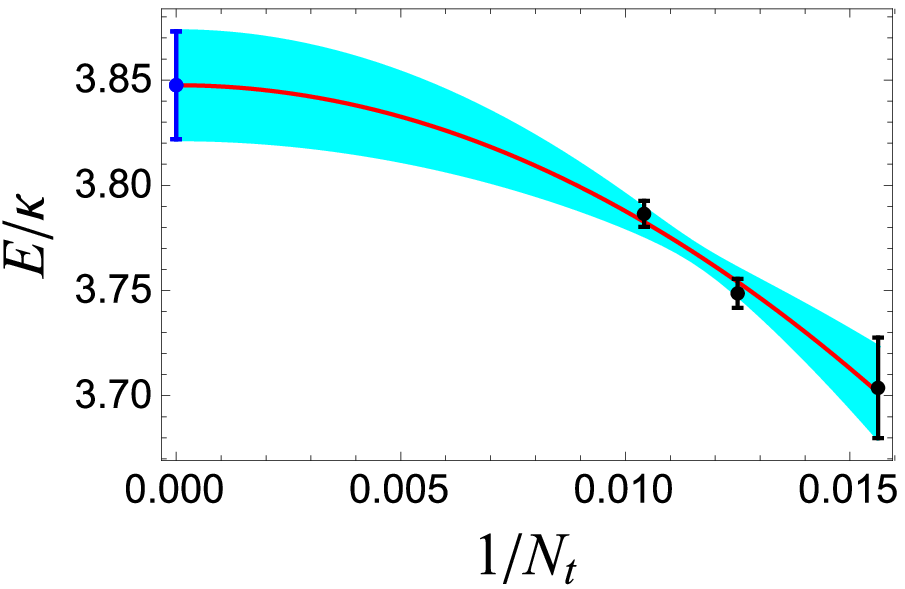}
\includegraphics[angle=0, width=.5\columnwidth]{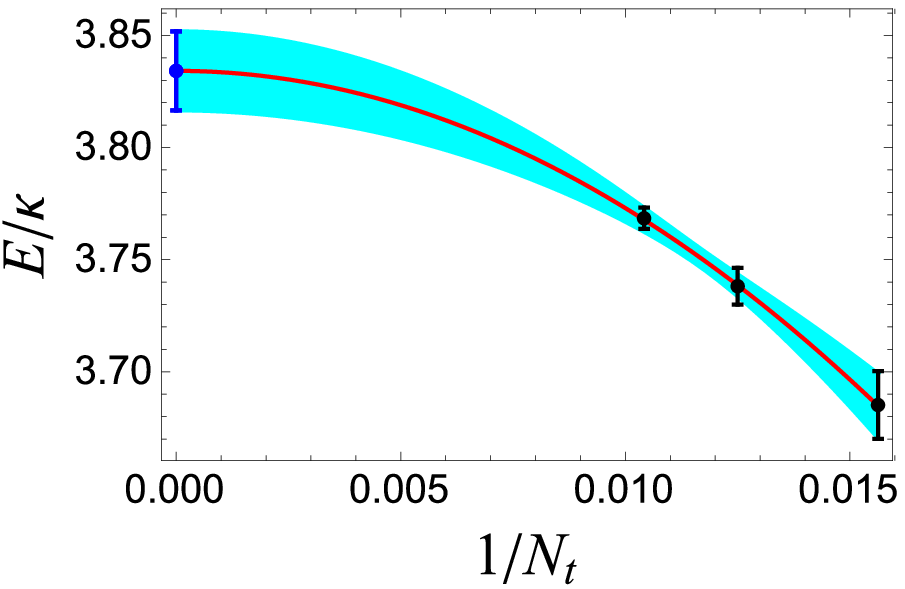}\includegraphics[angle=0, width=.5\columnwidth]{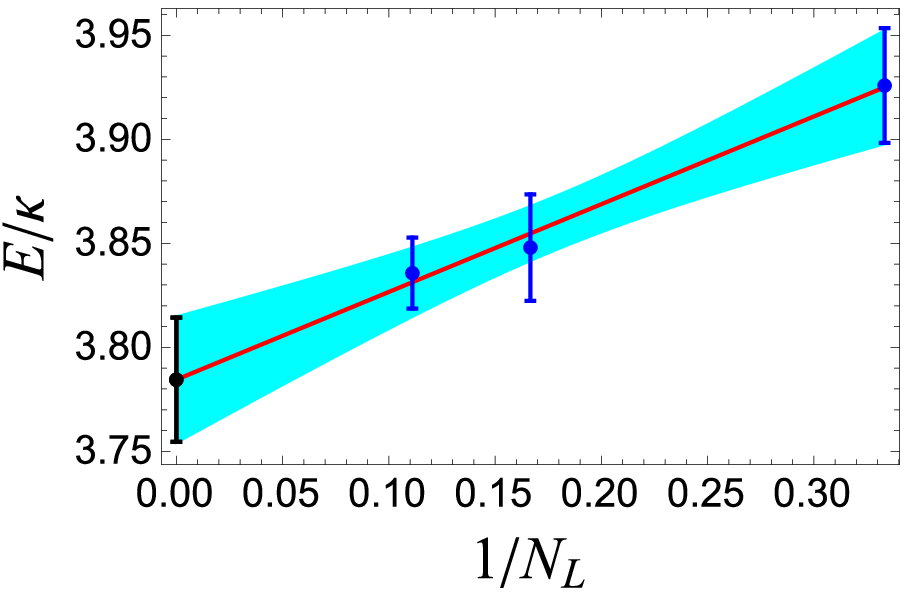}
\caption{Continuum and inifinite-volume extrapolations of the $\Gamma$ point.  The continuum-limit extrapolations for the $N_L=3$ (upper left), $N_L=6$ (upper right), and $N_L=9$ (bottom left) systems, using a scaling function quadratic in $\delta$.   The continuum-limit results are used in an infinite-volume extrapolation (bottom right), assuming a linear dependence in $1/N_L$.  \label{fig:extrapolations Gamma}}
\end{figure}

In tabs.~\ref{tab:nl3}-\ref{tab:nl9} we list all the extracted energies for each $N_L$, $N_t$, and momentum point.  The continuum-limit extrapolation is given in the last column of these tables.  Figure~\ref{fig:extrapolations Gamma} shows another example of this extrapolation but this time at the $\Gamma$ point.  As opposed to the Dirac point, the data points at different $\delta$ are statistically distinct, but still not sufficient to discern linear or quadratic in $\delta$ scaling.  To be consistent with the Dirac point analysis, we assume a quadratic scaling for the $\Gamma$ point as well as all other points on the dispersion.  Again, future studies with smaller values of $\delta$ should tell whether such an assumption is valid.

\begin{figure}
\includegraphics[angle=0, width=.5\columnwidth]{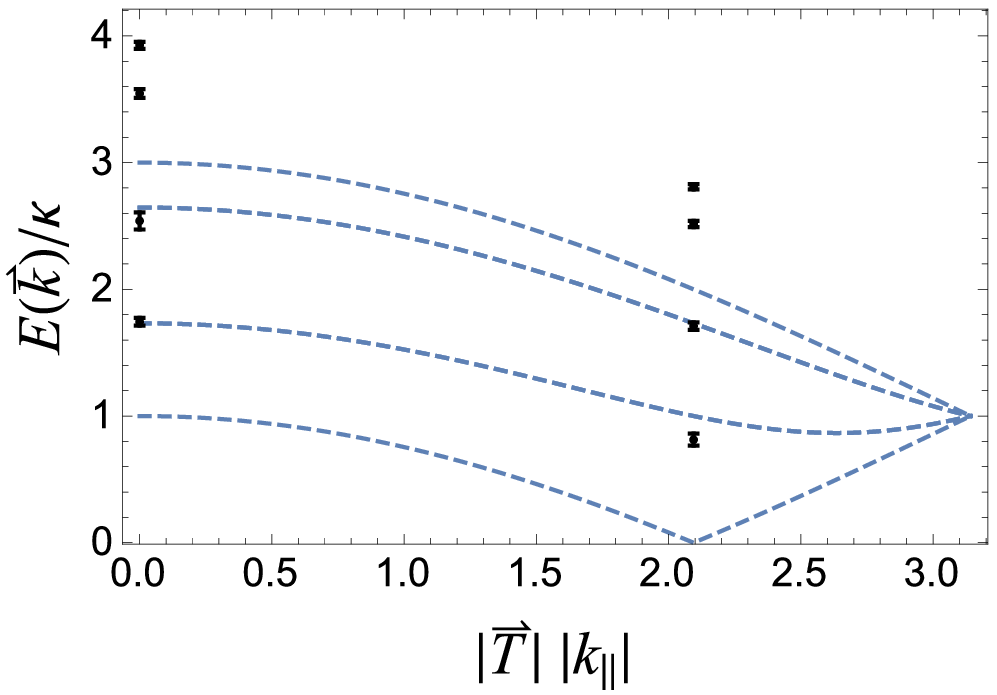}\includegraphics[angle=0, width=.5\columnwidth]{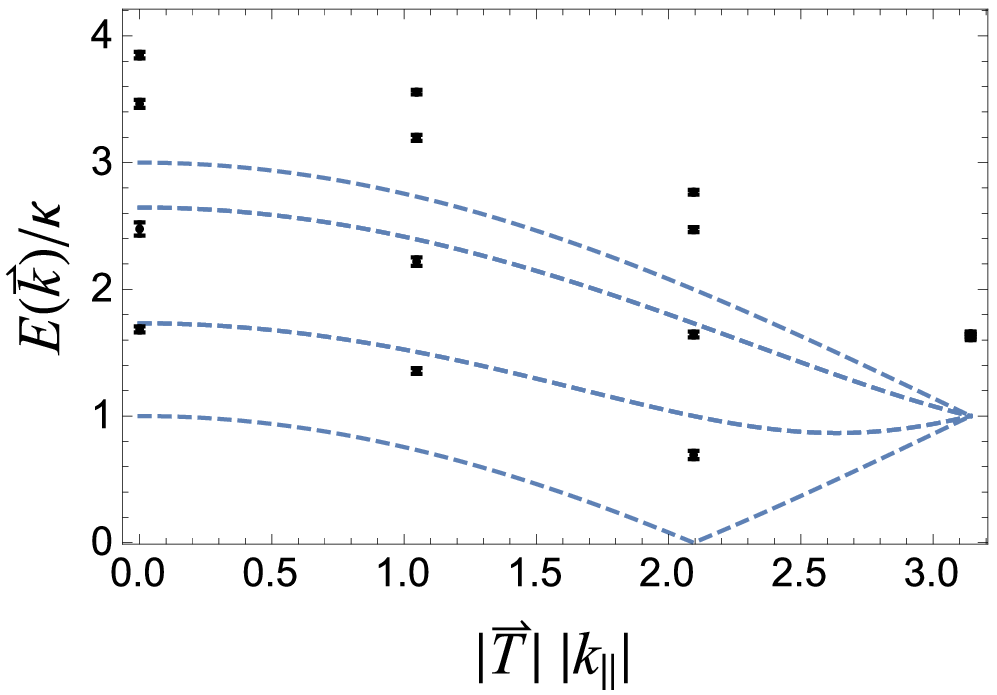}
\includegraphics[angle=0, width=.5\columnwidth]{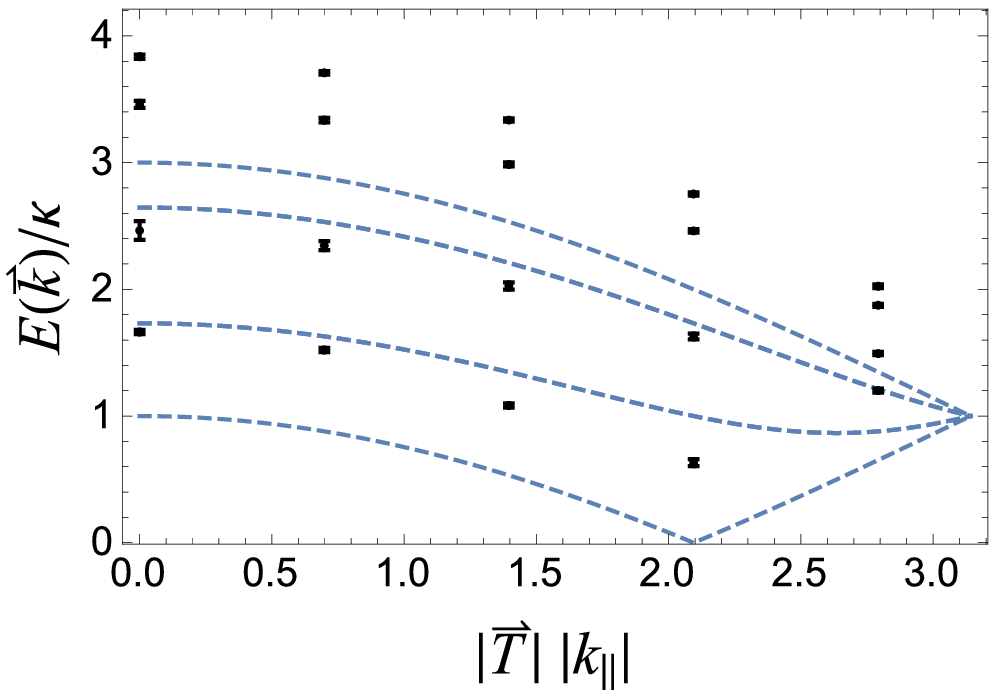}\includegraphics[angle=0, width=.5\columnwidth]{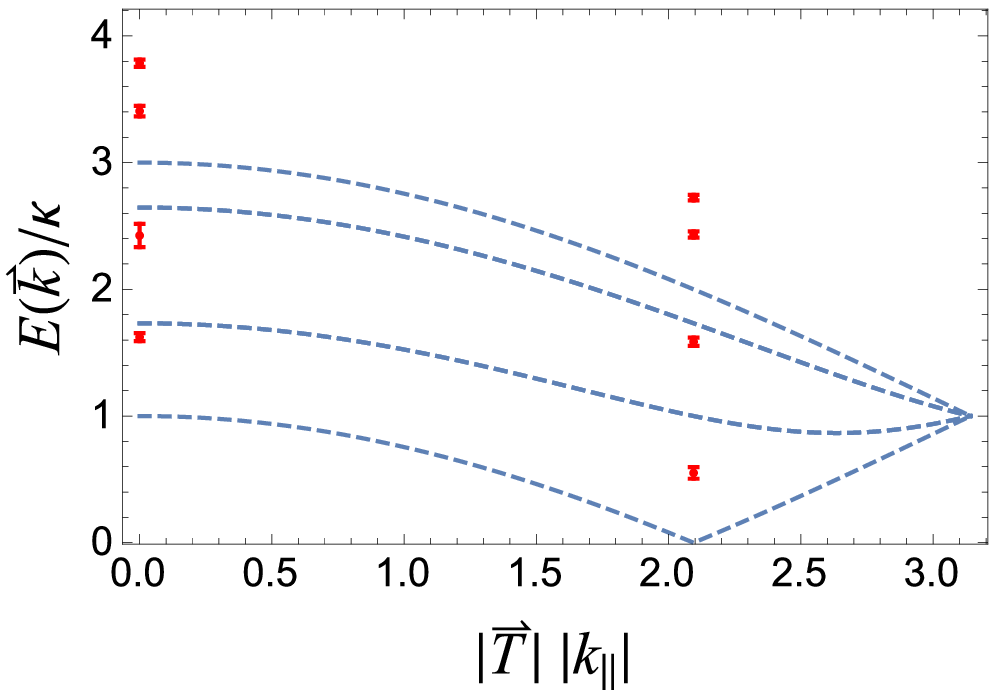}
\caption{Continuum-limit extrapolated spectrum of quasi-electron (black points) compared to non-interacting dispersion relation (dashed line) for the (3,3) tube with $N_L=3$ (upper left), 6 (upper right), and 9 (bottom left) calculations.  The bottom right panel shows the infinite-volume extrapolation of points common to all $N_L=3$, 6, and 9 systems, given by the last column in Table~\ref{tab:infinite volume points}. \label{fig:dispersions}}
\end{figure}

In Fig.~\ref{fig:dispersions} we show all continuum-limit results for each $N_L$ system, compared to the non-interacting dispersion (dashed lines).  As we only analyze the $G_-(\tau)$ correlators, and combine degenerate correlators when possible to increase statistics, we only show the upper-right portion of the dispersions in this figure (compare with Fig.~\ref{fig:dispersion33}).

\begin{table}
\center
\caption{Extracted energies for $N_L=3$ at different $N_t$ points.  All energies are given in units of the hopping parameter $\kappa$.  The first uncertainty is statistical, the second systematic.  The last column gives the continuum-limit extrapolation assuming a $\delta^2$ scaling.  Statistical and systematic errors were combined in quadrature to perform the extrapolation.\label{tab:nl3}}
\begin{tabular}{c|c|c|c|c}
\hline\hline
$|\vec{T}|\  (|k_{||}|,|k_{\bot,i}|)$ & $E(N_t=64)$ & $E(N_t=80)$ & $E(N_t=96)$ & $E(N_t=\infty)$ \\
\hline

$\left(0,0\right)$ & 3.758(15)(19) & 3.813(9)(8) & 3.853(7)(5) & 3.926(28)\\
$\left(0,\frac{2 \pi }{3 \sqrt{3}}\right)$ & 3.391(23)(19) & 3.447(14)(19) & 3.476(10)(4) & 3.546(35)\\
$\left(0,\frac{4 \pi }{3 \sqrt{3}}\right)$ & 2.413(44)(36) & 2.460(28)(11) & 2.483(22)(6) & 2.540(68)\\
$\left(0,\frac{2 \pi }{\sqrt{3}}\right)$ & 1.614(13)(25) & 1.655(8)(12) & 1.688(6)(5) & 1.744(31)\\
\hline
$\left(\frac{2 \pi }{3},0\right)$ & 2.658(9)(17) & 2.708(5)(12) & 2.744(4)(3) & 2.810(21)\\
$\left(\frac{2 \pi }{3},\frac{2 \pi }{3 \sqrt{3}}\right)$ & 2.376(9)(19) & 2.423(6)(16) & 2.455(5)(2) & 2.516(24)\\
$\left(\frac{2 \pi }{3},\frac{4 \pi }{3 \sqrt{3}}\right)$ & 1.591(16)(22) & 1.628(10)(11) & 1.661(7)(3) & 1.709(29)\\
$\left(\frac{2 \pi }{3},\frac{2 \pi }{\sqrt{3}}\right)$ & 0.808(25)(26) & 0.814(18)(9) & 0.810(11)(4) & 0.813(47)\\

\hline
\hline
\end{tabular}
\end{table}

\begin{table}
\center
\caption{Similar to Table~\ref{tab:nl3}, but for $N_L=6$.\label{tab:nl6}}
\begin{tabular}{c|c|c|c|c}
\hline\hline
$|\vec{T}| \ (|k_{||}|,|k_{\bot,i}|)$ & $E(N_t=64)$ & $E(N_t=80)$ & $E(N_t=96)$ & $E(N_t=\infty)$ \\
\hline

$\left(0,0\right)$ & 3.704(9)(22) & 3.749(6)(3) & 3.786(5)(3) & 3.848(26)\\
$\left(0,\frac{2 \pi }{3 \sqrt{3}}\right)$ & 3.322(16)(21) & 3.368(15)(8) & 3.403(9)(3) & 3.464(32)\\
$\left(0,\frac{4 \pi }{3 \sqrt{3}}\right)$ & 2.330(32)(29) & 2.371(29)(8) & 2.418(15)(3) & 2.477(53)\\
$\left(0,\frac{2 \pi }{\sqrt{3}}\right)$ & 1.540(8)(18) & 1.584(7)(8) & 1.623(5)(5) & 1.684(23)\\
\hline
$\left(\frac{\pi }{3},0\right)$ & 3.417(5)(15) & 3.462(5)(6) & 3.498(3)(1) & 3.556(18)\\
$\left(\frac{\pi }{3},\frac{2 \pi }{3 \sqrt{3}}\right)$ & 3.047(10)(18) & 3.094(9)(4) & 3.132(5)(2) & 3.195(23)\\
$\left(\frac{\pi }{3},\frac{4 \pi }{3 \sqrt{3}}\right)$ & 2.081(18)(21) & 2.121(17)(7) & 2.161(10)(2) & 2.219(33)\\
$\left(\frac{\pi }{3},\frac{2 \pi }{\sqrt{3}}\right)$ & 1.234(7)(19) & 1.272(5)(9) & 1.304(4)(5) & 1.355(22)\\
\hline
$\left(\frac{2 \pi }{3},0\right)$ & 2.617(6)(17) & 2.664(4)(4) & 2.702(3)(2) & 2.766(18)\\
$\left(\frac{2 \pi }{3},\frac{2 \pi }{3 \sqrt{3}}\right)$ & 2.325(7)(17) & 2.373(6)(6) & 2.408(4)(1) & 2.471(20)\\
$\left(\frac{2 \pi }{3},\frac{4 \pi }{3 \sqrt{3}}\right)$ & 1.513(12)(15) & 1.556(9)(8) & 1.588(6)(2) & 1.644(22)\\
$\left(\frac{2 \pi }{3},\frac{2 \pi }{\sqrt{3}}\right)$ & 0.666(16)(17) & 0.680(16)(9) & 0.679(13)(3) & 0.693(32)\\
\hline
$\left(\pi,0\right)$ & 1.499(7)(14) & 1.540(7)(7) & 1.572(4)(4) & 1.627(18)\\
$\left(\pi,\frac{2 \pi }{3 \sqrt{3}}\right)$ & 1.518(6)(16) & 1.560(5)(8) & 1.591(3)(4) & 1.647(19)\\
$\left(\pi,\frac{4 \pi }{3 \sqrt{3}}\right)$ & 1.518(6)(13) & 1.561(5)(9) & 1.590(3)(4) & 1.647(18)\\
$\left(\pi,\frac{2 \pi }{\sqrt{3}}\right)$ & 1.499(7)(18) & 1.543(6)(11) & 1.568(5)(5) & 1.622(23)\\

\hline
\hline
\end{tabular}
\end{table}

\begin{table}
\center
\caption{Similar to Table~\ref{tab:nl3}, but for $N_L=9$.\label{tab:nl9}}
\begin{tabular}{c|c|c|c|c}
\hline\hline
$|\vec{T}| \ (|k_{||}|,|k_{\bot,i}|)$ & $E(N_t=64)$ & $E(N_t=80)$ & $E(N_t=96)$ & $E(N_t=\infty)$ \\
\hline

$\left(0,0\right)$ & 3.685(7)(13) & 3.738(7)(5) & 3.768(4)(2) & 3.836(17)\\
$\left(0,\frac{2 \pi }{3 \sqrt{3}}\right)$ & 3.299(19)(11) & 3.351(11)(3) & 3.390(10)(2) & 3.459(27)\\
$\left(0,\frac{4 \pi }{3 \sqrt{3}}\right)$ & 2.305(39)(52) & 2.362(20)(7) & 2.393(19)(4) & 2.464(74)\\
$\left(0,\frac{2 \pi }{\sqrt{3}}\right)$ & 1.520(7)(11) & 1.576(7)(10) & 1.598(4)(4) & 1.663(17)\\
\hline
$\left(\frac{2 \pi }{9},0\right)$ & 3.554(6)(10) & 3.608(5)(3) & 3.640(3)(3) & 3.709(14)\\
$\left(\frac{2 \pi }{9},\frac{2 \pi }{3 \sqrt{3}}\right)$ & 3.177(13)(9) & 3.229(8)(3) & 3.266(6)(1) & 3.336(19)\\
$\left(\frac{2 \pi }{9},\frac{4 \pi }{3 \sqrt{3}}\right)$ & 2.193(27)(11) & 2.247(14)(5) & 2.278(14)(3) & 2.345(36)\\
$\left(\frac{2 \pi }{9},\frac{2 \pi }{\sqrt{3}}\right)$ & 1.382(5)(13) & 1.436(5)(11) & 1.458(3)(3) & 1.522(17)\\
\hline
$\left(\frac{4 \pi }{9},0\right)$ & 3.179(5)(10) & 3.232(5)(5) & 3.267(3)(3) & 3.336(13)\\
$\left(\frac{4 \pi }{9},\frac{2 \pi }{3 \sqrt{3}}\right)$ & 2.823(11)(7) & 2.879(6)(6) & 2.914(6)(2) & 2.986(16)\\
$\left(\frac{4 \pi }{9},\frac{4 \pi }{3 \sqrt{3}}\right)$ & 1.877(20)(9) & 1.931(11)(7) & 1.960(10)(3) & 2.026(28)\\
$\left(\frac{4 \pi }{9},\frac{2 \pi }{\sqrt{3}}\right)$ & 0.987(6)(11) & 1.028(6)(11) & 1.038(4)(3) & 1.083(17)\\
\hline
$\left(\frac{2 \pi }{3},0\right)$ & 2.596(5)(9) & 2.650(4)(6) & 2.684(3)(2) & 2.753(13)\\
$\left(\frac{2 \pi }{3},\frac{2 \pi }{3 \sqrt{3}}\right)$ & 2.299(8)(10) & 2.357(5)(7) & 2.390(4)(1) & 2.461(14)\\
$\left(\frac{2 \pi }{3},\frac{4 \pi }{3 \sqrt{3}}\right)$ & 1.490(14)(12) & 1.541(8)(8) & 1.565(7)(2) & 1.628(23)\\
$\left(\frac{2 \pi }{3},\frac{2 \pi }{\sqrt{3}}\right)$ & 0.622(16)(10) & 0.634(15)(5) & 0.624(7)(2) & 0.631(23)\\
\hline
$\left(\frac{8 \pi }{9},0\right)$ & 1.871(5)(11) & 1.926(4)(8) & 1.956(2)(3) & 2.025(15)\\
$\left(\frac{8 \pi }{9},\frac{2 \pi }{3 \sqrt{3}}\right)$ & 1.729(5)(9) & 1.781(3)(7) & 1.809(2)(1) & 1.874(12)\\
$\left(\frac{8 \pi }{9},\frac{4 \pi }{3 \sqrt{3}}\right)$ & 1.362(6)(10) & 1.412(4)(9) & 1.433(3)(1) & 1.493(13)\\
$\left(\frac{8 \pi }{9},\frac{2 \pi }{\sqrt{3}}\right)$ & 1.093(6)(11) & 1.135(5)(10) & 1.151(3)(3) & 1.201(15)\\

\hline
\hline
\end{tabular}
\end{table}

To perform an infinite-volume extrapolation, we must use momenta which are common in all $N_L=3$, 6, and 9 cases, which as can be seen from Fig.~\ref{fig:dispersions} occurs for points that have $|\vec{T}| |k_{||}|=$ 0 and $2\pi/3$.  We tabulate these points in Table~\ref{tab:infinite volume points}.
\begin{table}
\center
\caption{Infinite-volume extrapolations of momentum points common to $N_L=3$, 6, and 9 systems (last column) and their corresponding momenta (first column), and the points used to perform the extrapolation (middle columns).  Energies are shown in units of $\kappa$.  The functional form of the extrapolation is linear in $1/N_L$. \label{tab:infinite volume points}}
\begin{tabular}{c|ccc|c}
\hline
\hline
$|\vec{T}| \ (|k_{||}|,|k_{\bot,i}|)$ & $E(N_L=3)$ & $E(N_L=6)$ & $E(N_L=9)$ & $E(N_L=\infty)$ \\
\hline

$\left(0,0\right)$ & 3.926(28) & 3.848(26) & 3.836(17) & 3.784(29)\\
$\left(0,\frac{2 \pi }{3 \sqrt{3}}\right)$ & 3.546(35) & 3.464(32) & 3.459(27) & 3.406(41)\\
$\left(0,\frac{4 \pi }{3 \sqrt{3}}\right)$ & 2.540(68) & 2.477(53) & 2.464(74) & 2.424(91)\\
$\left(0,\frac{2 \pi }{\sqrt{3}}\right)$ & 1.744(31) & 1.684(23) & 1.663(17) & 1.623(31)\\
\hline
$\left(\frac{2 \pi }{3},0\right)$ & 2.810(21) & 2.766(18) & 2.753(13) & 2.723(21)\\
$\left(\frac{2 \pi }{3},\frac{2 \pi }{3 \sqrt{3}}\right)$ & 2.516(24) & 2.471(20) & 2.461(14) & 2.433(25)\\
$\left(\frac{2 \pi }{3},\frac{4 \pi }{3 \sqrt{3}}\right)$ & 1.709(29) & 1.644(22) & 1.628(23) & 1.586(32)\\
$\left(\frac{2 \pi }{3},\frac{2 \pi }{\sqrt{3}}\right)$ & 0.813(47) & 0.693(32) & 0.632(28) & 0.551(46)\\

\hline
\hline
\end{tabular}
\end{table}
In the bottom right panel of Fig.~\ref{fig:dispersions} we plot these points along with the non-interacting dispersion.  The effects of strongly-correlated electrons is clearly seen in the calculated spectrum of this system, and in general lifts all points above their non-interacting values.

\clearpage

\section{Conclusions}
We have demonstrated how lattice Monte Carlo methods can be applied to carbon nanotubes. 
We have derived the path-integral formalism for such systems, based on previous work for a planar hexagonal lattice, with appropriate (periodic) boundary conditions 
that depend on the nanotube chirality $(n,m)$. In so doing, we emphasized differences of our method with previous lattice Monte Carlo calculations of graphene, 
as well as with lattice QCD. We proceeded to benchmark our method for the (3,3) armchair nanotube, 
using the screened Coulomb interaction of Ref.~\cite{Smith:2014tha} which incorporates the values of $U_{00}$ through $U_{03}$
found by Ref.~\cite{PhysRevLett.106.236805}. Apart from the requirement that the potential matrix $V_{xy}$ be positive definite, 
we stress that our formalism is not dependent on any particular parametrization of the electron-electron interaction.  

As opposed to previous lattice Monte Carlo simulations, we extracted single quasi-particle energies by direct analysis of the momentum-projected one-body correlators, 
a method commonly used by LQCD calculations. This allowed us to not only extract the spectrum at the Dirac point, but at all allowed momentum modes. 
As the nanotube systems studied were relatively small, we were able to perform calculations at multiple time steps $N_t$ and multiple tube lengths $N_L$. The former 
allowed us to perform a continuum limit extrapolation, and the latter allowed us to consider nanotubes of infinite length. In all cases, we found that the non-interacting spectrum is
strongly modified by electron-electron correlation effects, in general raising the energies at all points in the Brillouin zone above their 
non-interacting (tight-binding) values. In particular, our result for the energy of the Dirac point in the (3,3) nanotube was found to be 
$E_K/\kappa=.551(46)$, consistent with a substantial interaction-induced energy gap in this (nominally) metallic nanotube.  

Our extrapolations in the temporal and spatial dimensions used simple scaling functions in $\delta^2$ and $1/N_L$. Though we have performed multiple Monte Carlo calculations 
for different $\delta$ and $N_L$, our preliminary results are not yet sufficient to exclude other possible power-law scalings. 
Systematic errors from other possible scalings have thus not been included in our analysis, although we note that future studies using a larger set of $\delta$ and $N_L$ values 
can address this issue. We also found that the small physical dimensions of our nanotubes, coupled with the long-range nature of the Coulomb interaction, induced a Gaussian term 
in our correlators, which we attributed to the zero-mode contributions of our auxiliary field. To account for this effect in our analysis of the large-time behavior of our
correlators, we performed simultaneous fits of both Gaussian and (leading) exponential terms.  

A possible application of our method would be to consider the energy gap at the Dirac point as a function of nanotube radius, for which considerable experimental data is available.
Such calculations would allow for a direct test of different models for the electron-electron interaction, which in turn could provide additional input to the problem of a possible Mott insulating
state in suspended graphene. Also, while we have so far only considered the single-quasiparticle dispersion relation, our formalism can easily be extended to the spectrum of multi-particle 
states. For example, the interacting $J=0$ electron-hole system can be represented by the ``interpolating operator'' 
$O^\dag(x,y) \equiv 1/\!\sqrt{2} \, [a_{-}^\dag(x)b_{-}^\dag(y)-b_{-}^\dag(x)a_{-}^\dag(y)]$,  where the operators $a^\dag_{\pm}$ and $b^\dag_{\pm}$ are defined in Eqn.~\eqref{eqn:apm}.  The
spectrum of such a system could be ascertained by analyzing the temporal behavior of the two-particle correlator 
$\left\langle O(x',y')O^\dag(x,y)\right\rangle$. In light of the results found in~\cite{PhysRevLett.106.037404}, such studies would be very interesting.

Our relatively large uncertainties can be traced back to the non-trivial contribution of zero-modes to our correlators and the fact that our system is physically very small.  Both contributions can be alleviated by increasing the volume (length) of our system.  As in LQCD, we anticipate a suppression of uncertainties that scale as $1/V^{3/2}$, where $V$ is the volume of the system~\cite{Luscher:2010ae}.  This would correspond to uncertainties that scale as $1/N_L^{1/2}$ for our system.  Such suppression is already evident when comparing the uncertainties between our $N_L$=3, 6, and 9 calculations (Tables~\ref{tab:nl3}-\ref{tab:nl9}).   However, the dimensions of such calculations would scale linearly in $N_L$.   We expect reduced uncertainties with larger diameter tube calculations as well.  The dimension of a (14,14) tube calculation with $N_L=9$, for example, is $\simeq$ 20 times larger than the (3,3) system with $N_L=9$ and same number of time steps.  Such a calculation would require resources beyond what we have committed to this paper, i.e. desktop workstation (we are currently modifying our codes to run on larger computer clusters), and would be ideally suited for GPUs \cite{Wendt:2010fq,Brower:2012zd,Smith:2014tha}. 

In conclusion, we emphasize that our Monte Carlo method is completely general and can be applied to other carbon nanostructures, such as graphene single- and multi-layers, 
multi-wall nanotubes and carbon nano-ribbons. The most significant restriction of our method is the requirement of a positive definite probability measure, the availability of which 
has to be assessed on a case-by-case basis. In addition to periodic boundary conditions, our method also allows for arbitrary boundary conditions (such as open or twisted boundary conditions). 
For instance, the latter choice could prove useful in studies of carbon nanotubes in an external magnetic field.  
 
\begin{acknowledgments}
We acknowledge financial support from the Magnus Ehrnrooth Foundation of the Finnish Society of Sciences and Letters, which enabled some of our numerical simulations.  The authors are indebted to A. Shindler for valuable discussions related to zero-modes, and to C. K\"orber and M. Hru$\check{\text{s}}$ka for their critical reading and discussions of the manuscript.  Additional thanks goes to C. K\"orber for aid in generation of figures.  Lastly, we thank L. von Smekal for discussions related to the normal ordering of interaction terms.
\end{acknowledgments}

\clearpage

\begin{appendices}
\section{Zero-mode contributions to the path integral\label{app:zero modes}}
We begin with the continuum limit (in time) expression for the expectation value of our fermion correlator in the quenched approximation (\emph{i.e.} setting $\det(M[\Phi]M^\dag[\Phi])=1$ in eq.~\eqref{eqn:Z function 3}),
\begin{equation}\label{eqn:quenched}
\langle M^{-1}(\vec{k}_\alpha,\tau)\rangle =\frac{1}{Z}
 \int \mathcal{D}\tilde{\Phi}\ e^{-S[\Phi]}
 \frac{1}{\beta}\sum_{n=-\infty}^{\infty}e^{-i\omega_n \tau}M^{-1}(\vec{k}_\alpha,\omega_n;\Phi_{x_0}(t_0))\ ,
\end{equation}
where 
\begin{eqnarray}
S[\Phi]&=&\frac{1}{2}\int_0^{\beta}dt\sum_{x,y\in\{X\}}\Phi^T_{x}(t)[V^{-1}]_{x,y} \Phi_{y}(t)\\ \label{eqn:phi action}
\omega_n&=&\pi(2n+1)/\beta\ ,
\end{eqnarray}
and
\begin{equation}\label{eqn:app1.5}
M^{-1}(\vec{k}_\alpha,\omega_n;\Phi_{x_0}(t_0))=\
\left((i\omega_n-\omega_+)(i\omega_n-\omega_-)\right)^{-1}m(i\omega_n,\vec{k}_\alpha,\Phi_{x_0}(t_0))\ ,
\end{equation}
where $m(i\omega_n,\vec{k}_\alpha,\Phi_{x_0}(t_0))$ is the following 2$\times$2 matrix\footnote{The appearance of $U_{00}$ in Eqn.~\eqref{eqn:app2} and subsequent equations comes from the `non-compact' formulation of our path integral (which we employ in this section), and the associated normal ordering of the onsite term.  See \cite{Brower:2012zd,Smith:2014tha} for a detailed discussion.  The results of this section do not depend on its appearance.}
\begin{equation}\label{eqn:app2}
m(i\omega_n,\vec{k}_\alpha,\Phi_{x_0}(t_0))=
\begin{pmatrix}
-(i\omega_n+i\phi^B_{x_0}(t_0)+U_{00}/2) & \kappa f(\vec{k}_\alpha)\\
\kappa f^*(\vec{k}_\alpha)& -(i\omega_n+i\phi^A_{x_0}(t_0)+U_{00}/2)
\end{pmatrix}\ .
\end{equation}
The matrix $m(i\omega_n,\vec{k}_\alpha,\Phi_{x_0}(t_0)$ contains no poles.  The frequencies $\omega_\pm$ are
\begin{eqnarray}\label{eqn:app3}
\omega_{\pm}&=&-i\frac{(\phi^A_{x_0}(t_0)+\phi^B_{x_0}(t_0))}{2}-\frac{U_{00}}{2}\pm
i\sqrt{\left(\frac{\phi^A_{x_0}(t_0)-\phi^B_{x_0}(t_0)}{2}\right)^2+|\kappa f(\vec{k}_\alpha)|^2}\\
&\equiv&-i\frac{(\phi^A_{x_0}(t_0)+\phi^B_{x_0}(t_0))}{2}-\frac{U_{00}}{2}\pm i\Delta\omega\ . \label{eqn:app3.5}
\end{eqnarray}
We first concentrate on the frequency sum in Eqn.~\eqref{eqn:quenched}, which using Eqns.~\eqref{eqn:app2}-\eqref{eqn:app3.5} can be written as
\begin{equation}\label{eqn:app4}
 \frac{1}{\beta}\sum_{n=-\infty}^{\infty}\frac{e^{-i\omega_n \tau}}{(i\omega_n-\omega_+)(i\omega_n-\omega_-)}m(i\omega_n,\vec{k}_\alpha,\Phi_{x_0}(t_0))\ .
 \end{equation}
Assuming that $0\le\tau\le\beta$, one can use the Matsubara weighting function $h(\omega)=(1+\exp(-\beta\omega))^{-1}$ and standard finite-temperature integration techniques \cite{abrikosov1975methods,mattuck2012guide} to show that the sum in Eqn.~\eqref{eqn:app4} is equal to 
\begin{multline}\label{eqn:app5}
\frac{1}{\beta}\frac{1}{(\omega_{-} -\omega_+)}\left(e^{-\omega_+\tau}m(\omega_+,\vec{k}_\alpha,\Phi_{x_0}(t_0))h(\omega_+)-e^{-\omega_-\tau}m(\omega_-,\vec{k}_\alpha,\Phi_{x_0}(t_0))h(\omega_-)\right)\\
=e^{i(\phi^A_{x_0}(t_0)+\phi^B_{x_0}(t_0))\frac{\tau}{2}}
\frac{1}{\beta}\frac{e^{U_{00}\frac{\tau}{2}}}{(\omega_{-} -\omega_+)}\times\\
\left(e^{-\Delta\omega\tau}m(\omega_+,\vec{k}_\alpha,\Phi_{x_0}(t_0))h(\omega_+)-e^{\Delta\omega\tau}m(\omega_-,\vec{k}_\alpha,\Phi_{x_0}(t_0))h(\omega_-)\right)\ .
\end{multline}

We concentrate on the small time dependence, $\tau\ll \beta$, of our expression and perform a small temperature, large $\beta$ expansion of the Matsubara regulator,
\begin{displaymath}
h(\omega_\pm)=1-e^{-\beta\omega_{\pm}}+e^{-2\beta\omega_{\pm}}+\ldots
\end{displaymath}
To leading order in this expansion, we have
\begin{multline}\label{eqn:app6}
\frac{1}{\beta}\frac{1}{(\omega_{-} -\omega_+)}\left(e^{-\omega_+\tau}m(\omega_+,\vec{k}_\alpha,\Phi_{x_0}(t_0))-e^{-\omega_-\tau}m(\omega_-,\vec{k}_\alpha,\Phi_{x_0}(t_0))\right)+\mathcal{O}(e^{-\beta\omega_{\pm}})\\
= e^{i(\phi^A_{x_0}(t_0)+\phi^B_{x_0}(t_0))\frac{\tau}{2}}
\frac{1}{\beta}\frac{e^{U_{00}\frac{\tau}{2}}}{(\omega_{-} -\omega_+)}\left(e^{-\Delta\omega\tau}m(\omega_+,\vec{k}_\alpha,\Phi_{x_0}(t_0))-e^{\Delta\omega\tau}m(\omega_-,\vec{k}_\alpha,\Phi_{x_0}(t_0))\right)\\
\equiv
e^{i(\phi^A_{x_0}(t_0)+\phi^B_{x_0}(t_0))\frac{\tau}{2}}F\left(\frac{\phi^A_{x_0}(t_0)-\phi^B_{x_0}(t_0)}{2},\tau\right)
\end{multline}
where the function $F$ can be determined by comparing the second and third lines of the equation above.  The argument structure of $F$ is written in such a manner as to stress the fact that its depencence on the auxiliary fields is through the \emph{difference} $\phi^A_{x_0}(t_0)-\phi^B_{x_0}(t_0)$,  which can be easily verified by analyzing Eqns.~\eqref{eqn:app2}-\eqref{eqn:app3.5}.

We now expand our auxiliary fields in momentum-frequency space,
\begin{eqnarray}
\phi^{A,B}_{x_0}(\tau_0)&=&\frac{1}{\beta}\sum_{n=-\infty}^{\infty}e^{-i\omega_n\tau_0}\frac{1}{N}\sum_{k_\alpha\in\{K\}}e^{-i\vec{k}_\alpha\cdot\vec{x}_0}\hat{\phi}^{A,B}_{k_\alpha,\omega_n}\\
&=&\frac{1}{\beta N}\hat{\phi}^{A,B}_{0}+\frac{1}{\beta N}\sum_{\{n,k_\alpha\}\ne\{0,0\}}e^{-i\omega_n\tau_0}e^{-i\vec{k}_\alpha\cdot\vec{x}_0}\hat{\phi}^{A,B}_{k_\alpha,\omega_n}\ ,
\end{eqnarray}
where in the second line we explicitly separate the zero-mode contribution.  Note that the frequency sum is over bosonic frequencies, $\omega_n=2\pi n/\beta$, and $N$ is the number of hexagons in our calculation. Further, it is convenient to define the fields $\hat{\phi}^{\pm}_0$ through the following linear combinations of the fields in momentum-frequency space,
\begin{eqnarray}
\frac{\phi^{A}_{x_0}(\tau_0)\pm\phi^{B}_{x_0}(\tau_0)}{2}&=&\frac{1}{\beta N}\frac{\hat{\phi}^{A}_{0}\pm\hat{\phi}^{B}_{0}}{2}+
\frac{1}{\beta N}\sum_{\{n,k_\alpha\}\ne\{0,0\}}e^{-i\omega_n\tau_0}e^{-i\vec{k}_\alpha\cdot\vec{x}_0}\frac{(\hat{\phi}^{A}_{k_\alpha,\omega_n}\pm
\hat{\phi}^{B}_{k_\alpha,\omega_n})}{2}\\
&\equiv&\frac{1}{\beta N}\hat{\phi}^{\pm}_0+\Delta^{\pm}[\Phi_{k_\alpha,\omega_n}]\ ,\label{eqn:app7}
\end{eqnarray}
where $\Delta^{\pm}[\Phi_{k_\alpha,\omega_n}]$ contains sums over terms that have non-zero momentum or frequency modes.

The action in Eqn.~\eqref{eqn:phi action} can be cast in momentum-frequency space,
\begin{eqnarray}
S[\Phi]&=&\frac{1}{2}\frac{1}{\beta N}\sum_{n,k_\alpha}\hat{\Phi}^T_{k_\alpha,\omega_n}[\hat{v}^{-1}]_{k_\alpha} \hat{\Phi}_{k_\alpha,\omega_n}\\
&=&\frac{1}{2}\frac{1}{\beta N}\hat{\Phi}^T_{0}[\hat{v}^{-1}]_{0} \hat{\Phi}_{0}+
\frac{1}{2}\frac{1}{\beta N}\sum_{\{n,k_\alpha\}\ne\{0,0\}}\hat{\Phi}^T_{k_\alpha,\omega_n}[\hat{v}^{-1}]_{k_\alpha} \hat{\Phi}_{k_\alpha,\omega_n}\\
&\equiv&\frac{1}{2}\frac{1}{\beta N}\hat{\Phi}^T_{0}[\hat{v}^{-1}]_{0} \hat{\Phi}_{0}+S[\hat{\Phi}_{k,\omega}]\ ,\label{eqn:app8}
\end{eqnarray}
where $\hat{v}_{k_\alpha}$ is the discrete fourier transform of the screened Coulomb potential and we have again separated out the zero-mode contribution and defined the remainder as $S[\hat{\Phi}_{k,\omega}]$. In terms of $\hat{\phi}^{A,B}_0$ we have that
\begin{eqnarray}
\frac{1}{2}\frac{1}{\beta N}\hat{\Phi}^T_0[\hat{v}^{-1}]_{0}\hat{\Phi}_0
&=&\frac{1}{2}\frac{1}{\beta N}\frac{1}{(\hat{v}^{AA}_0)^2-(\hat{v}^{AB}_0)^2}\left(\hat{v}^{AA}_0 [(\hat{\phi}^A_0)^2+(\hat{\phi}^B_0)^2]-2\hat{v}^{AB}_0\hat{\phi}^A_0\hat{\phi}^B_0\right)\\ \nonumber
&=&\frac{1}{\beta N}\frac{1}{(\hat{v}^{AA}_0)^2-(\hat{v}^{AB}_0)^2}\left((\hat{\phi}^+_0)^2(\hat{v}^{AA}_0-\hat{v}^{AB}_0)+(\hat{\phi}^-_0)^2(\hat{v}^{AA}_0+\hat{v}^{AB}_0)\right)\\ \nonumber
&\equiv&\frac{1}{\beta N}\frac{(\hat{v}^{AA}_0-\hat{v}^{AB}_0)}{(\hat{v}^{AA}_0)^2-(\hat{v}^{AB}_0)^2}(\hat{\phi}^+_0)^2+S[\hat{\phi}^-_0]
\ , \label{eqn:app9}
\end{eqnarray}
where
\begin{eqnarray}\label{eqn:ft V}
\hat{v}^{AA}_0&=&\sum_{x\in\{X\}}V(|\vec{x}|)\\
\hat{v}^{AB}_0&=&\sum_{x\in\{X\}}V(|\vec{x}+\vec{a}|)\ ,
\end{eqnarray}
and $\vec{a}$ is the basis unit vector.  Finally we factor out the zero mode measures in the integration measure,
\begin{equation}\label{eqn:integration measure}
\mathcal{D}\Phi=d\hat{\phi}^A_0d\hat{\phi}^B_0\mathcal{D}\Phi_{(k_\alpha,\omega)\ne(0,0)}=d\hat{\phi}^+_0d\hat{\phi}^-_0\mathcal{D}\Phi_{(k_\alpha,\omega)\ne(0,0)}\ .
\end{equation} 
The Jacobian from the change of variables in the last expression is unity.  Combining Eqns.~\eqref{eqn:app6}, \eqref{eqn:app7}, \eqref{eqn:app8}, \eqref{eqn:app9}, and \eqref{eqn:integration measure}, one gets
\begin{multline}
\int d\hat{\phi}^+_0\exp\left\{-\frac{1}{\beta N}\left(\frac{(\hat{v}^{AA}_0-\hat{v}^{AB}_0)}{(\hat{v}^{AA}_0)^2-(\hat{v}^{AB}_0)^2}(\hat{\phi}^+_0)^2-i \hat{\phi}^+_0\tau\right)\right\}\times\\
\int d\hat{\phi}^-_0\mathcal{D}\Phi_{(k_\alpha,\omega)\ne(0,0)}e^{-S[\hat{\phi}^-_0]-S[\hat{\Phi}_{k_\alpha,\omega}]+i\Delta^{+}[\Phi_{k_\alpha,\omega}]\tau}F\left(\hat{\phi}^-_0+\Delta^{-}[\Phi_{k_\alpha,\omega}],\tau\right)\ .
\end{multline}
We can now perform the integral over $\hat{\phi}^+_0$ explicitly.  Up to an irrelvant multiplicative factor, the result is
\begin{multline}
\exp\left\{-\frac{(\hat{v}^{AA}_0+\hat{v}^{AB}_0)}{4\beta N}\tau^2\right\}\\
\int d\hat{\phi}^-_0\mathcal{D}\Phi_{(k_\alpha,\omega)\ne(0,0)}e^{-S[\hat{\phi}^-_0]-S[\hat{\Phi}_{k_\alpha,\omega}]+i\Delta^{+}[\Phi_{k_\alpha,\omega}]\tau}F\left(\hat{\phi}^-_0+\Delta^{-}[\Phi_{k_\alpha,\omega}],\tau\right)\ ,
\end{multline}
which shows the Gaussian dependence in $\tau$.  The remaining functional integrals over $\hat{\phi}^-_0$ and $\mathcal{D}\Phi_{(k_\alpha,\omega)\ne(0,0)}$ produce exponential depencence in the spectrum $E_i(\vec{k}_\alpha)$ of the system.  Thus for low temperatures, small time regime, and quenched approximation, our correlator behaves as
\begin{equation}
\langle M^{-1}(\vec{k}_\alpha,\tau)\rangle\sim
\exp\left\{-\frac{(\hat{v}^{AA}_0+\hat{v}^{AB}_0)}{4\beta N}\tau^2\right\}\sum_i A_i e^{-E_i(\vec{k}_\alpha)\tau}\ .
\end{equation}

\section{Benchmark calculations of the 2-site and 4-site Hubbard model\label{sect:benchmark}}
We provide details of our benchmark calculations of correlators calculated with our lattice code compared to analytic calculations of the 2- and 4-site Hubbard model.  As the Hubbard model has onsite interactions only (\emph{i.e.} no long-range interaction) we do not have zero-mode induced Gaussian dependence in our correlators.  The 4-site model has two momentum modes that allow us to test our momentum projection routines.

\subsection{2-site Hubbard model}
The simplest case that one can consider that includes interactions is the 2-site Hubbard model.  The Hamiltonian for the Hubbard model at half-filling is
\begin{equation}\label{eqn:hubbard hamiltonian}
\hat{H}=-\kappa\sum_{\langle i,j\rangle}c^{\dag}_{i,\sigma}c_{j,\sigma}+U_{00}\sum_i n_{i,\uparrow}n_{i,\downarrow}-\frac{U_{00}}{2}\sum_i\left(n_{i,\uparrow}+n_{i,\downarrow}\right)+\mbox{const.}\ ,
\end{equation}
where $\langle i,j\rangle$ denotes nearest neighbor summation, $c^{\dag}_{i,\sigma}$ ($c_{i,\sigma}$) is the creation (annihilation) operator for an electron of spin $\sigma$ at site $i$, $n_{i,\sigma}=c^{\dag}_{i,\sigma}c_{i,\sigma}$ is the number operator for spin $\sigma$ at site $i$, and $U_{00}$ is the onsite repulsive interaction parameter.  We note that the relevant dimensionless parameter in this model is simply the ratio $\lambda=U_{00}/\kappa$.

\begin{figure}
\includegraphics[height=.45\columnwidth,angle=-90]{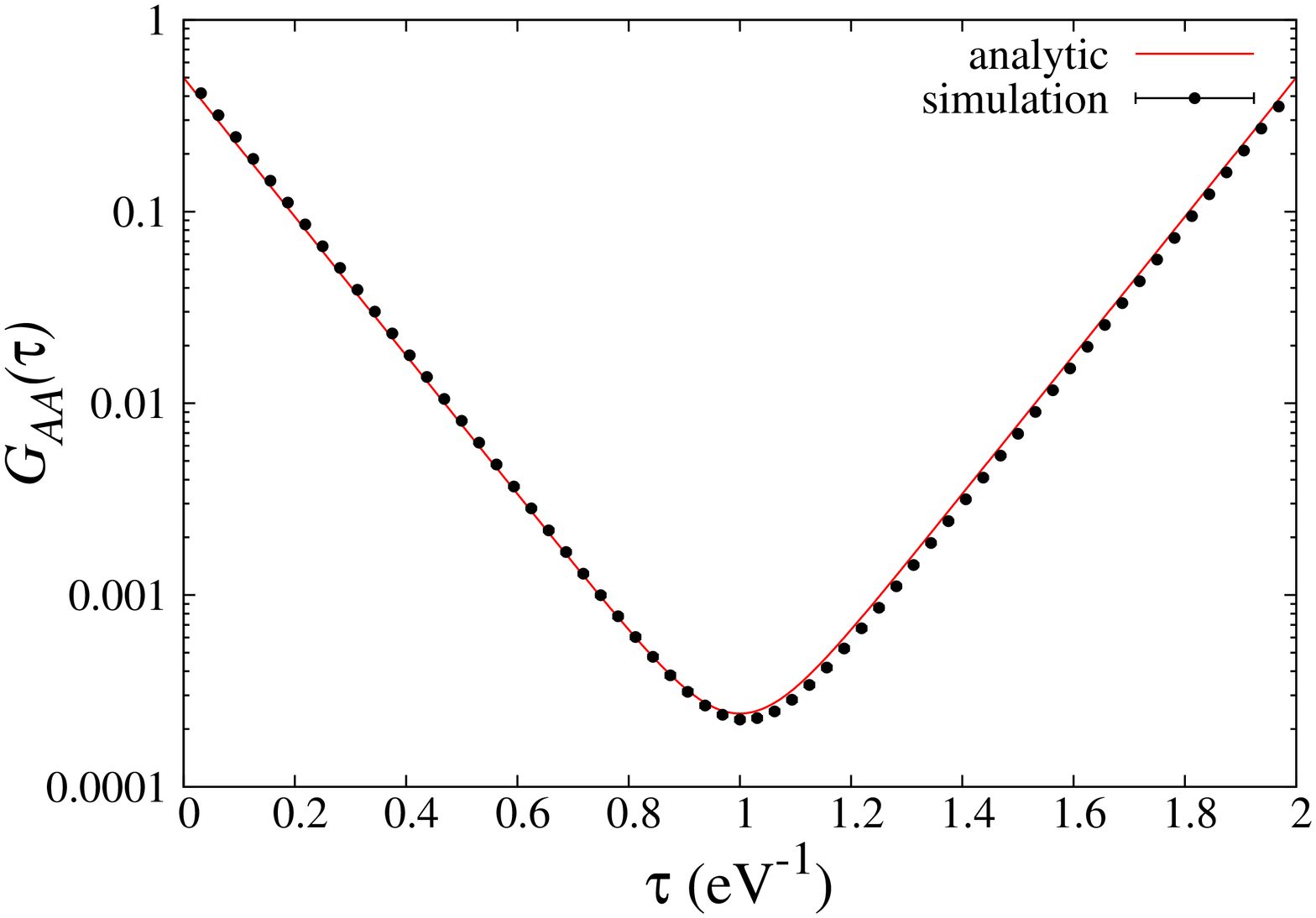}\ \includegraphics[height=.45\columnwidth,angle=-90]{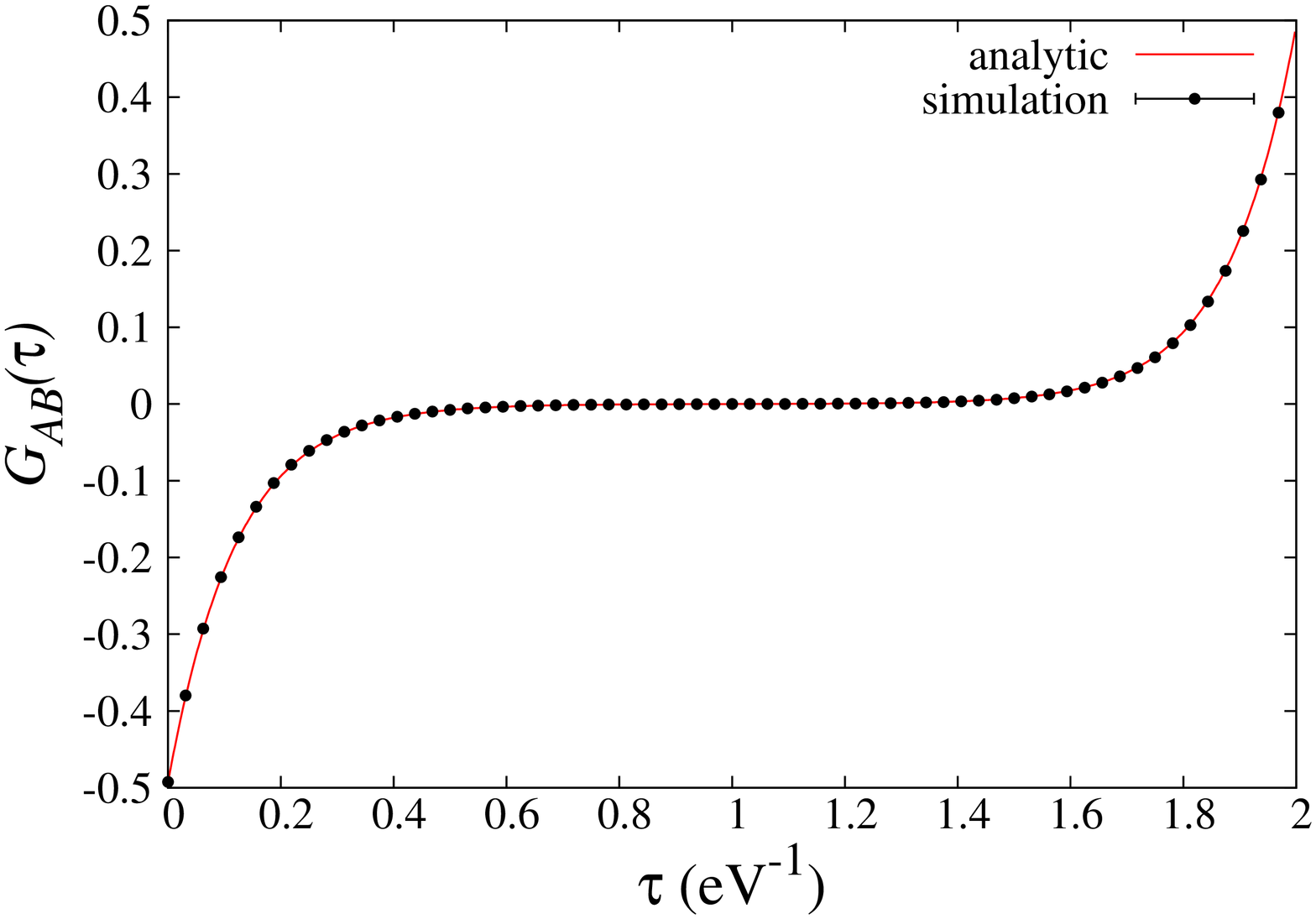}\\
\includegraphics[height=.45\columnwidth,angle=-90]{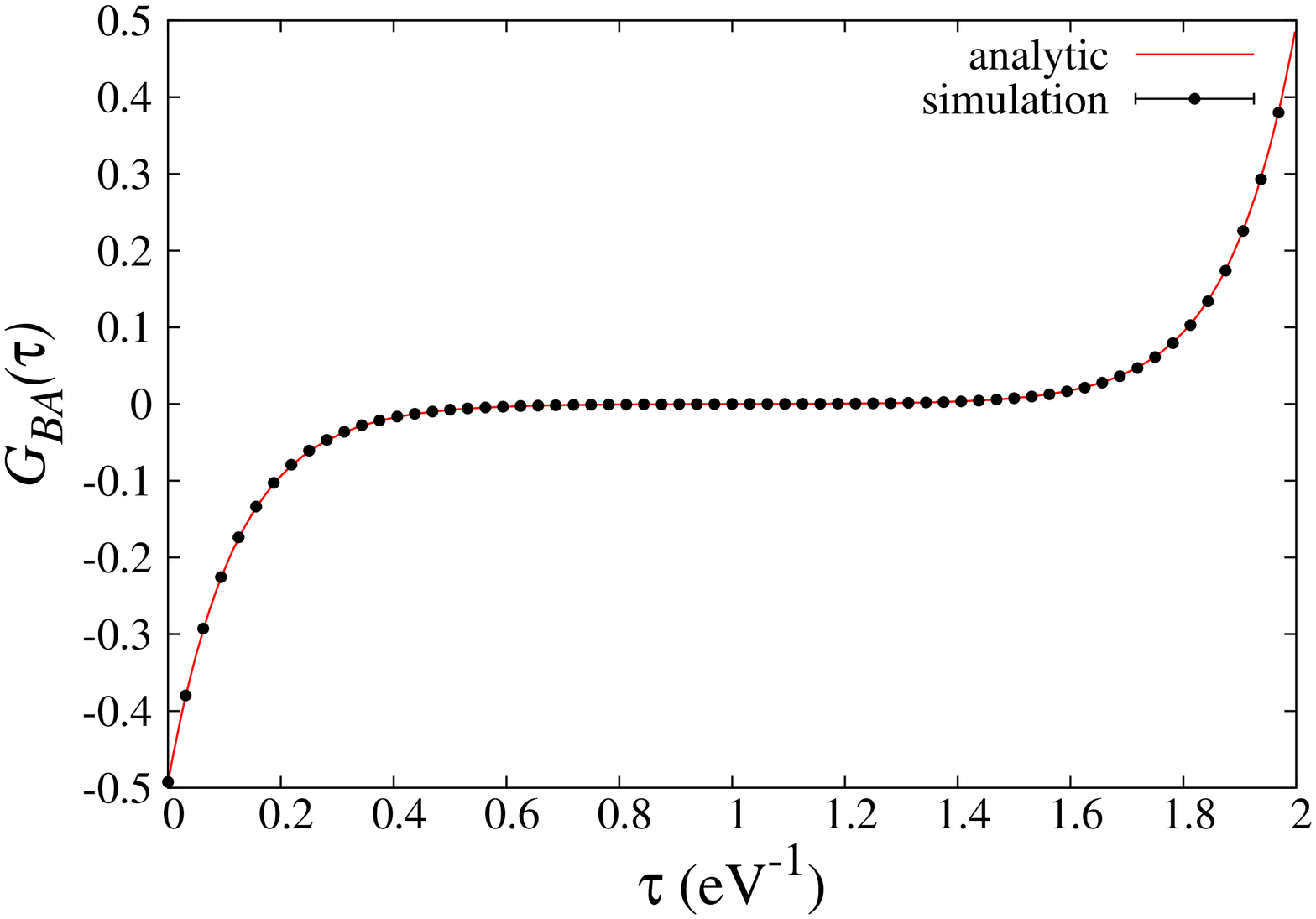}\ \includegraphics[height=.45\columnwidth,angle=-90]{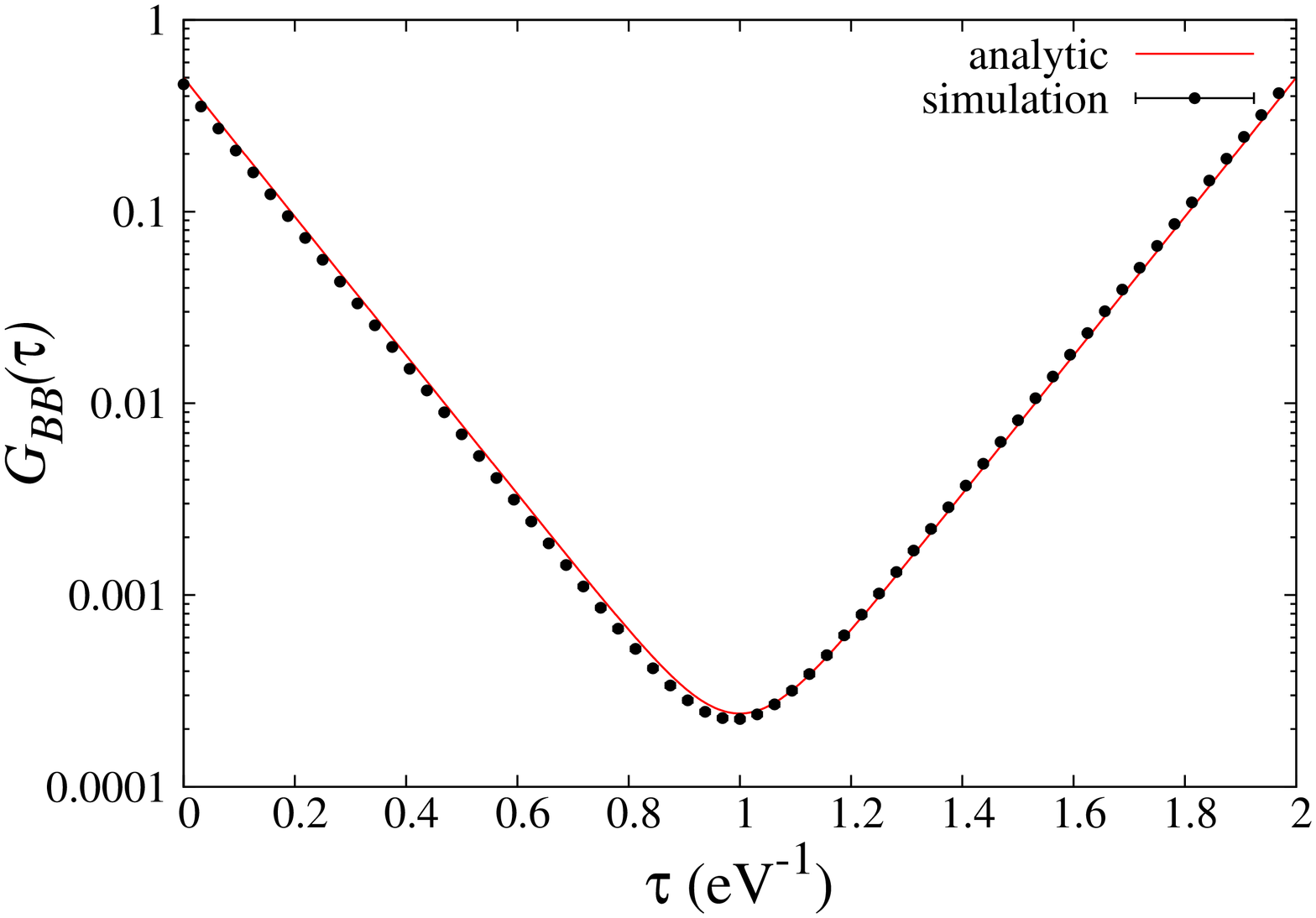}
\caption{The matrix of correlators $G_{ij}(\tau)$ for 2-site Hubbard model with $\beta=2$ eV$^{-1}$ and $\kappa=2.7$ eV, and $U/\kappa=2$.  The solid lines are the analytical results.  The points are from a full lattice calculation with $N_t=64$ timesteps.  Errorbars, obtained via boostrap, are too small to be visible.\label{fig:first interaction} }
\end{figure}

The eigenvalues of the system can be obtained by direct diagonalization, and the single-electron correlation function can be obtained using the expression
\begin{equation}
G_{ij}^{\sigma\sigma'}(\tau)\equiv\langle c_{i,\sigma}(\tau)c^{\dag}_{j,\sigma'}(0)\rangle=\frac{1}{\mathcal{Z}}
\sum_i\langle i|c_{i,\sigma}(\tau)c^{\dag}_{j,\sigma'}(0)|i\rangle e^{-\beta E_i}\ ,
\end{equation}
where the sum is over eigenstates $|i\rangle$ of the system, $E_i$ is the eigenvalue for state $|i\rangle$, and 
\begin{displaymath}
\mathcal{Z} =\sum_i e^{-\beta E_i}\ .
\end{displaymath}
For a given $\beta$, $U$, and $\kappa$, we perform our lattice calculations and compare our calculated correlators with those derived analytically.  In Fig.~\ref{fig:first interaction} we compare our lattice results with analytic results for the case when $U/\kappa=2$.  Details of lattice calculation are given in the caption.

\begin{figure}
\includegraphics[height=.5\columnwidth,angle=-90]{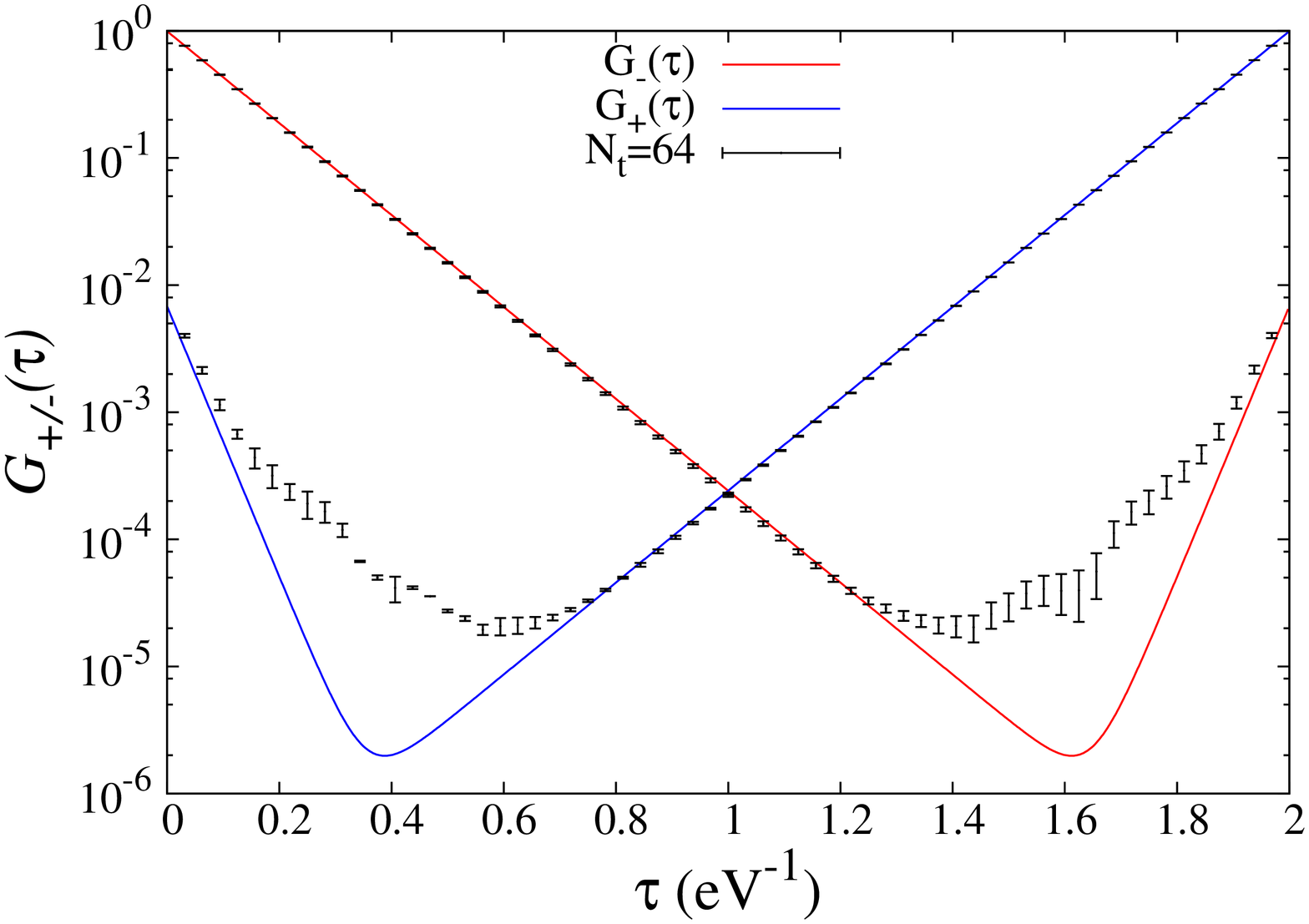}\includegraphics[height=.5\columnwidth,angle=-90]{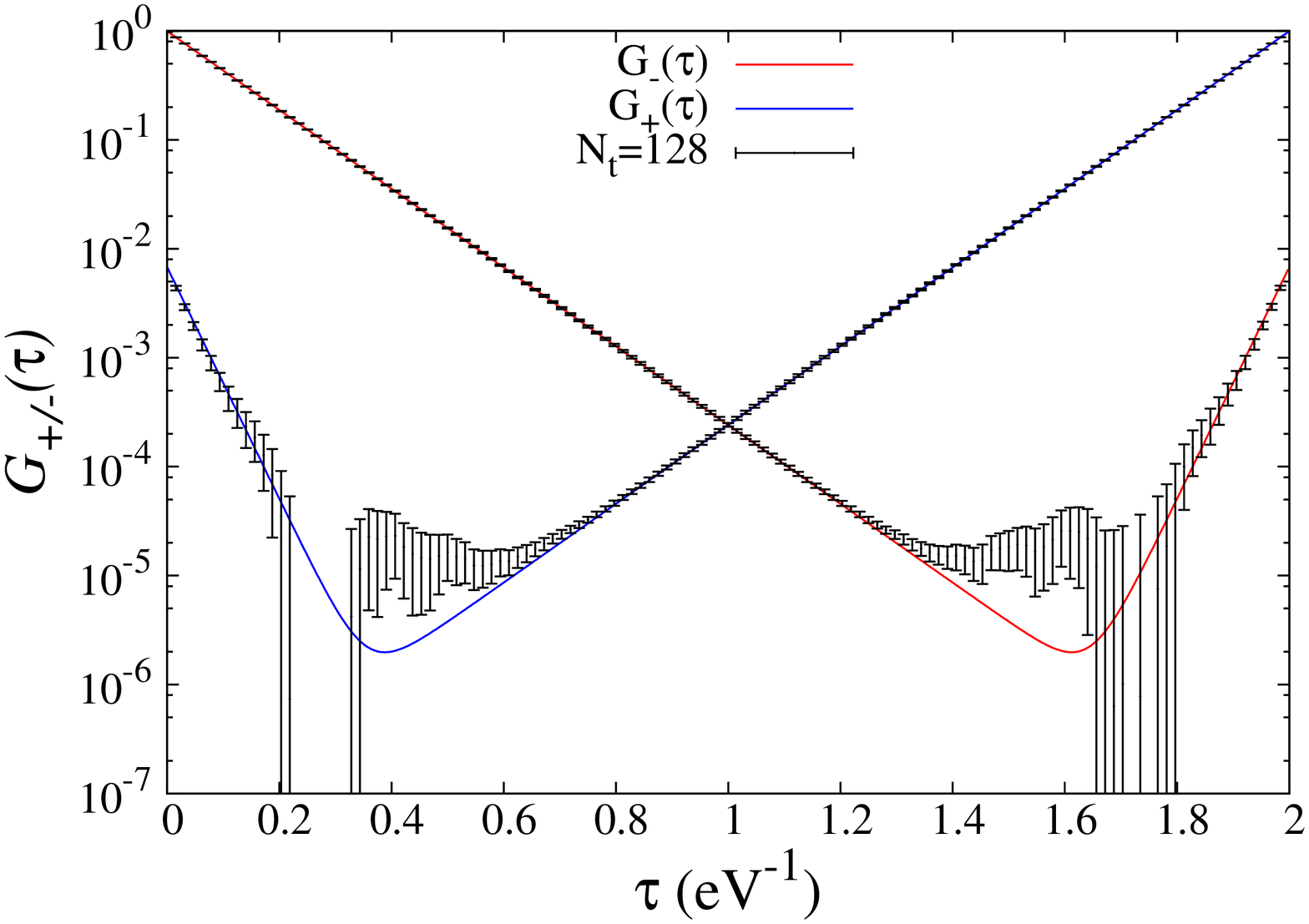}
\caption{$G_{\pm}(\tau)$ correlators for 2-site Hubbard model with $\beta=2$ eV$^{-1}$ and $\kappa=2.7$ eV, and $U/\kappa=2$.  The solid lines are the analytical results.  Left (right) plot has $N_t=64\ (128)$ timesteps.\label{fig:G plusminus} }
\end{figure}

\begin{figure}
\includegraphics[height=.5\columnwidth,angle=-90]{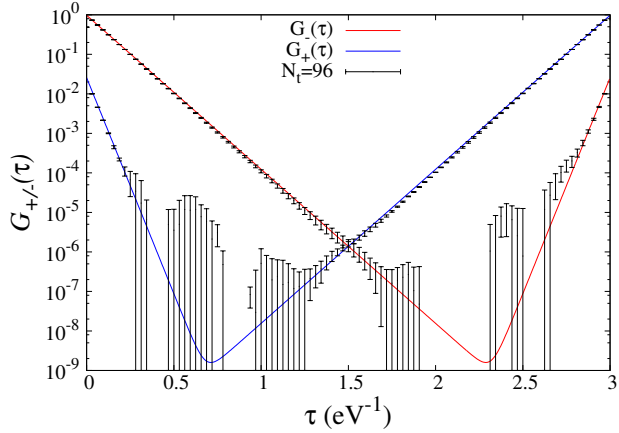}\caption{Comparison of numerical calculation of $G_{\pm}(\tau)$ with analytic result for $\beta=3$ eV$^{-1}$, $\kappa=2.7$ eV,  $U/\kappa=4$, and $N_t=96$ timesteps.  The solid line is the analytical result. \label{fig:G plusminus 2} }
\end{figure}

The relevant correlators to extract energies are given by 
\begin{equation}\label{eqn:G plusminus}
G_{\pm}(\tau)\equiv \frac{1}{2}\left[G_{AA}(\tau)+G_{BB}(\tau)\pm\left(G_{AB}(\tau)+G_{BA}(\tau)\right)\right]\ .
\end{equation}
Figure~\ref{fig:G plusminus} shows comparison of lattice results using  $N_t=64$ and $N_t=128$ compared with analytic results.  Clear convergence with the analytic results is seen, particularly for the $G_+(\tau)$ correlator.
In Fig.~\ref{fig:G plusminus 2} we show results for $G_{\pm}(\tau)$ for the case of $\beta=3$ eV$^{-1}$, $U/\kappa=4$, and $N_t=96$.  

\subsection{4-site Hubbard model}
\begin{figure}
\includegraphics[width=.9\columnwidth,angle=270]{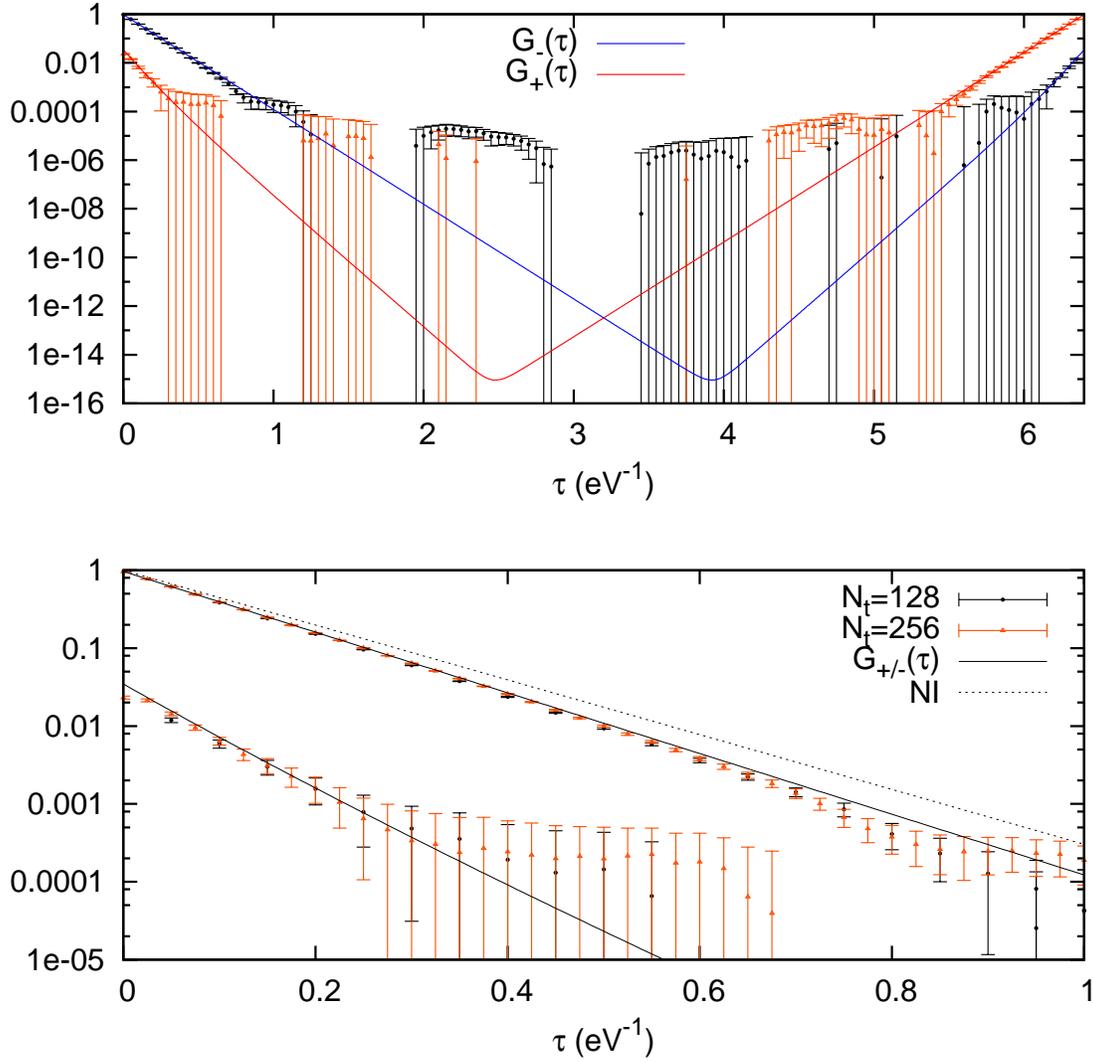}
\caption{4-site Hubbard calculation of $\vec{k}_1$ momentum correlators with $U_{00}=9.3$ eV and $\beta=6.4$ eV$^{-1}$.  Bottom plot is a close up and shows the non-interacting results (dashed line).\label{fig:4siteU00k1}}
\end{figure}

\begin{figure}
\includegraphics[width=.9\columnwidth,angle=270]{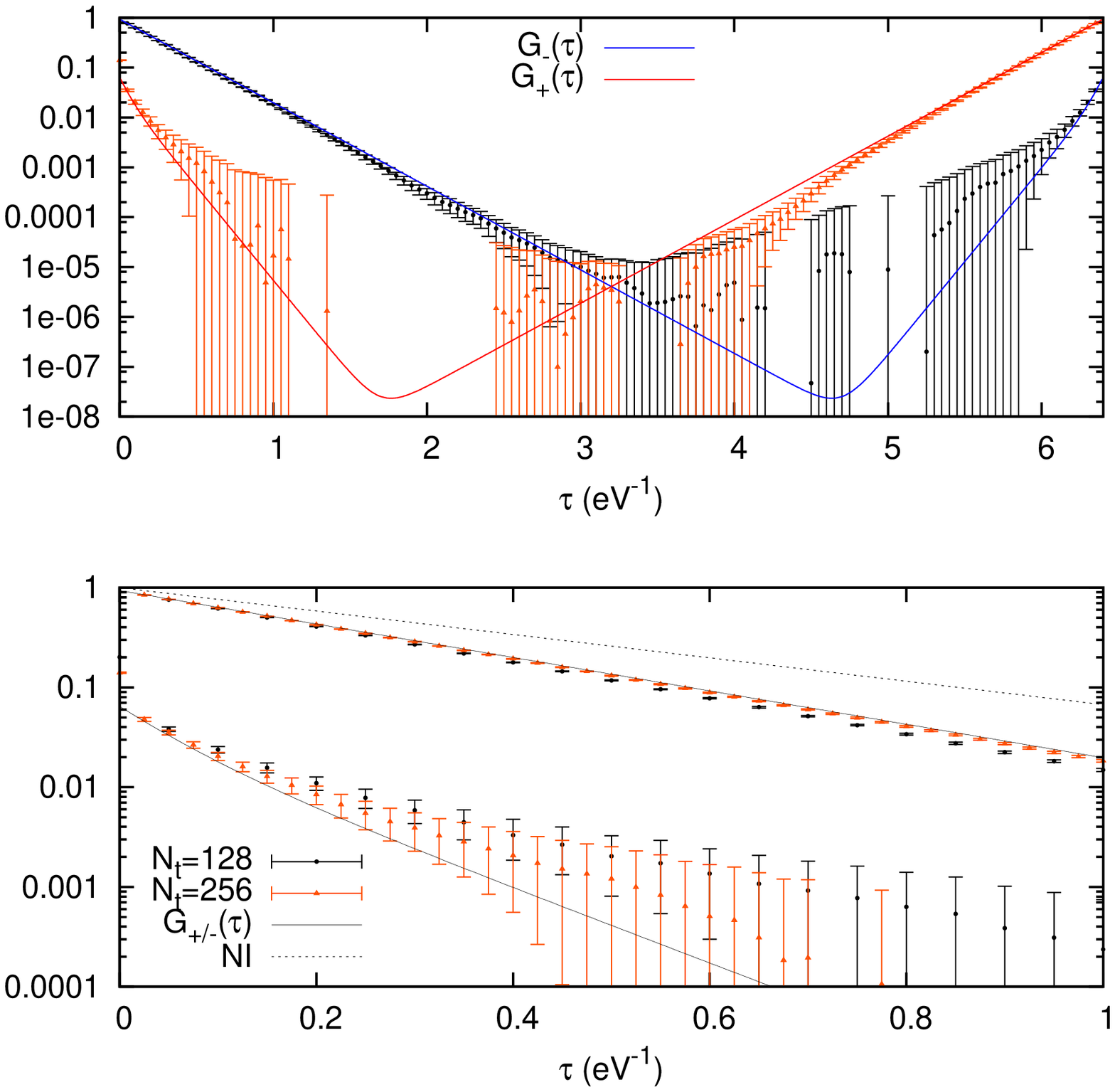}
\caption{4-site Hubbard calculation of $\vec{k}_2$ momentum correlators with $U_{00}=9.3$ eV and $\beta=6.4$ eV$^{-1}$.  Bottom plot is a close up and shows the non-interacting results (dashed line).\label{fig:4siteU00k2}}
\end{figure}

The 4-site Hubbard model is equivalent to the (1$\times$2) graphene lattice.  There are two unit cells in this case, giving 4 ion positions in total.  The Hamiltonian is the same as in Eqn.~\eqref{eqn:hubbard hamiltonian}, however construction of the correlators is a little trickier since there are now two allowed momenta within the first BZ,
\begin{displaymath}
\vec{k}_1=(0,0)\quad,\quad\vec{k}_2=\left(\frac{\pi}{3a},-\frac{\pi}{\sqrt{3}a}\right)\ ,
\end{displaymath}
Momentum projection on $G_{\pm}$ is given by
\begin{multline}
G_{\pm}(\vec{k}_i,;\tau)=\frac{1}{4}\frac{1}{\mathcal{Z}}\times\\
\sum_{l,m=1}^2e^{i\vec{k}_i\cdot(\vec{x}_l-\vec{x}_m)}
\left[\langle c_l^A(\tau)c_m^{A\dag}(0)\rangle+\langle c_l^B(\tau)c_m^{B\dag}(0)\rangle
\pm\left(\langle c_l^A(\tau)c_m^{B\dag}(0)\rangle+\langle c_l^B(\tau)c_m^{A\dag}(0)\rangle\right)\right]
\end{multline}
where the sum is over the unit cells (not ion sites).  In figs.~\ref{fig:4siteU00k1} and ~\ref{fig:4siteU00k2} we show calculations compared to exact results (determined via diagonalization) for the different momentum projections, using $U_{00}=9.3$ eV and $\beta=6.4$ eV$^{-1}$.  Calculations were done with $N_t$=128 and 256, and shows definitive convergence.   Also shown are the non-interacting (NI) solutions.

In addition to onsite interactions, we have also benchmarked our codes to 2- and 4-site systems with onsite, nearest-neighbor, and next-to-nearest neighbor interactions.  Though we do not show results of these studies here, we find our code gives equally good agreement with analytic and direct diagonalization methods.  We note that for systems that have only onsite $U_{00}$ and nearest neighbor $U_{01}$ interactions, our Monte Carlo code fails due to instability of the Hubbard transformation.

\end{appendices}

\clearpage
%\bibliography{referencesv2}


\end{document}